\documentclass[a4paper,12pt]{article}
\usepackage[utf8]{in
    putenc}
\usepackage{cancel}
\usepackage{ulem}
\usepackage{amsfonts}
\usepackage{amssymb}
\usepackage{graphicx}
\usepackage{amsmath,bm}
\usepackage{enumerate}
\usepackage{mathtools}
\usepackage{tikz} \usetikzlibrary{calc}
\usepackage[countmax]{subfloat}
\usepackage{xcolor}
\usepackage{float}
\usepackage{color}
\usepackage{here}
\usepackage{cite}
\usepackage{subcaption}
\usepackage{mwe}
\usepackage[outline]{contour}
\usepackage{tabularx}
\usepackage{multirow}
\usepackage{mathrsfs}
\usepackage{float,epsfig}
\usepackage{dcolumn}
\usepackage{pgfplots}
\pgfplotsset{compat=1.18}
\usepackage{graphicx}
\usepackage{bm}
\usepackage{amsmath,amssymb,amsthm}
\usepackage[colorlinks=true,linkcolor=blue,citecolor=red]{hyperref}
\definecolor{lime}{HTML}{A6CE39}
\newcommand{\orcidicon}{%
    \begin{tikzpicture}
    \draw[lime, fill=lime] (0,0)
        circle [radius=0.16]
        node[white] {{\fontfamily{qag}\selectfont \tiny ID}};
    \draw[white, fill=white] (-0.0625,0.095)
        circle [radius=0.007];
    \end{tikzpicture}   \hspace{-2mm}
}

\newcommand\orcidHasan{{\href{https://orcid.org/0000-0001-7408-0910}{\orcidicon}}}
\newcommand\orcidKarima{{\href{https://orcid.org/0000-0001-5419-8516}{\orcidicon}}}
\newcommand\orcidFaical{{\href{https://orcid.org/0000-0002-2977-0821}{\orcidicon}}}

\title{\bf Rényi Topology of Charged-flat Black Hole: Hawking-Page and Van-der-Waals Phase Transitions}

\author{
F. Barzi\orcidFaical\!\!$^{1,3}$\thanks{faical.barzi@edu.uiz.ac.ma},  
 H.  El Moumni\orcidHasan\!\!$^1$\thanks{h.elmoumni@uiz.ac.ma (Corresponding author)}, K. Masmar\orcidKarima\!\!$^2$\thanks{karima.masmar@gmail.com}
\\
{\small $^{1}$ LPTHE, Physics Department, Faculty of Sciences, Ibnou Zohr University, Agadir, Morocco. }\\
{\small $^{2}$Laboratory of  High Energy Physics and Condensed Matter
HASSAN II University,}\\{\small Faculty of Sciences Ain Chock, Casablanca, Morocco.}\\
{\small $^{3}$CRMEF, Regional Center for Education and Training Professions Marrakesh, Morocco.
}
}

\date{\today}
\begin{document} 
\maketitle
\begin{abstract}
In this paper, we extend the proposed setup in \cite{Yerra:2022coh,Wei:2022dzw} for finding the topological charges associated with the Hawking-Page and Van-der-Waals transition points as well as equilibrium phases to catch the nonextensive nature of the black hole entropy,  Rigorously speaking we incorporate the Rényi statistics formalism in off-shell Bragg-Williams free energy landscape to examine topologically the Hawking-Page phase transition related to the uncharged/charged-flat black hole in grand canonical and the Van-der-Waals transition in the canonical ensemble and where a vortex/anti-vortex structure is found. For this purpose, we introduce three mappings, the $\psi$- and $\xi$-mapping, for phase transitions classification and the $\eta$-mapping for equilibrium phases classification. We found that Hawking-Page and Van-der-Waals phase transitions belong to different topological classes and exhibit an interplay of total charge values hinting to a possible new correspondence.

Our topological study provides further substantiation for a possible conjecture positing a correspondence between the thermodynamic characteristics of black holes in asymptotically flat spacetime using Rényi statistics, and those in asymptotically Anti-de-Sitter spacetime employing Gibbs-Boltzmann statistics.
{\noindent}
\end{abstract}

\tableofcontents

\section{Introduction}

Black hole thermodynamics is an intriguing subject that investigates the relationships between black holes and the standard thermodynamic concepts \cite{Hawking:1976de,Hawking:1982dh,Bardeen:1973gs}. Jacob Bekenstein and Stephen Hawking's pioneering works \cite{Bekenstein:1973ur,Hawking:1975vcx} laid the groundwork for our understanding of black hole entropy and its relationship to the event horizon area \cite{Bekenstein:1974ax,Srednicki:1993im}. The Bekenstein-Hawking entropy raises fascinating questions about the fundamental nature of entropy and its connection to a black hole's underlying microscopic degrees of freedom. It implies that black holes have an enormous number of microstates that give rise to the macroscopic properties observed.

The non-extensive nature of black hole entropy is a peculiar aspect. In fact, extensive quantities in standard thermodynamics scale linearly with system size or volume. However, black hole entropy defies this rule. A black hole's entropy is proportional to its surface area, not its volume. This indicates a departure from the standard thermodynamic framework and suggests a more complex underlying theory. Several approaches have been proposed to explain the non-extensive nature of black hole entropy, including string theory \cite{Crossley:2014oea,DeHaro:2019gno,Tsallis:2019giw}, loop quantum gravity \cite{Alonso-Serrano:2020hpb,Headrick:2015gba}, and holography \cite{Hung:2011nu,Giveon:2015cgs,Dong:2016fnf}. These theories seek to reconcile black hole thermodynamics and quantum mechanics, as well as to provide a microscopic understanding of black hole entropy. They propose that a black hole's entropy arises from the entanglement of quantum states associated with its horizon.
Indeed, the traditional Gibbs-Boltzmann approach, which assumes the extensive scaling of entropy with the system's size, may not be directly applicable to self-gravitating systems like black holes. This discrepancy arises due to the non-local and non-extensive nature of black holes and is not relevant in their vicinity where long-range interactions like gravity in the strong scheme stem.

Exploring the non-extensive nature of black hole entropy is an active area of research, bridging concepts from general relativity, thermodynamics, and quantum mechanics. It offers valuable insights into the fundamental nature of space, time, and information \cite{Hirunsirisawat:2022fsb,Promsiri:2021hhv,Nojiri:2021czz,Barzi:2022ygr,Cimdiker:2022ics,Barzi:2023mit,Cimidiker:2023kle,Chunaksorn:2022whl,Samart:2020klx}.

Another avenue of exploration involves considering the black hole phase transition  from the notion of the extended phase space point of view,   where the cosmological constant is treated as a variable, black holes can undergo phase transitions analogous to those observed in ordinary thermodynamic systems \cite{Gunasekaran:2012dq,Kubiznak:2012wp,Belhaj:2015hha,Kubiznak:2016qmn,Kubiznak:2014zwa}  including first-order phase transitions, second-order phase transitions, supercritical structure, and reentrant phase transitions, depending on the values of the relevant parameters \cite{Wu:2022plw,Altamirano:2013ane,Frassino:2014pha}. It is worth noting that these investigations are made via various thermodynamic tools such as standard thermodynamics, geothermodynamics \cite{Bravetti:2012hd,Chabab:2019mlu,Bhattacharya:2017hfj}, quasinormal modes \cite{Liu:2014gvf,Chabab:2016cem}, and effective thermodynamics \cite{Simovic:2019zgb,Chabab:2020xwr,Ali:2020bgc}. Complementary tools like the free energy formalism \cite{Li:2020khm,Ali:2023wkq,Li:2020nsy}, heat capacity analysis, black hole chemistry \cite{Kubiznak:2016qmn,Kubiznak:2014zwa}, and holographic methods \cite{Nguyen:2015wfa,ElMoumni:2018fml,Li:2018aax} further contribute to understanding black hole thermodynamics, phase transitions, and stability criteria.

Topology serves as a highly effective means to analyze the intrinsic characteristics of physical systems. Its application in particle physics and condensed matter research has been extensive and longstanding. In the 1980s, Duan proposed a comprehensive theoretical framework to systematically investigate the behavior of topological current and charge in physical systems \cite{Duan:1984ws}. This pioneering theory yielded fascinating predictions into the dynamics of topological defects \cite{Duan:1998kw,Fu:2000pb,Guo:2020qwk,Bargueno:2022vkf}, elucidating phenomena such as defect collisions, coalescence, splitting, as well as the generation and annihilation of vortex/anti-vortex pairs \cite{}.

Lately, such a theory has found application in the examination of specific characteristics pertaining to gravitational entities. For instance, it has been employed to investigate properties like light rings \cite{Cunha:2017qtt,Cunha:2020azh} and the thermodynamic behavior of black holes \cite{Fan:2022bsq,Fang:2022rsb,Wei:2021vdx,Yerra:2022alz,Ahmed:2022kyv,Yerra:2022coh,Wei:2022dzw,Ali:2023zww,Sadeghi:2023aii,Gogoi:2023qku,Zhang:2023uay,Wu:2023sue,Wu:2022whe,Bai:2022klw,Liu:2022aqt,Chatzifotis2023}.
Nevertheless, {\it the topology of black hole phase transitions in non-Boltzmaniann statistics has not yet been  studied}.
Henceforth, proposing an investigation of the phase portrait of charged-flat black holes within Rényi statistics by incorporating the topological concept aims to uncover the fundamental basis of the non-extensive nature of black hole entropy and the deep connection between  the cosmological constant and the Rényi non-extensive parameter.
Such non-Boltzmaninan alternative approaches go beyond the traditional extensive GB formalism and provide a framework that can account for the non-local and non-extensive information of black hole entropy. By incorporating these characteristics and recalling the off-shell free energy topological tool, we gain deeper insights into the black hole phase structures.

The Hawking-Page phase transition \cite{Hawking:1982dh} is a significant phenomenon occurring in asymptotically AdS space. By considering the black hole as a state within the thermodynamic ensemble, two stable thermodynamic phases are observed: the thermal AdS space phase and the large Schwarzschild-AdS black hole phase.
At a specific temperature, a first-order phase transition has been identified between the thermal AdS space and the large Schwarzschild-AdS black hole. In AdS/CFT conjecture framework \cite{Maldacena:1997re,Witten:1998qj}, the Hawking-Page transition can be aptly interpreted as the confinement/deconfinement transition in quantum chromodynamics \cite{Witten:1998zw}. This pioneer work by Hawking and Page has also been extended to incorporate other gravity theories \cite{Cai:2007wz,Gursoy:2010jh,Zhang:2015wna,Adams:2014vza,Banados:2016hze,Su:2019gby}, while its kinetics  based on the underlying of  free energy landscape investigated   has been subject of investigations in \cite{Li:2020khm} through the Fokker-Planck equation and its topological charge evaluated in \cite{Yerra:2022coh}.
 Moreover, in the context of four-dimensional charged black holes in asymptotically flat spacetime, under Rényi statistics, it is observed that the charged black hole can achieve thermodynamic equilibrium with the surrounding thermal radiation. Interestingly, similar to the case of AdS-charged black holes, these systems also exhibit a Hawking-Page phase transition \cite{Promsiri:2020jga,dilaton}.

Moreover, the strong similitude between black holes and the Van-der-Waals $(VdW)$ systems becomes prominent in the context of charged-AdS black holes in the canonical ensemble\cite{Kubiznak:2012wp,Majhi:2016txt,Li:2014ixn,Bhattacharya:2017hfj}, where a rich phase structure reminiscent of the $(VdW)$ fluid manifests in black hole backgrounds. Concretely, by interpreting the cosmological constant as the thermodynamic pressure in the extended phase space, variations in the charge of a black hole and the cosmological constant can induce phase transitions analogous to those occurring in the Van-der-Waals $(VdW)$ systems. In a most intriguing manner, asymptotically flat black holes via Rényi thermodynamics exhibit also a $(VdW)$ phase structure with the emergence of small $(SBH)$, intermediate $(IBH)$ and large $(LBH)$ black hole phases, conjointly with first and second order phase transitions\cite{Promsiri:2020jga,Barzi:2023mit,DEMAMI2023116316}.

With these underlying motivations, our endeavor in this study is to make a meaningful contribution to these activities. Specifically, we undertake an exploration of the topological aspects of the Hawking-Page and the Van-der-Waals phase transitions within the framework of Rényi statistics. Our goal is to unveil the impact of the nonextensive nature of the black hole entropy on the topological charges associated with each black hole phase and consolidate the bridge linking the nonextensive parameter to the cosmological constant.


To enhance clarity, we predefine the methodology utilized in our investigation. Within the grand canonical ensemble, the governing parameters encompass the scaled potential $\phi$ and the coexistence temperature $t$. In order to classify phase transitions that are {\it temperature-independent}, representing global phase transitions, and to determine their associated equilibrium phases, we employ two mappings denoted as the $\psi$-mapping and the $\eta$-mapping. These mappings are functions of $\phi$ and $t$. We first conduct this analysis for the uncharged case ($\phi=0$) and subsequently for the charged case ($\phi\neq 0$).
Shifting to the canonical ensemble, in addition to the previously mentioned mappings, which now depend on the scaled electric charge $Q$ and temperature $t$, we introduce an additional mapping known as the $\xi$-mapping. Its purpose, as will become evident later, is to unveil the topological structure of {\it temperature-dependent} phase transitions, i.e., those phase transitions that become apparent at a specific coexistence temperature within the black hole system. This analysis is performed across three regimes: the subcritical regime ($Q<Q_c$), the critical regime ($Q=Q_c$), and the supercritical regime ($Q>Q_c$).
Within each of these regimes, we explore and classify the transition phase structure using the $\psi$-mapping and the $\xi$-mapping, followed by an examination of the equilibrium phase profiles through the $\eta$-mapping. By manipulating the control parameters, namely ($\phi$, $t$) in the grand canonical ensemble and ($Q$, $t$) in the canonical ensemble, we provide a two-dimensional classification. This classification is achieved vertically by varying the electric charge and potential and horizontally by altering the scaled coexistence temperature.

This paper is organized as follows. In Section 2, we present a concise overview of the thermodynamics that underlies flat-charged black holes when analyzed through  Rényi statistics. Moving into Section 3, we take the initiative to extend the Bragg-Williams formalism into the realm of Rényi statistics. In doing so, we proceed to evaluate the topological charge associated with both uncharged and charged black hole phases, all within the grand canonical ensemble framework.
The subsequent section, Section 4, is dedicated to a comprehensive topological examination of the Van-der-Waals phases as they manifest within the canonical ensemble. 
As our investigation draws to a close, Section 5 presents a succinct conclusion where we summarize our findings and provide insightful concluding remarks.
Lastly, the appendix offers essential details regarding the key quantities of the Rényi-Bragg-Williams formalism, specifically tailored to charged black holes existing within an asymptotically flat spacetime of arbitrary dimensions.



\section{Review of Rényi thermodynamics of a charged-flat black hole in four dimensions.}

%
As introduced in \cite{PhysRevE.83.061147}, the Rényi entropy $S_R$ is logarithmically related to the Bekenstein-Hawking entropy $S_{BH}$  as follows:
\begin{equation}
S_R=\frac{1}{\lambda}\ln(1+\lambda S_{BH}). \label{bh17}
\end{equation}
Where the $\lambda$  denotes the non-extensive parameter.
It is important to note that when $\lambda$ approaches zero, the standard Gibbs-Boltzmann statistics, $\big(S_R\underset{\lambda\rightarrow0}{\longrightarrow}S_{BH}\big)$, is recovered. 
 The Rényi entropy generalizes the Gibbs-Boltzmann entropy by considering non-extensive effects, resulting in a first-order correction in $\lambda$ to the temperature of the Reissner-Nordstrom-flat black hole. The Rényi temperature $T_R$ can be expressed as \cite{Promsiri:2020jga},
\begin{eqnarray}
T_R = \frac{1}{\partial{S_R/\partial{M}}} &=& T_H(1+\lambda S_{BH})\label{Tr}\\ 
&=& \frac{(r_h^{2}-Q^2)(1+ \lambda \pi r_h^{2})}{4\pi r_h^{3}}, \label{bh25}\\
&=&\displaystyle  \frac{ \left(\pi \lambda r_{h}^{2} + 1\right)\left( 1-\phi^{2}\right)}{4 \pi r_{h}}. \label{T_R}
\end{eqnarray}
where $	M = \frac{r_h}{2}\left( 1 + \frac{Q^2}{r_h^2} \right)$, $T_H= \frac{r_h^{2}-Q^2}{4\pi r_h^{3}}$ and $S_{BH}=\pi r_h^{2}$ are the mass the Hawking temperature and the Bekenstein-Hawking entropy of RN-flat black hole, respectively.  While $r_h$ is nothing then the black hole event horizon radius, 
$Q$ stands for its charge and the electric potential is $\phi = \frac{Q}{r_{h}}$. 
Combining all the relevant quantities, we can present the first law of Rényi thermodynamics along with its corresponding Smarr formula as follows:
\begin{eqnarray}
dM = T_RdS_R  + \phi dQ, \qquad\qquad 
M = 2T_RS_R + \phi Q. \label{Smarr_rényi_mod}
\end{eqnarray}

Having explored some essential thermodynamic quantities of the RN-flat black hole within the framework of Rényi statistics, our attention now shifts to the Bragg-Williams setup. This setup allows us to uncover the non-Boltzmannian topological features of its landscape, both in the grand canonical ensemble and the canonical ensemble. 

\section{Rényi topological formalism in  the grand canonical ensemble: Hawking-Page phase transition.}
\subsection{Definition of the topological charge.}

As in Ref.\cite{Yerra:2022coh}, the starting point is defining the so-called Bragg-Williams free energy as an off-shell one \cite{Wei:2022dzw}  extended to Rényi framework such as,

\begin{align}
\bar{f_R}(r,t,\phi)&=M-tS_R-\phi Q, \\
&=\displaystyle  \frac{r}{2} - \frac{\phi^{2} r}{2}  - \frac{t \log{\left(\pi \lambda r^{2} + 1 \right)}}{\lambda}.\label{f_R_4}
\end{align}

\begin{figure}[!ht]
	\centering
	\begin{tabbing}	
		\centering
		\hspace{-1.6cm}
		\includegraphics[scale=.52]{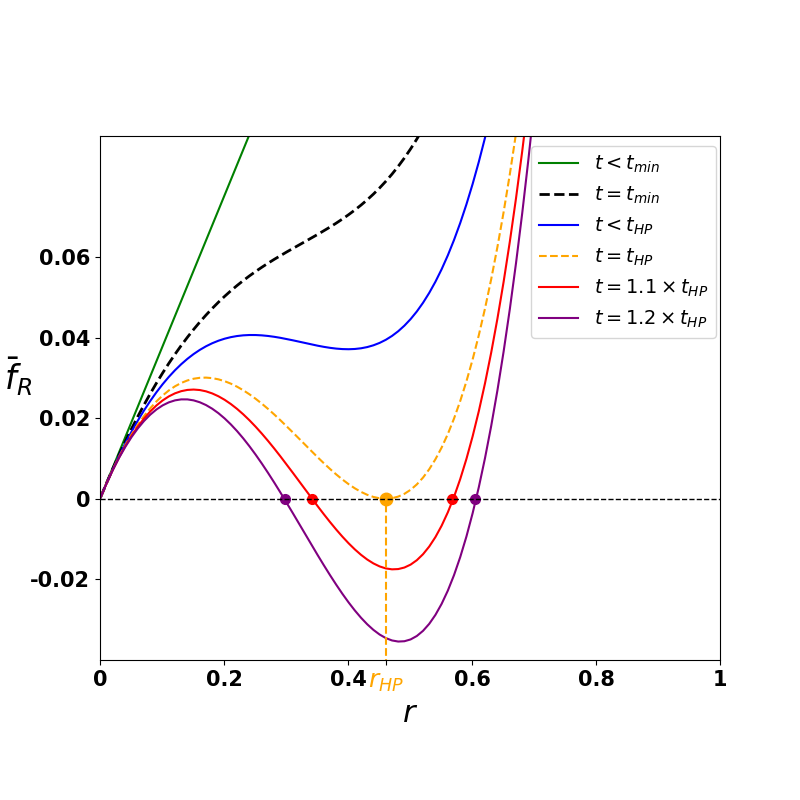}
  		\hspace{-.3cm}
		\includegraphics[scale=.5]{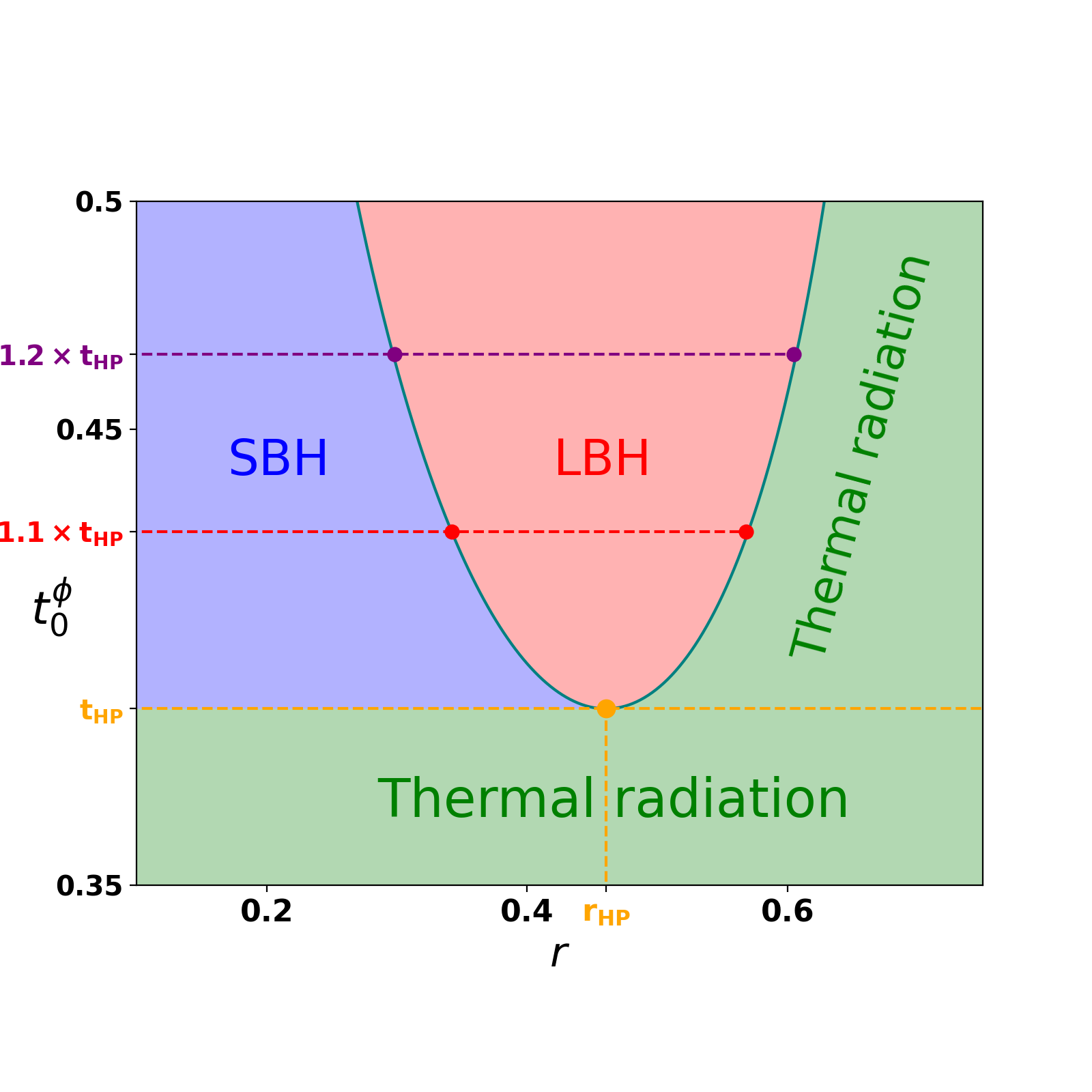}
	\end{tabbing}
	\vspace{-2.cm}
	\caption{\footnotesize\it {\bf Left:} Behavior of the Bragg-Williams free energy $ \bar{f}_R $, as a function of $r$ at different temperatures $ t $. {\bf Right:} The coexistence temperature $t_0^\phi$ of the black hole and thermal phases, as a function of $r$, shows the HP transition point at its minimum (orange dot). Other colored
		dots correspond to respective colored curves in the left panel. The potential $\phi$ is set to $0.5$}\label{fig1}
\end{figure}

Where $r$, $t$, and $\phi$ are treated as free parameters that only coincide with physical parameters $T_R$, \textcolor{red}{$\phi_h$}, and $r_h$, respectively, for thermodynamic equilibrium solutions. For $0<\lambda<<1$, and scaling $r \rightarrow r/\sqrt{\lambda}$, $t\rightarrow \sqrt{\lambda} t$ and $ \bar{f_R}\rightarrow  \bar{f_R}/\sqrt{\lambda}$, the Bragg-Williams free energy without the nonextensive parameter dependency is turned to be
\begin{equation}\label{f_R}
\bar{f_R}(r,t,\phi)=\displaystyle \frac{r \left(1- \phi^{2} - 2 \pi r t + \pi^{2} r^{3} t \right)}{2}.
\end{equation}
 In the left panel of Fig.\ref{fig1}, we depict the behavior of the free energy $\bar{f_R}(r,t,\phi)$ concerning the order parameter $r$ for various values of the scaled Rényi temperature $t$, while the electric potential is fixed at $\phi=0.5$. It's commonly known that the equilibrium states are reached at the minima and maxima of the function $\bar{f_R}(r,t,\phi)$. We generally associate the minima to stable or metastable states/phases, whereas the maxima are unstable states. In what follows and through out this paper a great deal of attention is given to these extrema.

The flat space phase ($r = 0$) acts as the zero point for the free energy. When the temperature reaches a critical value $t_{HP}$, the black hole phase becomes globally unstable, as its free energy surpasses that of the Minkowski phase. However, at temperatures above $t_{HP}$, the black hole phase becomes globally stable, possessing lower free energy compared to the Minkowski phase. As a result, a first-order phase transition, known as the Hawking-Page (HP) transition \cite{Promsiri:2020jga}, occurs at $t_{\text{HP}}$, wherein the preferred phase passes from flat space background to black holes, in a similar manner to what occurs in the black holes in the AdS background via Gibbs-Boltzmann statistics formalism. The determination of this transition point involves satisfying two simultaneous conditions $\bar{f_R}=0$ and  $\partial_{r}\bar{f_R}=0$.

The  first condition, $\bar{f_R}=0$, gives the coexistence temperature between the Minkowski and black holes phases,
\begin{equation}\label{t0_grand_canonical}
\bar{f_R}(r,t^\phi_0,\phi)=0 \implies t^\phi_0=\displaystyle \frac{r \left(1 - \phi^{2}\right)}{2 \log{\left(\pi  r^{2} + 1 \right)}}\approx\displaystyle \frac{\phi^{2} - 1}{\pi r \left(\pi r^{2} - 2\right)}.
\end{equation}

We highlight the significance of the coexistence temperature, denoting it as the temperature at which a phase transition occurs, satisfying the requisite condition $\Bar{f}_R=0$. The Hawking-Page phase transition, representing the initial possible transition from the pure thermal radiation phase in Minkowski spacetime to the black hole phase, corresponds to the smallest achievable coexistence temperature. It is important to distinguish the coexistence temperature from the minimal black hole one, which is the lowest temperature attainable by an already formed black hole in the grand canonical ensemble; notably, at this temperature, the free energy does not vanish. Determining this minimal temperature involves minimizing the thermodynamic temperature of the black hole rather than the coexistence temperature. In our analysis, we have plotted the free energy for both $t=t_{min}$ (indicated by the black dashed line) and $t=t_{HP}$ (shown as the orange dashed line) in the left panel. An essential observation emerges as we increase the temperature from $t_{min}$ to approach $t_{HP}$: local extrema of $\Bar{f}_R$ emerge, signifying the onset of black hole phases or states. However, these states remain largely dominated by the more stable pure radiation phase. It is only upon crossing the threshold at $t_{HP}$, as depicted in the right panel of Fig.\ref{fig1}, that the black hole phase begins to dominate. In addition, the right panel of the Fig.\ref{fig1} shows the phase profile of the charged black hole in the grand canonical. As can be seen two coexistence curves which delimits the three phase regions, Namely, (LBH), (SBH) and thermal radiation. The positive slope curve marks at each temperature, the maximum stable black hole's radius beyond which a black hole cannot possibly exist in thermal equilibrium. On the other hand the negative slop curve delineates the coexistence between the large stable and small unstable black hole phases.

By minimizing the coexistence temperature $t_0^\phi$, we found, the Hawking-Page temperature $t_{HP}^\phi$ and its associated event horizon radius  $r_{HP}^\phi$,
\begin{equation}\label{r_hp}
\frac{\partial t^\phi_0}{\partial r}=0 \implies r^\phi_{HP}=\displaystyle \frac{\sqrt{6}}{3 \sqrt{\pi}},
\end{equation}
\begin{equation}\label{t_hp}
t^\phi_{HP}=\displaystyle \frac{3 \sqrt{6} \cdot \left(1 - \phi^{2}\right)}{8 \sqrt{\pi}}.
\end{equation}
We remark that the $r_{HP}^\phi$, is independent of $\phi$, thus it remains the same for the uncharged and charged cases. We also compute the scaled minimal temperature by minimizing and scaling the Rényi black hole temperature $T_R$, Eq.\eqref{T_R}.

\begin{equation}\label{t_min}
r^\phi_{min}=\displaystyle \frac{1}{\sqrt{\pi}},\quad t^\phi_{min}=\displaystyle \frac{1- \phi^{2}}{2 \sqrt{\pi}}.
\end{equation}


In the forthcoming steps, we extend  the topological setup to determine the topological charges associated with the Hawking-Page transition point as well as with the equilibrium phases for the charged-flat black hole in the grand canonical ensemble to the Rényi framework. To this end, we introduce two scalar field scalar fields or mappings. First, the scalar field $\Psi_\phi$ as in \cite{Wei:2021vdx}
\begin{equation}\label{Psi_t_0_grand_canonical}
\Psi_\phi(r,\theta)=\frac{1}{\sin\theta} t^\phi_0=\displaystyle \frac{\phi^{2} - 1}{\pi r \left(\pi r^{2} - 2\right) \sin{\left(\theta \right)}},
\end{equation}
and its gradient components
\begin{equation}\label{topological_field_grand_canonical}
\begin{cases}
\psi^\phi_r=\partial_r\Psi_\phi=\displaystyle  \frac{\left(1-\phi^{2}\right) \left(3 \pi r^{2} - 2\right)}{\pi r^{2} \left(\pi r^{2} - 2\right)^{2} \sin{\left(\theta \right)}},\\
\psi^\phi_{\theta}=\partial_{\theta}\Psi_\phi=\displaystyle \frac{\left(1 - \phi^{2}\right) \cos{\left(\theta \right)}}{\pi r \left(\pi r^{2} - 2\right) \sin^{2}{\left(\theta \right)}}.
\end{cases}
\end{equation}
The vector field $\psi^\phi$ permits the visualization and computation of the topological charge of the Hawking-Page transition. Second, employing  Bragg-Williams free energy constructed on Rényi statistics, i.e. Eq.(\ref{f_R_4}), we can define the mapping $\eta^\phi=(\eta^\phi_r,\eta^\phi_\theta)$ as,

\begin{equation}\label{xi_phi}
\begin{cases}
\eta^\phi_{r}=\displaystyle\partial_r\bar{f}_R=\displaystyle \frac{ \left(\pi r^{2}+1\right) \left(1-\phi^{2} \right) - 4 \pi r t }{2 \left(\pi r^{2} + 1\right)}\\
\eta^\phi_{\theta}=\displaystyle-\cot\theta \cos\theta,
\end{cases}
\end{equation}

Which is suitable for associating topological charges to equilibrium phases of Rényi charged-flat black holes in the grand canonical ensemble. This choice comes from the property that the local extrema of a system's free energy indicates the stable and unstable thermodynamic phases.

The existence of a topological current, satisfying the usual conservation law $\partial_\mu j^\mu=0$, is a direct consequence of the definition of the vector fields $\psi^\phi$ and $\eta^\phi$ (and also of $\xi$ which will be defined later)\cite{Wei:2022dzw,Ali:2023zww} at the coexistence temperature $t^\phi_0$ and Bragg-Williams free energy $\Bar{f}_R$, as given by Eqs. (\ref{Psi_t_0_grand_canonical}), (\ref{topological_field_grand_canonical}), \eqref{f_R_4} and \eqref{xi_phi}, respectively. Notably, these topological currents are found to be zero everywhere except at the zeroes of the $\psi^\phi$  and $\eta^\phi$ mappings. The formal definition of the topological charge within a given region $(\Sigma)$ follows from the above-mentioned formalism and is given as,
\begin{equation}
\displaystyle\mathcal{Q}_{t}=\int_{{\Sigma}} j^0 d\Sigma=\sum_{k} w_k.
\end{equation}
Here $w_k$ represents the winding number of the $k$-th point, corresponding to the $k$-th zero of $\psi^\phi$ or  $\eta^\phi$, which precisely corresponds to special points in the phase profile of the thermodynamic system under consideration\footnote{With a special emphasis given to the scenario where these vector fields precisely vanish at the Hawking-Page transition and equilibrium phases points.}. As $\Sigma$ sweeps the available parameter space of the thermodynamic system, the critical points, or a subset of them, can belong to different topological classes. This is evident from the fact that $\mathcal{Q}_{t}$ can take positive, negative, or null values, and it is also possible for the total topological charge to be zero. By the above construction, all critical points are positioned at $\theta=\frac{\pi}{2}$ as can be seen from the $(\cos(\theta))$ dependency of the components $\psi^\phi_\theta$ and $\eta^\phi_\theta$, Eq.\eqref{topological_field_grand_canonical} and \eqref{xi_phi}, respectively.

In order to compute these topological quantities, we adopt the following parametrization for a positively (anti-clockwise) oriented contour $\mathcal{C}$ around a given point $(r_0,\frac{\pi}{2})$ in the $r-\theta$ plane,
\begin{equation}
\begin{cases}
r=r_0+a_1\cos\alpha \\
\theta= \frac{\pi}{2}+a_2\sin\alpha .
\end{cases}
\end{equation}

Where $\alpha$ spans the interval $[0,2\pi[$ and $(a_1,a_2)$ are positive real numbers defining the contour $\mathcal{C}$. The deflection angle $\Omega(\alpha)$ of each vector field is evaluated to be
\begin{equation}
\Omega(\alpha)=\int_{0}^{\alpha} \left(n^r\partial_{\tilde{\alpha}}n^\theta-n^\theta\partial_{\tilde{\alpha}}n^r\right) d\tilde{\alpha}.
\end{equation}
Where $n$ is the normalized vector field computed from $\psi^\phi$ or  $\eta^\phi$. We note that a clockwise orientation of the contours would inverse the signs of topological charges, therefore only relative signs are meaningful. The integration of the deflection angle along an oriented contour $\mathcal{C}$, i.e. $\alpha=2\pi$, around a point or sets of points gives the topological charge such as,
\begin{equation}\label{k`ey}
\displaystyle\mathcal{Q}_{t}=\frac{\Omega(2\pi)}{2\pi}.
\end{equation}

Through the mappings generated from the vector fields $\psi^\phi$ and $\eta^\phi$, we aim at a topological classification of the phase transitions and equilibrium phases of the asymptotically flat uncharged (Sch) and charged (RN) black holes.

After establishing the topological charge setup related to the Hawking-Page transition in the grand canonical ensemble within Rényi statistics, we will proceed to apply it to specific examples of flat black holes.
\subsection{Uncharged-flat black hole.}
\begin{figure}[!ht]
	\centering
	\begin{tabbing}
		\centering
		\hspace{9.3cm}\=\kill
		\includegraphics[scale=.4]{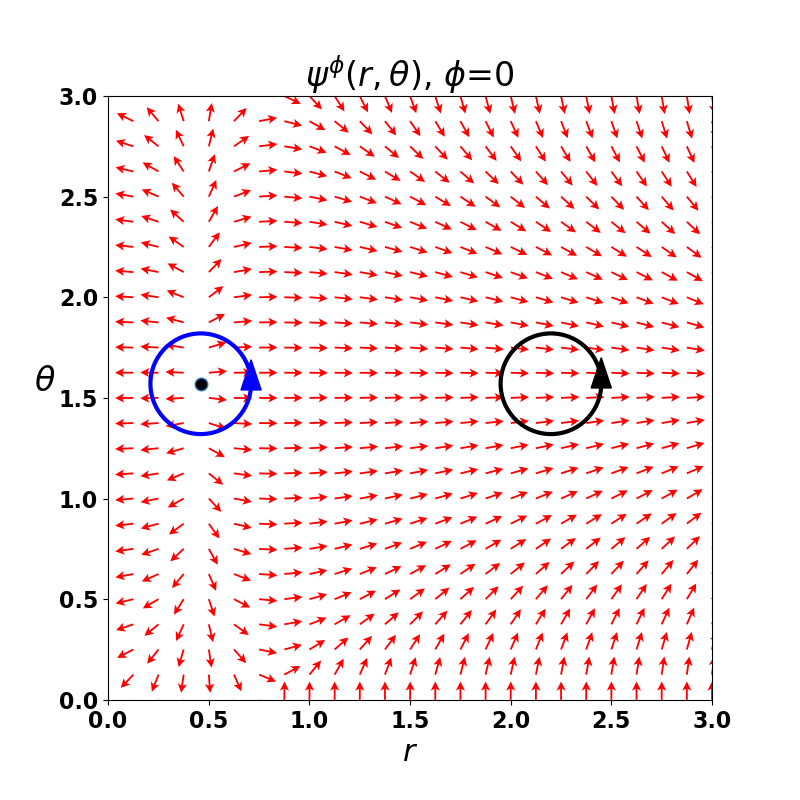}\>
		\includegraphics[scale=.4]{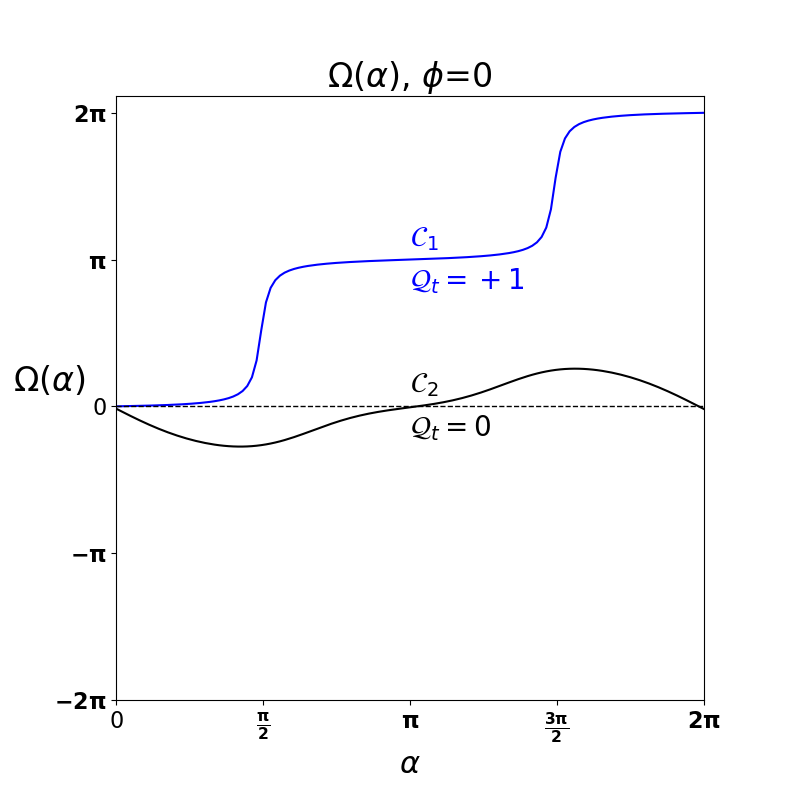}\\
	\end{tabbing}
	\vspace{-1.4cm}
	\caption{\footnotesize\it Topological charge of Hawking-Page phase transition for an uncharged-flat black hole in the grand canonical ensemble within Rényi formalism ($\phi=0$). {\bf Left:}  the flow of the normalized vector field  of $\psi^{\phi}$, in the $r-\theta$ plane.
	 {\bf Right:}  the deflection angle for the two contours on the left panel in terms of the angle $\alpha$.   }
	\label{fig2}
\end{figure}

We begin with the case of the uncharged-flat black hole in Rényi formalism ($\phi=0$).
In order to assign the topological charge to HP transition, we employ the temperature $t^0_0$, Eq.(\ref{t0_grand_canonical}), obtained from the Bragg-Williams free energy. Through the definition of the topological field $\psi^0=(\psi^0_r,\psi^0_\theta)$, Eq.(\ref{topological_field_grand_canonical}), with the substitution $\phi=0$. In Fig.\ref{fig2}, we illustrate  the normalized vector field $n$ in $r-\theta$ plane and the behaviour of the deflection $\Omega$ in terms of $\alpha$.

From the left panel, the flow of the normalized vector field  of $\psi^{0}$, in the $r-\theta$ plane shows that the Hawking-Page phase transition point (black dot)  is located at $r=r_{HP}=\frac{\sqrt{6}}{3\sqrt{\pi}}\approx 0.46065$, the blue contour contains the HP transition point while the black contour is void.  Their associate deflection angles lead to the topological charge of the Hawking-Page transition (blue contour)  $\mathcal{Q}_{t}=\mathbf{+1}$,  whereas the black void contour (black) has zero topological charge.

\begin{figure}[!ht]
	\centering
	\begin{tabbing}
		\centering
		\hspace{8.3cm}\=\kill
		\includegraphics[scale=.36]{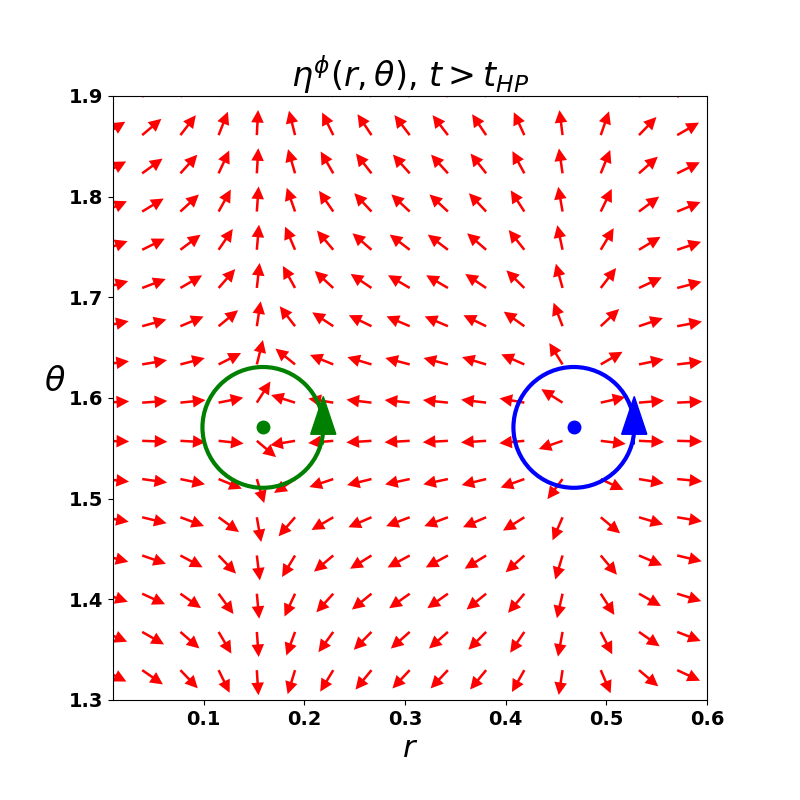}\>
		\includegraphics[scale=.36]{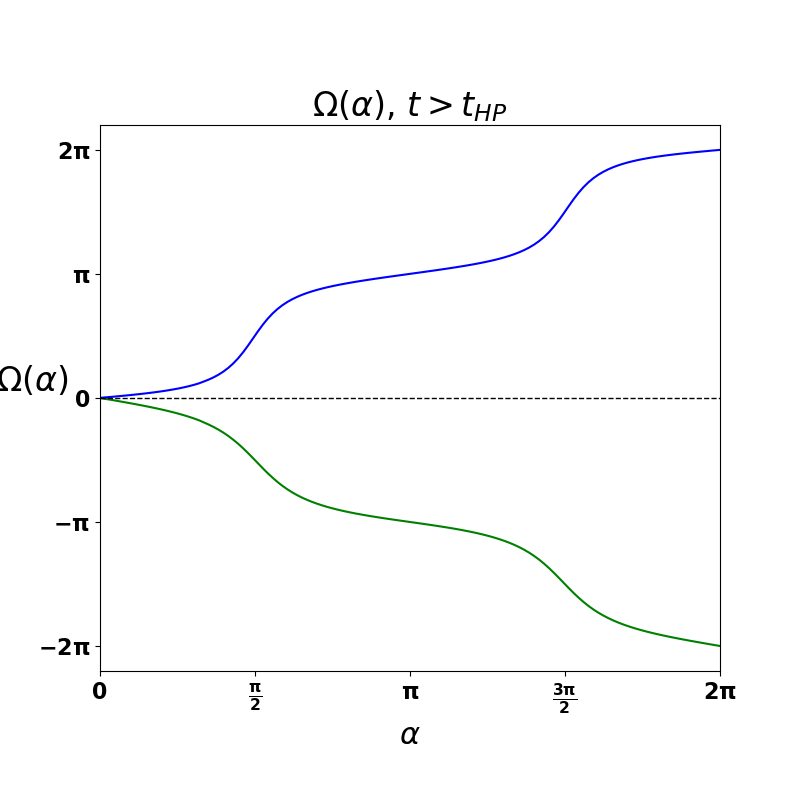}\\
		\includegraphics[scale=.36]{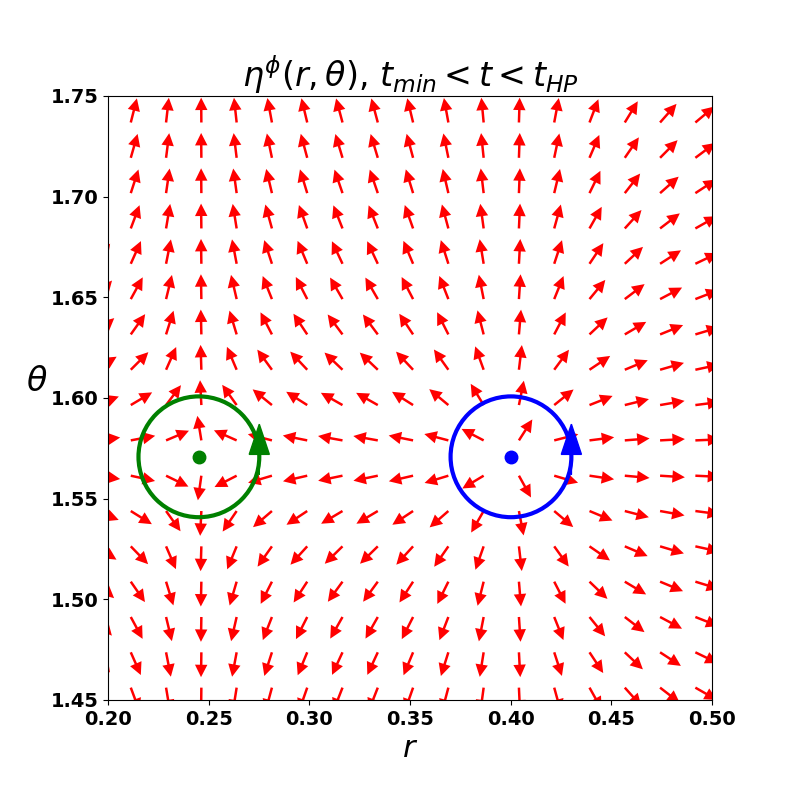}\>
		\includegraphics[scale=.36]{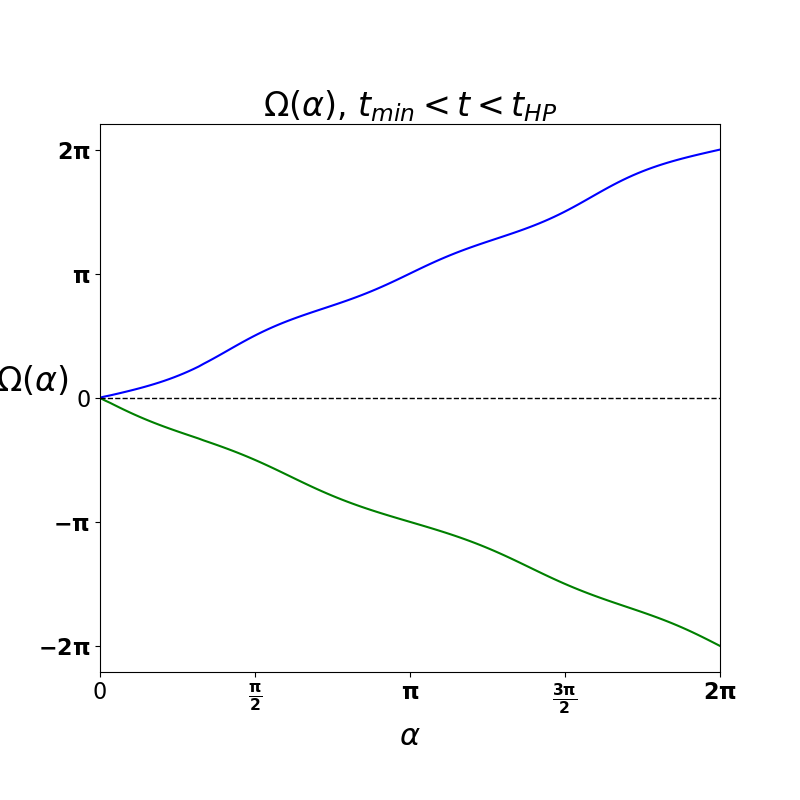}\\
		\includegraphics[scale=.36]{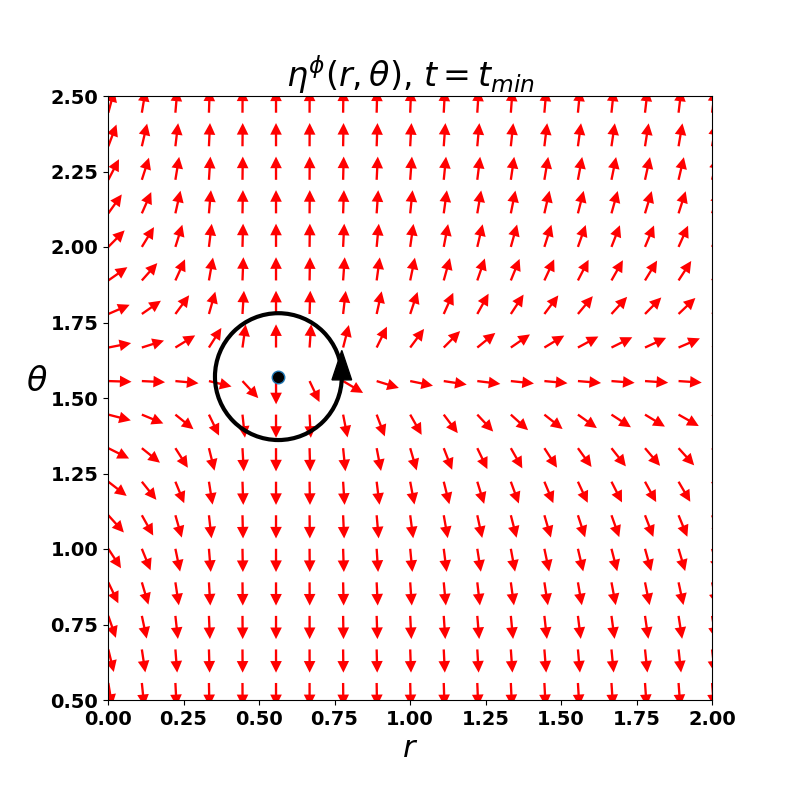}\>
		\includegraphics[scale=.36]{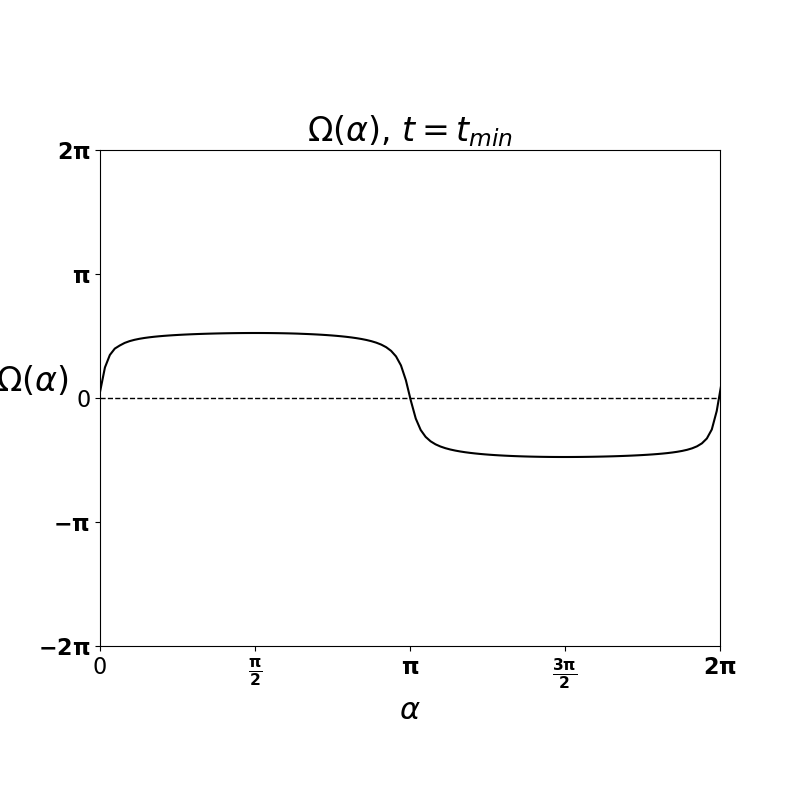}\\
	\end{tabbing}
	\vspace{-1.5cm}
	\caption{ \footnotesize\it Phases of the uncharged-flat black hole in Rényi formalism for different scaled temperatures $t$. {\bf Left:} the flow of the normalized field of $\eta^\phi$ at $\phi=0$. {\bf Right:} the behavior of
    deflection angle for corresponding black hole phases on the left panel.
    {\bf Top}, {\bf middle} and {\bf bottom} panels are associated with $t\geq t_{HP}$, $t_{min}<t<t_{HP}$, and $t=t_{min}$ respectively.}
	\label{fig3}
\end{figure}

Next, we move to evaluate the topological charge of equilibrium phases for the uncharged-flat black hole case. 
The expression of the topological field $\eta^0$, Eq.\eqref{xi_phi}, depends on the scaled Rényi temperature $t$. By changing the values of $t$, one is able to probe the topological phase structure of the uncharged-flat black hole in the regions above and below the Hawking-Page phase transition. For such a reason and by also taking $\phi=0$ we depict in Fig.\ref{fig3} the normalized vector field $n$ flow in $r-\theta$ plane and the variation of the deflection $\Omega$ in terms of the angle $\alpha$ for different Rényi temperature

This figure unveils several remarks:
\begin{itemize}
    \item Top panels associated with $t\geq t_{HP}$ situation show two black hole phases, small black hole (SBH) (the green dot at $r\approx0.15883$) and large black hole LBH (the blue dot at $r\approx0.46775$, it's also noticeable that the two-phase process  opposite topological charges. Namely, $\mathcal{Q}_{t}=\mathbf{+1}$ for LBH and $\mathcal{Q}_{t}=\mathbf{-1}$ for SBH.
    \item Middle panels corresponding to $t_{min}<t<t_{HP}$ case revel also the existence of two black hole phases with opposite topological charges, SBH (the green dot at $r\approx0.24514$) and LBH (the blue dot at $r\approx0.40015$).
    \item From the bottom panel, where the temperature reaches $t=t_{min}$, the situation is quite different, and just one black hole phase (black dot at $r\approx0.56418$) corresponding to the inflection point at the minimal location. In such a case the topological charge is evaluated to zero $\mathcal{Q}_{t}=\mathbf{0}$.
\end{itemize}

That is to say that the appearance of the small black hole (SBH) and the large black hole (LBH) phases as topological defects \cite{Wei:2022dzw} within the $\eta^0$ field is clear in the top and middle panels, whereas the emergence of the minimal temperature uncharged black hole is shown in the bottom one. In addition, the values of the topological charges at these points confirm on one hand the emergence of these phases, on the other hand, that the minimal temperature black hole presents a vanishing topological charge while (SBH) and (LBH) indicate opposite values whatever the convention chosen to orient the contours. This difference in topological charges (null/ no null) consolidates the previous discussion on the Hawking page temperature and the minimum temperature.

Having topologically probed the uncharged black hole solution, we put our emphasis on the charged one, to disclose the charge effect on the topological behavior.
\subsection{Charged-flat black hole in the grand canonical ensemble.}
\begin{figure}[!ht]
	\centering
	\begin{tabbing}
		\centering
		\hspace{8.3cm}\=\kill
	\includegraphics[scale=.4]{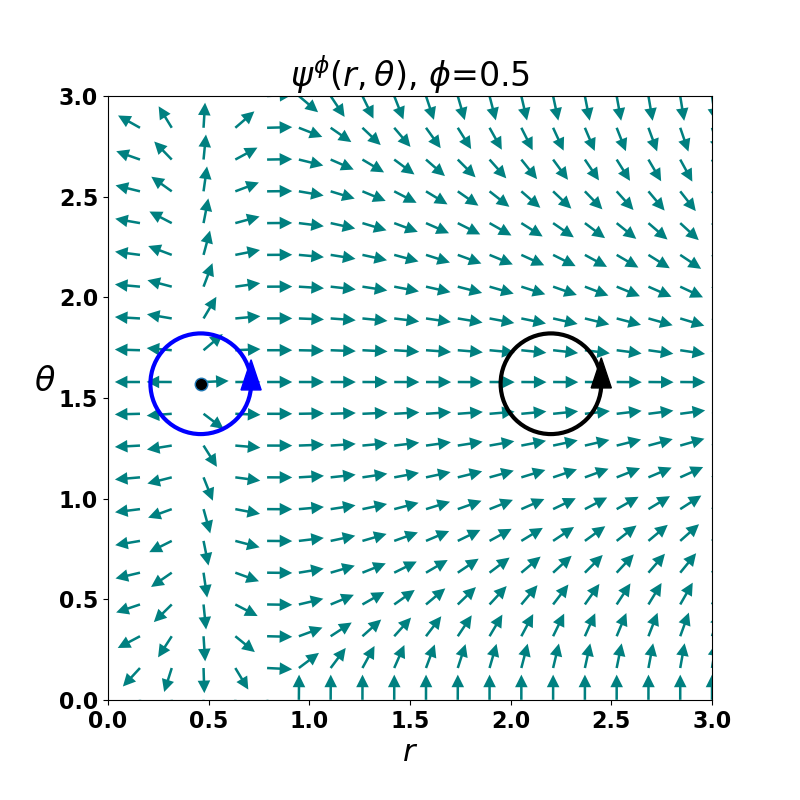}\>
		\includegraphics[scale=.4]{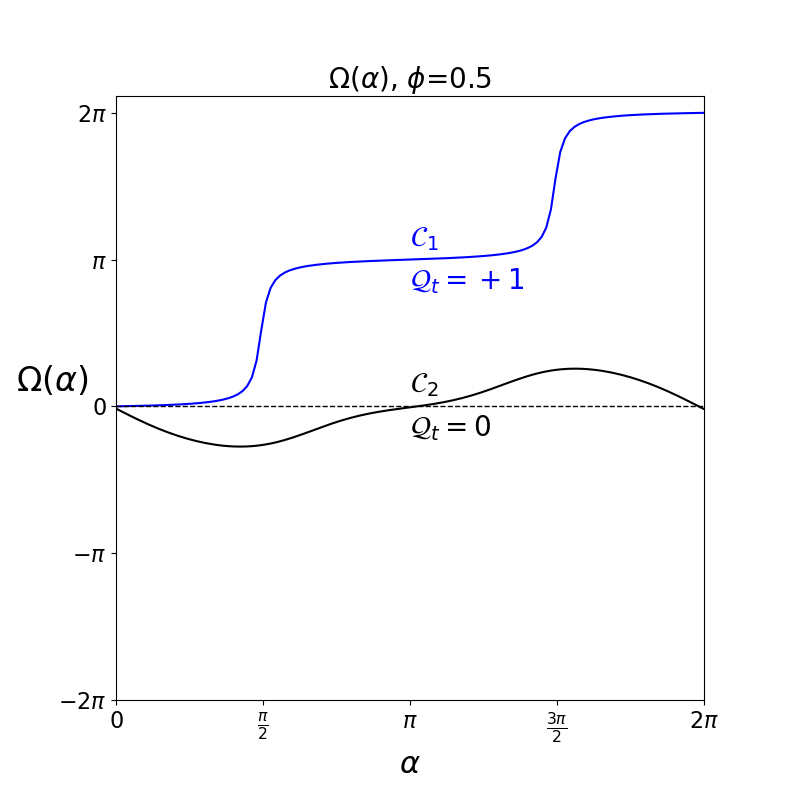}\\
	\end{tabbing}
	\vspace{-1.4cm}
	\caption{\footnotesize\it Topological charge of Hawking-Page phase transition for a charged-flat black hole in the grand canonical ensemble within Rényi formalism ($\phi=0.5$). {\bf Left}: The flow of the normalized vector field  of $\psi^{\phi}$, in the $r-\theta$ plane {\bf Right}: deflection angle for the two contours on the left panel }
	\label{fig4}
\end{figure}

For a charged-flat black hole in the grand canonical ensemble, the electric potential is fixed $(|\phi|\leq1)$, while the thermodynamical system may exchange quanta of electric charge.
To assign the topological charge to the Hawking-Page transition point, we use in the same manner, the temperature $t^\phi_0$ obtained from the Bragg-Williams free energy, and we depict the  flow of the normalized vector field  of $\psi^{\phi}$ and deflection angle for the chosen contours within a nonvanishing electric potential in Fig.\ref{fig4}
It's remarked that   the HP phase transition point (black dot)  located  at $r=\frac{\sqrt{6}}{3\sqrt{\pi}}\approx0.46065$) and surrounded by the blue contour contains the HP transition have a topological charge $\mathcal{Q}_{t}=\mathbf{+1}$
 while the void black contour shows $\mathcal{Q}_{t}=\mathbf{0}$.

\begin{figure}[!ht]
	\centering
	\begin{tabbing}
		\centering
		\hspace{8.3cm}\=\kill
		\includegraphics[scale=.36]{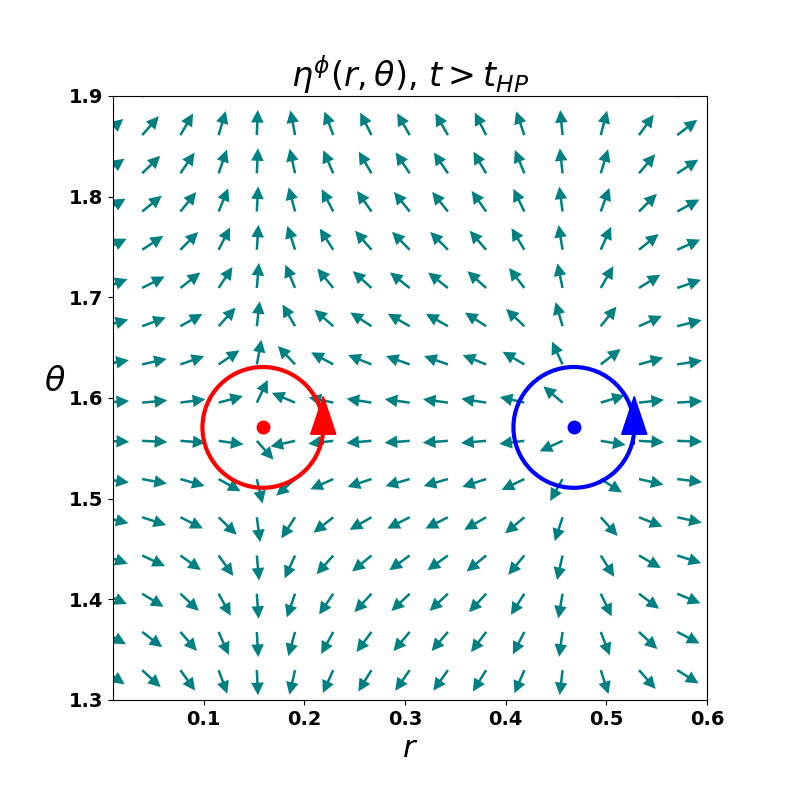}\>	\includegraphics[scale=.36]{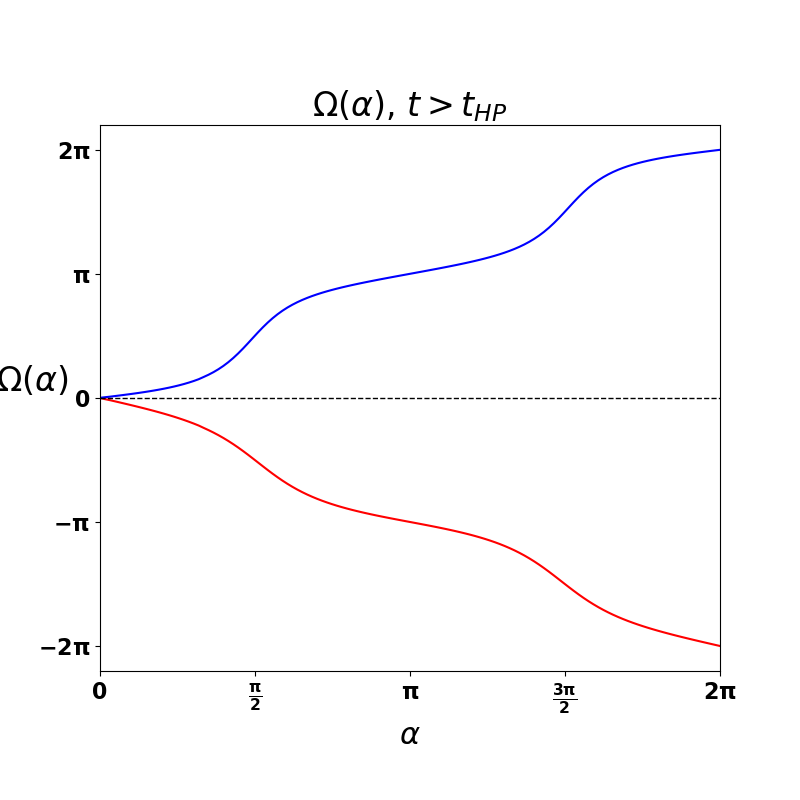}\\
  \includegraphics[scale=.36]{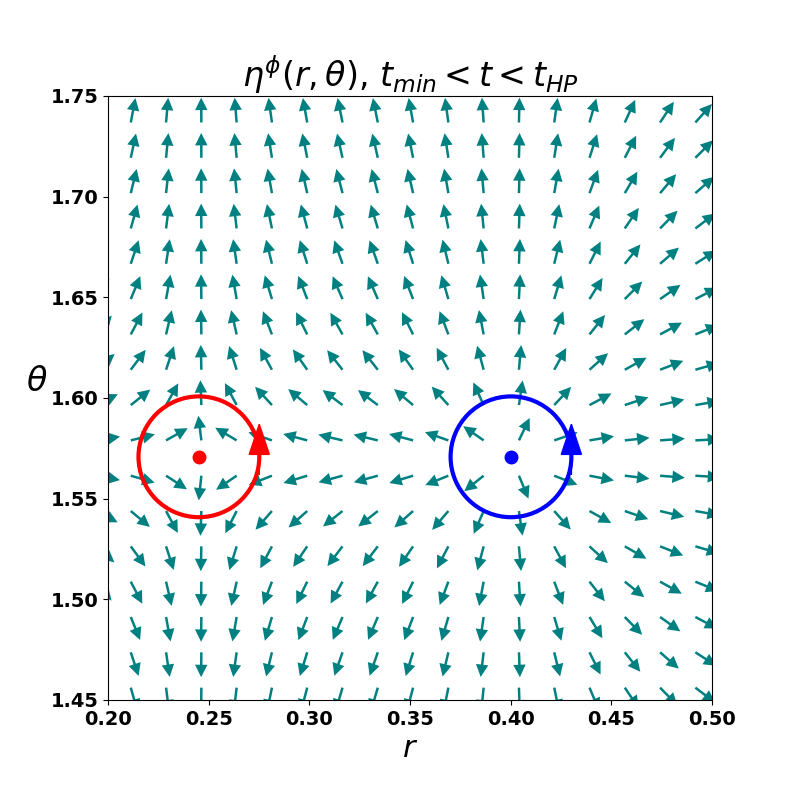}\>
		\includegraphics[scale=.36]{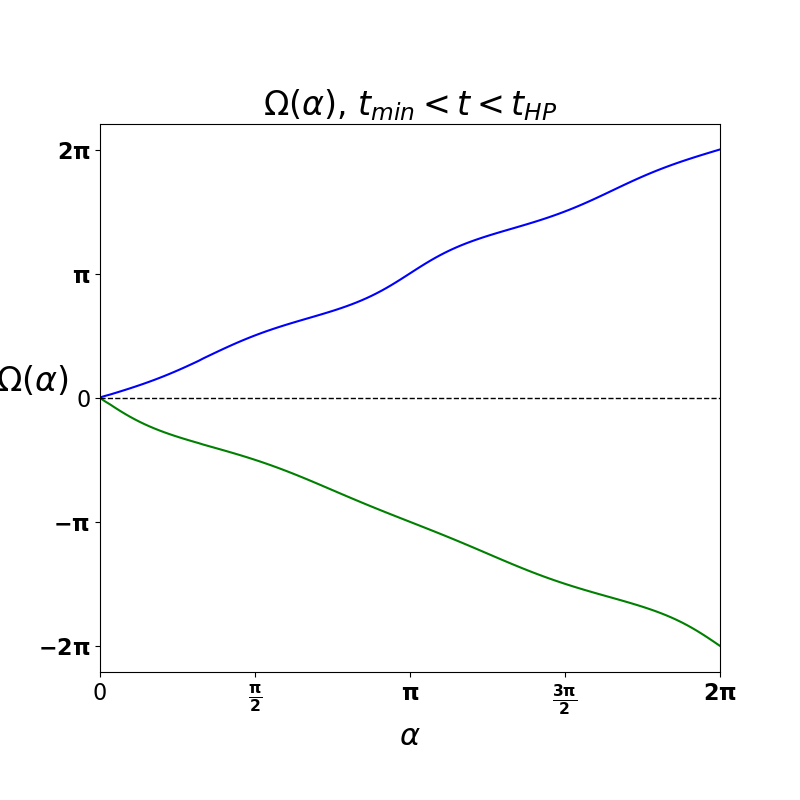}\\
	\includegraphics[scale=.36]{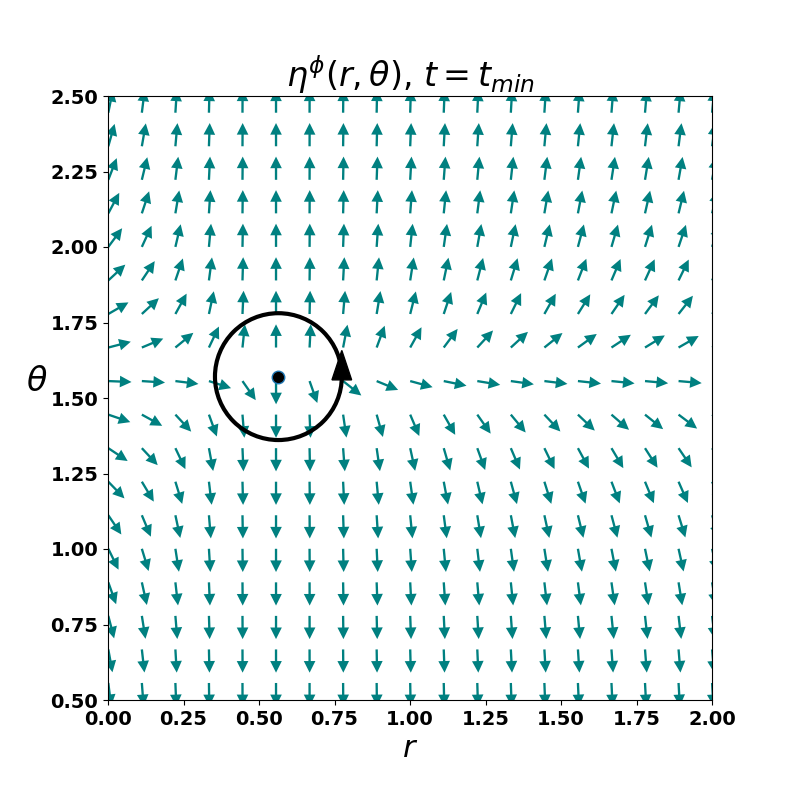}\>	
		\includegraphics[scale=.36]{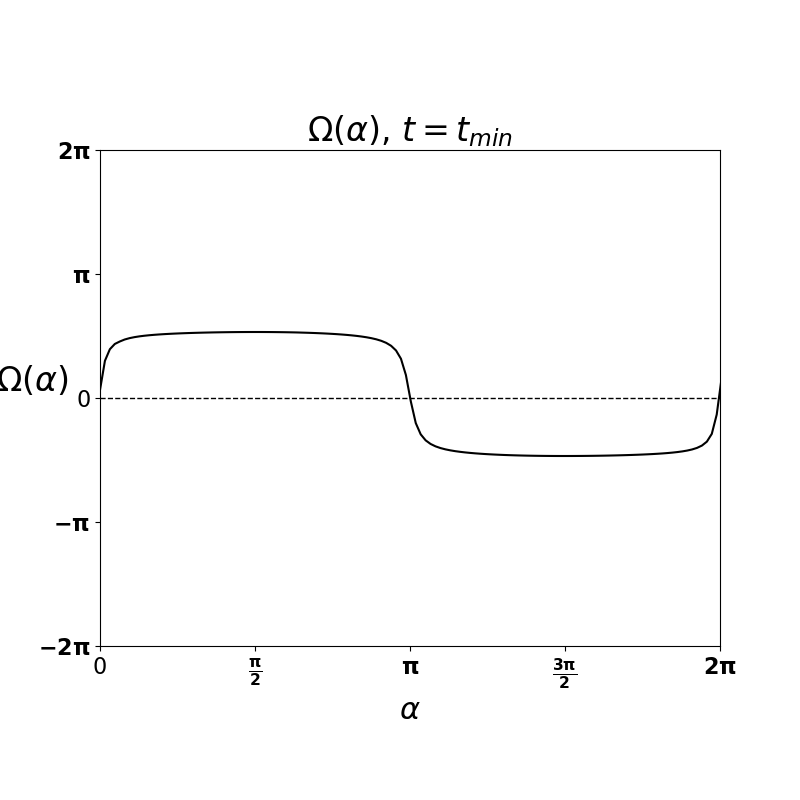}\\
 \end{tabbing}
	\vspace{-1.5cm}
	\caption{\footnotesize \it  Phases of the charged-flat black hole in Rényi formalism for scaled temperatures $t \geq t_{HP}$ (top), $t_{min}<t<t_{HP}$ (middle), and $t=t_{min}$ (bottom). {\bf Left:} The flow of the normalized field of $\eta^\phi$ at $\phi=0.5$. {\bf Righ:} the behavior of
    deflection angle for corresponding black hole phases. In all panels, the electric potential is set to $\phi=0.5$.}
	\label{fig5}
\end{figure}

As previously, and 
to understand black hole phases as topological defects of the $\eta^\phi$ field in Fig.\ref{fig5}, we plot, at $\phi=0.5$, the flow of such a field and the behavior of its associated deflection angle for corresponding black holes phases
$t \geq t_{HP}$ (top), $t_{min}<t<t_{HP}$ (middle) and $t=t_{min}$ (bottom), respectively.
At a temperature higher than Hawking-Page one, we observe from the right panel, two black hole phases, SBH (red dot at $r\approx0.15883$) and LBH (the blue dot at $r\approx0.46775$), while,  the left panel, it's noticeable that there are  opposite topological charges of the two black hole phases.
Down to a temperature under $t_{HP}$ and higher to $t_{min}$, both the small and large black hole phases persist in the flow of the normalized field portrait (red and blue dots at $r\approx0.24514$ and  $r\approx0.40015$, respectively). The topological charges of such phases are evaluated to $\mathcal{Q}_{t}=\mathbf{-1}$ and $\mathcal{Q}_{t}=\mathbf{+1}$.
Finally, when the temperature reaches its minimum, the flow of the normalized field of $\eta^\phi$ at $\phi=0.5$ shows one black hole phase (black dot in bottom panel at $r\approx0.56418$), and its associated topological charge is null.

Now we turn our attention to investigating  the topological charge in the canonical ensemble. Such an  ensemble entails a fixed-charge ensemble, which allows us to explore the Rényi statistical mechanics of charged-flat black holes, unveiling their intricate interplay between charge and other thermodynamic variables from the topological point of view. 

\section{Rényi topological formalism in  the canonical ensemble: Van-der-Waals phase transition.}

In the canonical ensemble, electric charge $Q$ is fixed, which gives the Bragg-Williams free energy such as,
\begin{equation}\label{f_R_Q_4}
\bar{f_R}(r,t,Q)=\displaystyle  \frac{r}{2} - \frac{Q^{2}}{2 r}  - \frac{t \log{\left(\pi \lambda r^{2} + 1 \right)}}{\lambda}.
\end{equation}
Where $r$, $t$, and $Q$ are treated as free parameters. The remarkable analogy between the asymptotically flat black hole phase structure in the canonical ensemble within Rényi thermodynamics and that of the $VdW$ fluid prompts one to consider the extended phase space and the above free energy in terms of the Rényi pressure by the replacement \cite{Promsiri:2020jga}, 
\begin{equation}\label{lam_p}
\lambda=\displaystyle \frac{32 p r^{2}}{3 \left(r^{2}- Q^{2} \right)},
\end{equation}

here $p$ denotes the Rényi pressure. The Bragg-Williams free energy becomes,
\begin{equation}
  \bar{f_R}(r,t,Q)=  \displaystyle \frac{16 p r \left(- Q^{2} + r^{2}\right) + 3 t \left(Q^{2} - r^{2}\right) \log{\left(\frac{3 Q^{2} - 32 \pi p r^{4} - 3 r^{2}}{3(Q^{2} - r^{2})} \right)} }{32 p r^{2}}.
\end{equation}

For $0<p<<1$, and in order to absorb the $p$-dependency, we use the following scaling definitions, $r \rightarrow r/\sqrt{p}$, $t\rightarrow \sqrt{p} t$, $Q\rightarrow Q/\sqrt{p}$, and $ \bar{f_R}\rightarrow  \bar{f_R}/\sqrt{p}$. Leading to

\begin{equation}\label{keyx}
    \bar{f_R}(r,t,Q)=\displaystyle \frac{3 \left(Q^2-r^2\right)^2+2 \pi  r^3 t \left(3 Q^2+16 \pi  r^4-3 r^2\right)}{6 r\left(r^2-Q^2\right)}
\end{equation}

The vanishing condition of the free energy gives,

\begin{equation}
   \bar{f_R}(r,t^Q_0,Q)=0 \implies t^Q_0=\displaystyle \frac{16 r}{3 \log{\left(\frac{32 \pi r^{4} + 3 r^{2}- 3 Q^{2}}{3(r^{2}-Q^{2}) } \right)} }\approx \displaystyle \frac{8 r}{3} - \frac{Q^{2}}{2 \pi r^{3}}  + \frac{1}{2 \pi r}. 
\end{equation}
In the following, we distinguish three regimes in the phase space of the charged-flat black hole according to its electric charge, namely, subcritical ($Q<Q_c$), critical ($Q=Q_c$) and supercritical ($Q>Q_c$) regimes, where the  scaled critical charge is $\mathbf{Q_c\approx0.087269}$. The value of $Q_c$ is computed by solving the following system of equations,

\begin{equation}
    \left(\frac{\partial t_0^Q}{\partial r}\right)_{Q_c}=\left(\frac{\partial^2 t_0^Q}{\partial r^2}\right)_{Q_c}=0
\end{equation}

\subsection{Topological charge in the subcritical regime : \texorpdfstring{$Q<Q_c$}.}
\begin{figure}[!ht]
	\centering
	\begin{tabbing}	
		\centering
		\hspace{-1.4cm}
		\includegraphics[scale=.5]{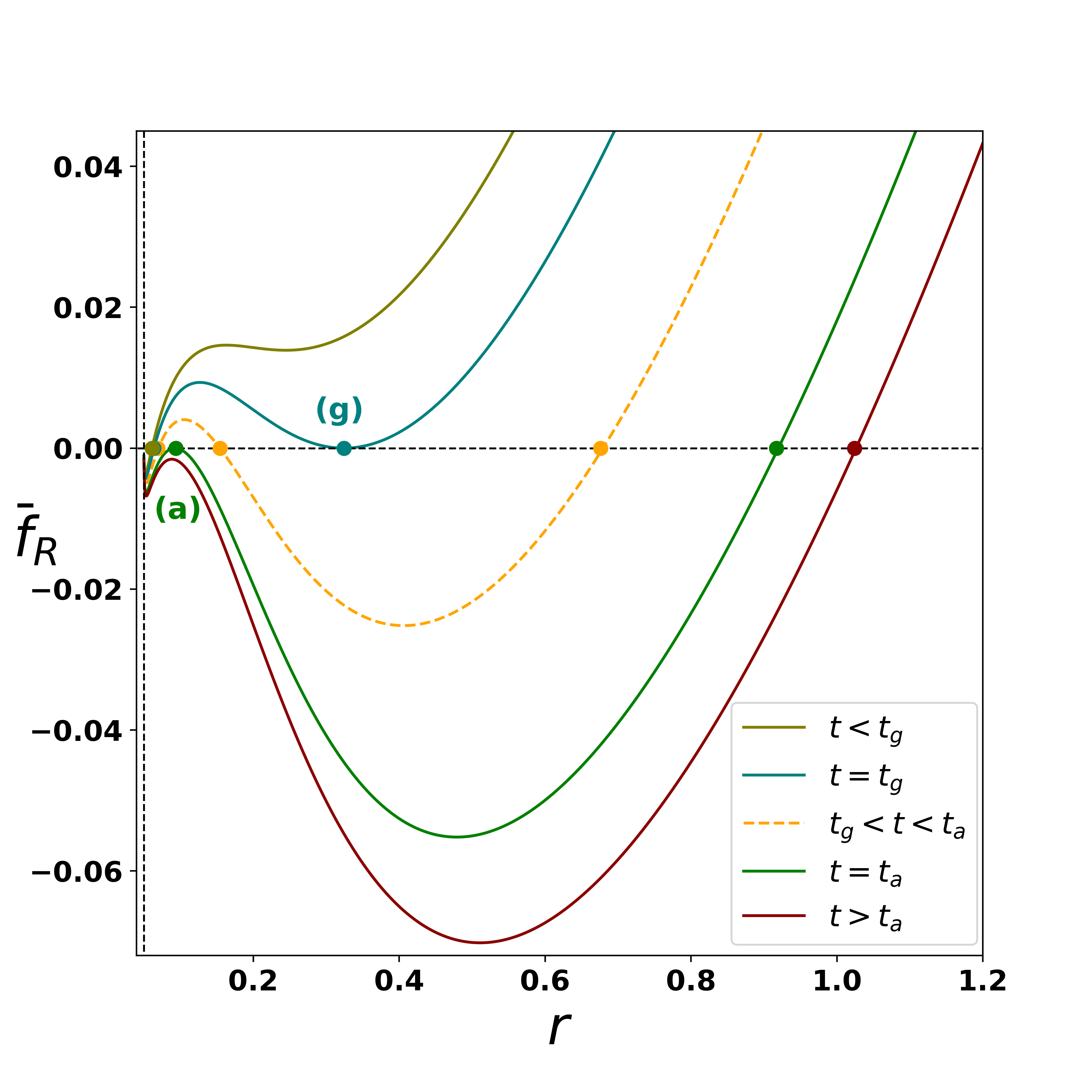}
       \hspace{-1.cm}
		\includegraphics[scale=.5]{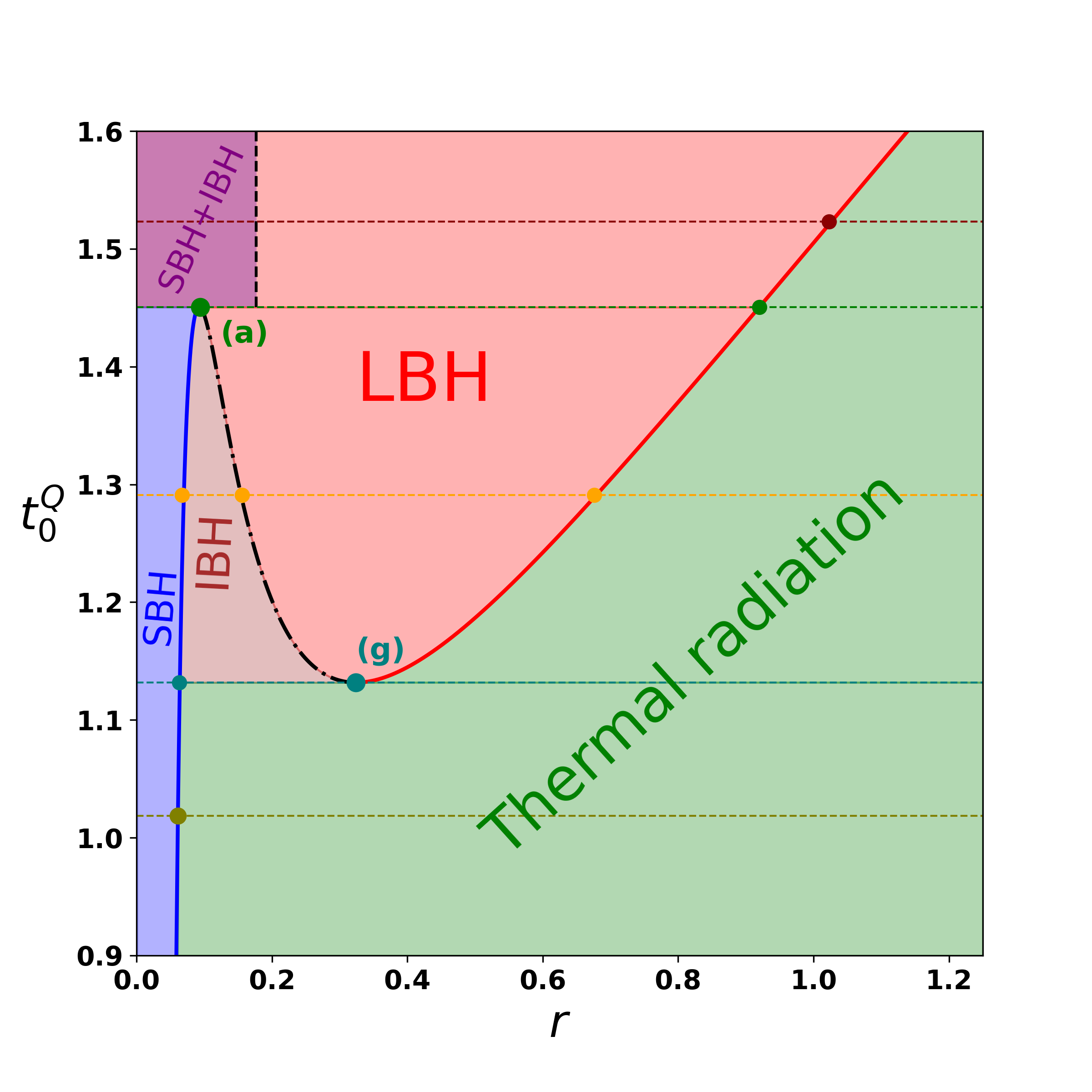}
	\end{tabbing}
	\vspace{-1.1cm}
	\caption{\footnotesize\it {\bf Left:} Behavior of the Bragg-Williams free energy $ \bar{f_R} $, as a function of $r$ at different temperatures $ t $ for the subcritical regime. {\bf Right:} The coexistence temperature $t_0^Q$ of the small, intermediate, and large black hole phases, as a function of $r$, shows the $SBH\longleftrightarrow IBH\longleftrightarrow LBH$ transition points at the various temperatures chosen in the left panel. Colored
		dots  and horizontal lines correspond to respective colored curves in the left panel.}\label{fig1_Q}
\end{figure}

In this regime, as shown in the right panel of Fig.\ref{fig1_Q}, the charged-flat black hole possesses three phases, the small black hole phase (SBH), the intermediate black hole phase (IBH), and the large black hole phase (LBH). Furthermore, The black hole undergoes two global phase transitions. Starting from the (SBH) phase, as the temperature is increased towards $t_g$, we see the generation of the (IBH) and (LBH) phases at the point $(g)$, thus called {\it a generation point} of the charged-flat black hole. A further increase of temperature, reaching $t_a$, reveals an inverse process where an annihilation of the (SBH) and (IBH) occurs at the point $(a)$, therefore $(a)$ is an {\it annihilation point} of the present black hole. From a topological point of view, these two processes can be interpreted as the creation and subsequent annihilation of vortex/anti-vortex pairs \cite{Wei2022,Ahmed:2022kyv} caused by the arising instability in the intermediate region between $t_g\leq t\leq t_a$. The stable phases are represented by vortices while unstable phases are anti-vortices. Within this picture, the (SBH) and (LBH) are vortices while the (IBH) is a anti-vortex. 

\paragraph{}Depending on the temperature of the phase system we will topologically probe three domains:

\begin{itemize}
    \item Below the generation temperature, $t<t_g$ (the green olive line on left panel), the (SBH) phase dominates largely and is stable for small order parameter $r$, since it is a  global minimum of $\Bar{f}_R$  and the slop ($\partial t^\phi_0/\partial r$) of this branch is positive. On the other hand, the pure thermal radiation phase dominate for larger radii. As the temperature reaches $t_g$, the (IBH) and (LBH) phases are generated and become distinguishable, with the (LBH) becoming stable (a local minimum of $\Bar{f}_R$) while the (IBH) is unstable (a local maximum of $\Bar{f}_R$).
    \item The coexistence domain, $t_g\leq t\leq t_a$ (the orange dashed line on left panel), where the three phases exist simultaneously. (SBH) and (LBH) phases are stable (they are minima of $\Bar{f}_R$ and have a positive slop positive) whereas (IBH) branch develops a negative slop and is unstable (a maximum of $\Bar{f}_R$). As the temperature increases, the (SBH) and (IBH) become more and more indistinguishable and annihilate at the point $(a)$.
    \item  Above the annihilation temperature, $t>t_a$ (the dark red line on the left panel),  the strongly dominant stable phase is the (LBH) one (a global minimum of $\Bar{f}_R$). The (SBH) is a faint local minimum of the $\Bar{f}_R$ almost indistinguishable from the (IBH) phase (a local maximum of $\Bar{f}_R$). The black dashed line marks a slowly shrinking {\it spinodal region} where (LBH)  phase decomposes to the (SBH)/(IBH) pair, in the absence of any phase transition point with the (LBH) state. This region is found by determining where the curvature of the free energy curves become negative, that is, $\frac{\partial^2_r\Bar{f}_R}{\partial r^2}=0$, thus, the dashed black curve is the locus of inflection points known as spinodes. Within the spinodal domain, the (LBH) phase is unstable under small fluctuations of the order parameter $r$, and since there is no transition point, no thermodynamic barrier to the appearance of the other phases exists. It is also noticeable that as the temperature increases, the stability domain of charged large black holes becomes wider overwhelming the phase profile.
\end{itemize}

Minimizing the Rényi coexistence temperature $t^Q_0$ in Eq.\eqref{keyx} gives the coordinates of the points $(a)$ and $(g)$, one can easily determine numerically, for example, at $Q=0.05<Q_c$, 

\begin{equation}
\frac{\partial t^Q_0}{\partial r}=0 \implies
\begin{cases}\label{diff_t_0_phi}
 (a) : r_a= 0.09379 \quad and \quad t_a=1.45042,\\
 (g) : r_g=0.32421 \quad and \quad t_g=1.13175.\\
 \end{cases}
\end{equation}
Now, and by taking the charge fixed, we define the pertinent scalar field, 
\begin{equation}
   \Psi_Q(r,\theta)=\frac{t^Q_0}{\sin\theta}\approx\displaystyle \frac{- 3 Q^{2} + 16 \pi r^{4} + 3 r^{2}}{6 \pi r^{3} \sin{\left(\theta \right)}},
\end{equation}

and likewise its gradient components,

\begin{equation}\label{psi_Q}
\begin{cases}
\psi^Q_r=\partial_r\Psi_Q\approx\displaystyle \frac{9 Q^{2} + 16 \pi r^{4} - 3 r^{2}}{6 \pi r^{4} \sin{\left(\theta \right)}},\\

\psi^Q_{\theta}=\partial_{\theta}\Psi_Q\approx\displaystyle \frac{\left(3 Q^{2} - 16 \pi r^{4} - 3 r^{2}\right) \cos{\left(\theta \right)}}{6 \pi r^{3} \sin^{2}{\left(\theta \right)}}.
\end{cases}
\end{equation}

\begin{figure}[!ht]
	\centering
	\begin{tabbing}
		\centering
		\hspace{8.3cm}\=\kill
		\includegraphics[scale=.4]{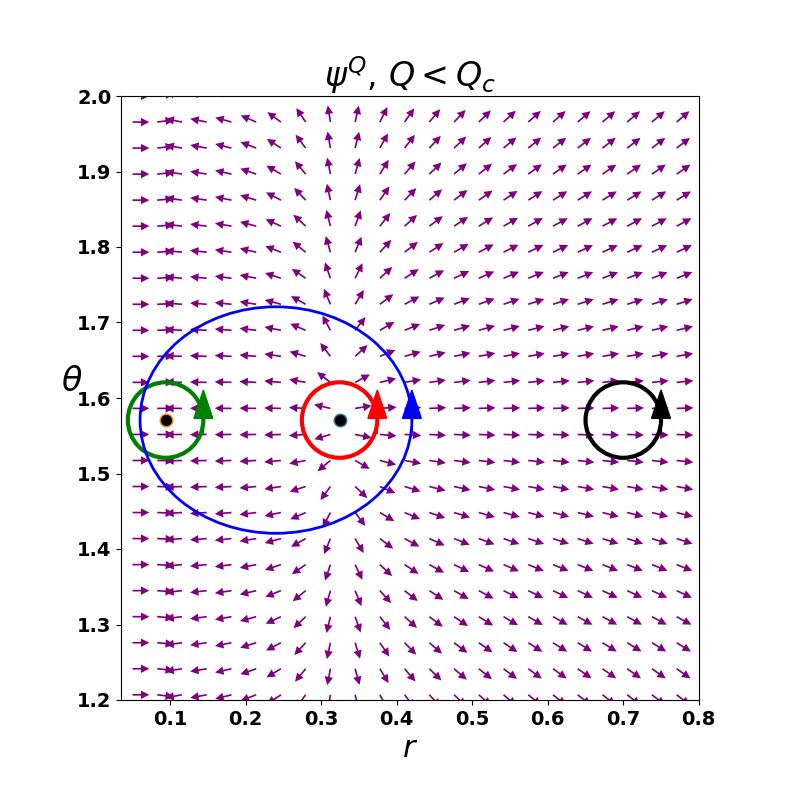}\>
		\includegraphics[scale=.4]{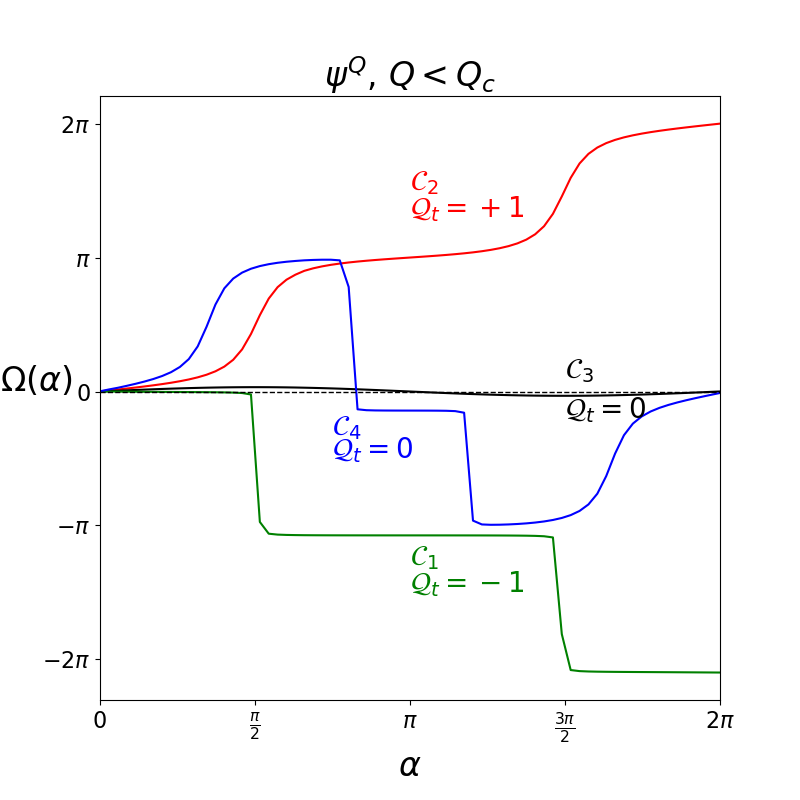}\\
	\end{tabbing}
	\vspace{-1.5cm}
	\caption{\footnotesize\it Topological charge of Van-der-Waals phase transition for a charged-flat black hole in the canonical ensemble within Rényi formalism at $Q=0.05<Q_c$. {\bf Left:} The flow of the normalized vector field  of $\psi^{Q}$, in the $r-\theta$. {\bf Right:} deflection angle for the contours on the left panel.   }\label{fig6}
\end{figure}

To evaluate the topological charge associated with $VdW$-like phase transitions in this canonical context, we replot now, in Fig.\ref{fig6} the normalized flow  of $\psi^Q$ in
$r-\theta$ plane and the behavior of the deflection angle $\Omega$ as a function of the angle $\alpha$ for different contours, for $Q=0.05$.
One can observe from the right panel, the existence of two-phase transition points (black dots at $r=r_g$ and $r=r_a$). The green, red, and blue contours contain the transitions corresponding to the generation and annihilation points, respectively, while the black contour is void. The curves of the deflection angle $\Omega$ from the right panel, reveal,  that the transition points have a topological charge of $\mathcal{Q}_{t}=\mathbf{\pm1}$ whereas the black void contour has zero topological charge. It is not surprising to note that the two transition points $(g)$ and $(a)$ acquire opposite topological charges since in one an instability ensues, while in the other the instability terminates.

We can reveal more topological structure of the phase profile of the charged-flat black hole as a Van-der-Waals system by studying the temperature dependent phase transitions or coexistence points indicated by the graph of the scaled temperature $t_0^Q$ in the right panel of Fig.\ref{fig6}. In addition to global transition points $(g)$ an $(a)$, topologically characterised by means of the $\psi$-mapping, a deeper investigation of all transition points can be achieved by the following $\xi$-mapping defined by,
\begin{align}\label{xi_Q}
\begin{cases}
\xi^Q_{r}&=\displaystyle\bar{f}_R,\\
\xi^Q_{\theta}&=\displaystyle-\cot\theta \cos\theta.
\end{cases}
\end{align}
This choice stems from the well known feature that free energy of a thermodynamic system vanishes at transition points. 
\paragraph{}With the help of the $\xi$-mapping, we move to evaluate the topological charge of transition points for the charged-flat black hole in the subcritical regime ($Q<Q_c$), the electric charge is fixed at $Q=0.05<Q_c$.

As for the $\eta$-mapping, the topological field $\xi^Q$ depends on the scaled Rényi temperature $t$. By changing the values of $t$, keeping the electric charge $Q$ fixed and below the critical value, one is able to map the topological phase transition structure of the charged-flat black hole in the regions above and blow the generation and annihilation transition temperatures $t_g$ and $t_a$. To achieve this goal, we depict in Fig.\ref{fig_trans_sub_1} the normalized flow of $\xi^Q$ in $r-\theta$ plane and the variation of the deflection $\Omega$ in terms of the angle $\alpha$ for different scaled temperatures, and expose black hole (VdW) phase transitions as topological defects of the $\xi^Q$ field. The top, middle  and bottom panels correspond to scaled temperatures above $t_a$, between $t_a$ and $t_g$ and below $t_g$, respectively. One can make the following observations:

\begin{figure}[!ht]
\vspace{0.3cm}
	\centering
	\begin{tabbing}
		\centering
		\hspace{8.3cm}\=\kill
		\includegraphics[scale=.37]{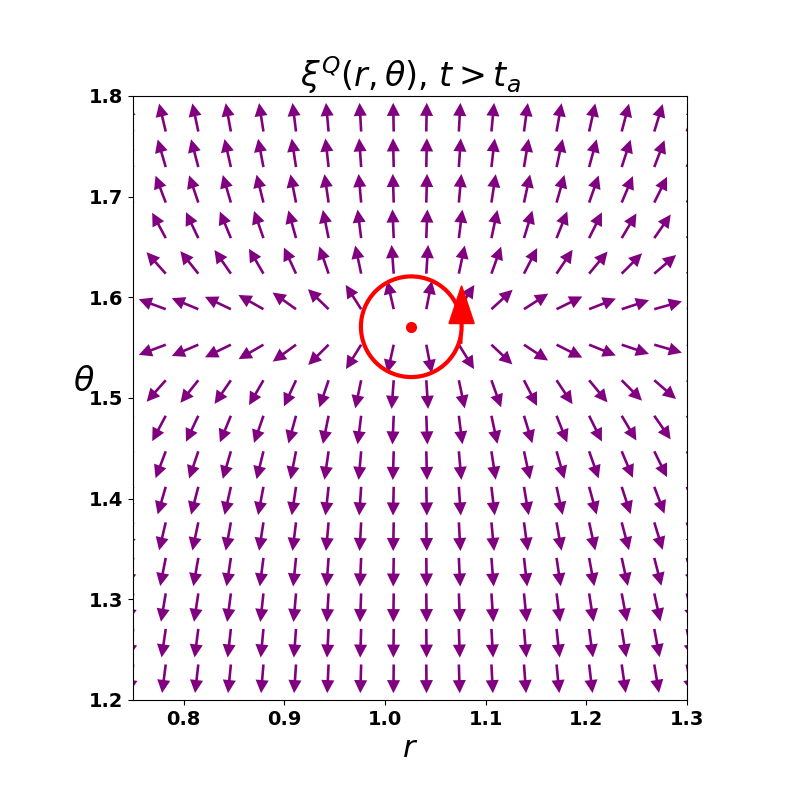}\>
		\includegraphics[scale=.36]{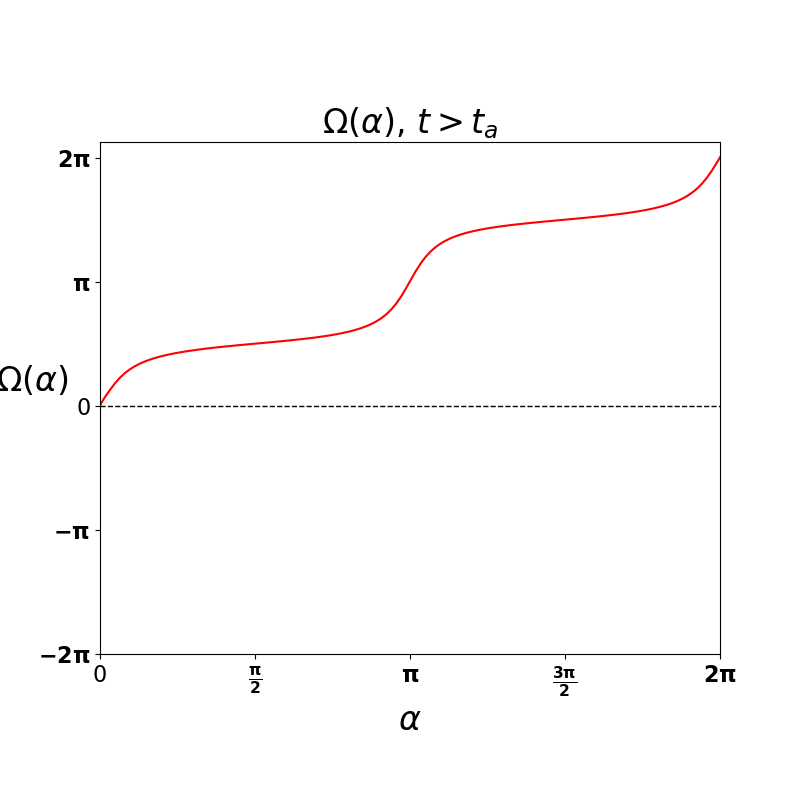}\\
  \includegraphics[scale=.36]{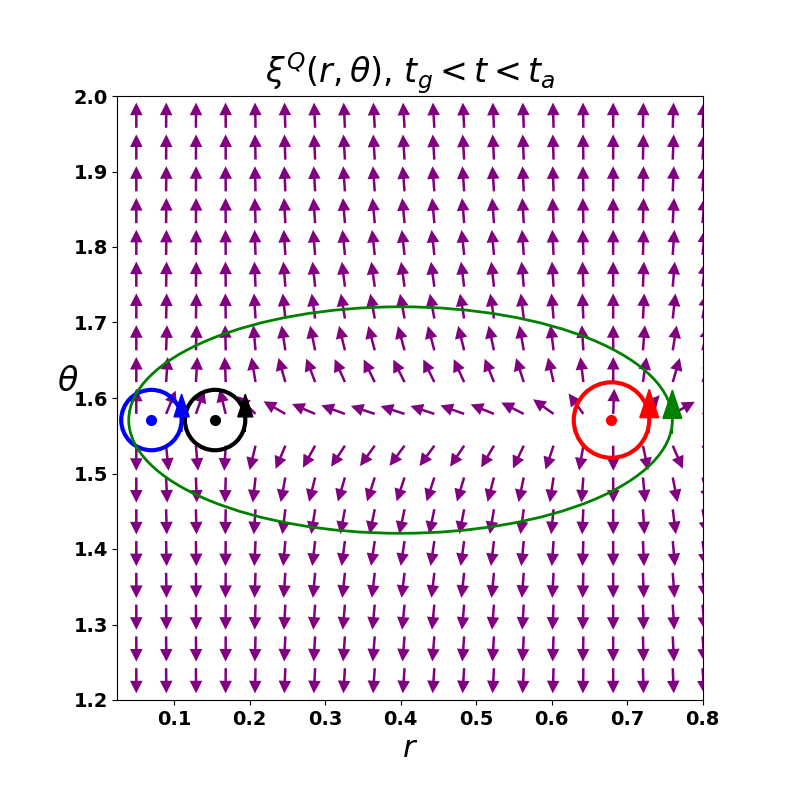}\>
		\includegraphics[scale=.36]{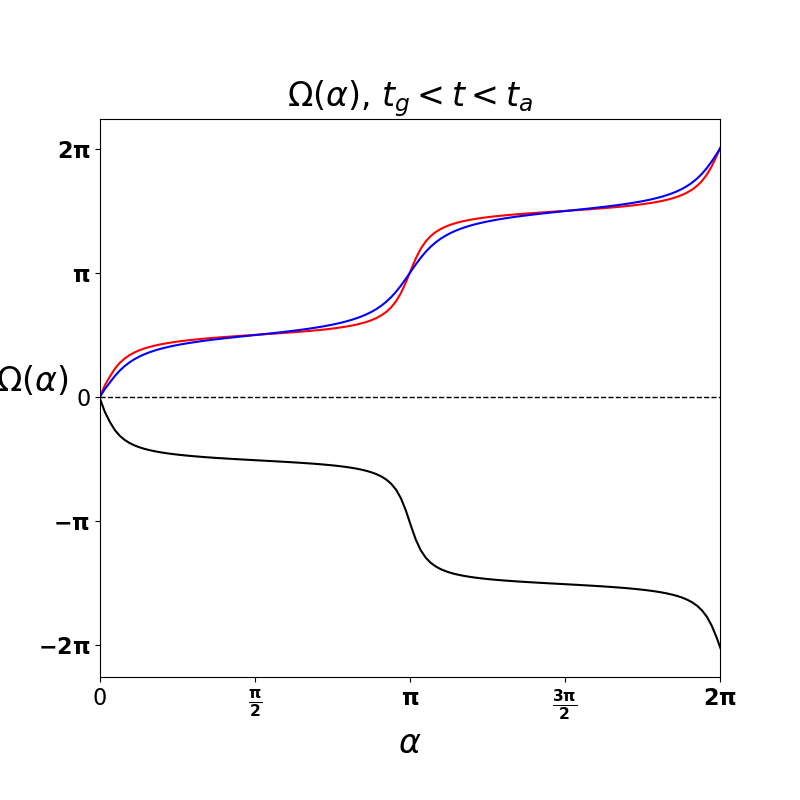}\\
  \includegraphics[scale=.36]{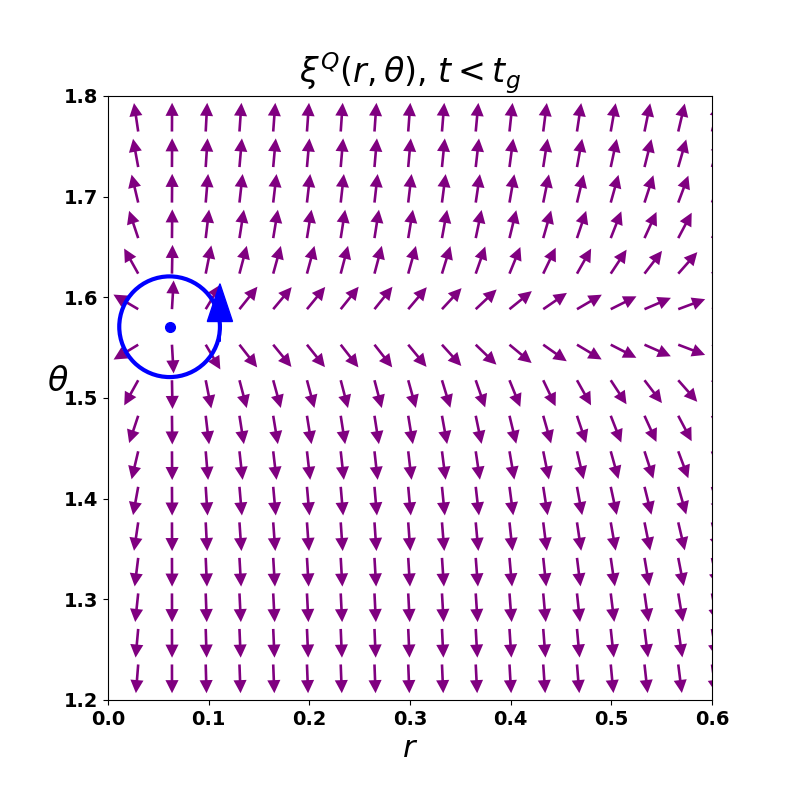}\label{}\>
		\includegraphics[scale=.36]{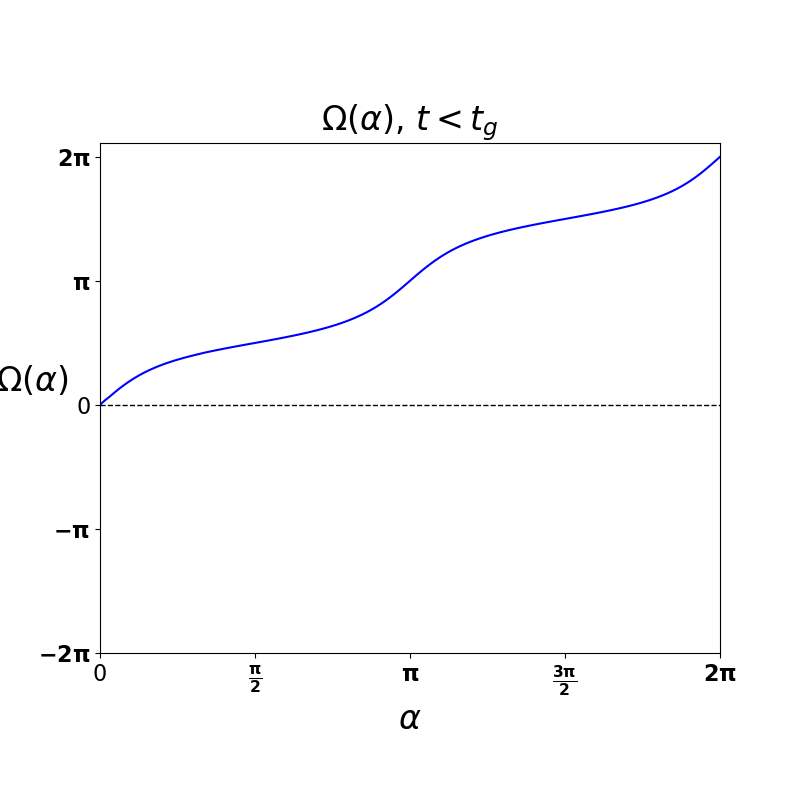}\\
	\end{tabbing}
	\vspace{-1.7cm}
	\caption{\footnotesize\it Transition points of the charged-flat black hole in the canonical ensemble within Rényi formalism in the subcritical regime for scaled temperatures $t > t_{a}$ (top), $t_{g}<t<t_{a}$ (middle), and $t<t_{g}$ (bottom). {\bf Left:} The normalized flow of $\xi^{Q}$ at $Q<Q_c$. {\bf Right:} the behavior of
    deflection angle for corresponding black hole phases. In all panels, the electric potential is set to $Q=0.05$.}\label{fig_trans_sub_1}
\end{figure}

\begin{itemize}
    \item Whenever the black hole temperature exceeds the $t_a$ temperature ($t\geq t_{a}$) only one coexistence or phase transition point manifests itself (the red dot at $r\approx1.026249$) where the transformation $LBH\longleftrightarrow Thermal\;radiation$ takes place with a strong shift towards the more stable (LBH) phase as the order parameter $r$ decreases. As a stable vortex, It is again clear that this transition possesses a nonvanishing topological charge. Namely, $\mathcal{Q}_{t}=\mathbf{+1}$.
    \item For, intermediate black hole temperatures between $t_g$ and $t_a$, ($t_{g}<t<t_{a}$), three coexistence points appear with opposite topological charges, the blue dot at $r\approx0.06999$, the black dot at $r\approx0.15435$ and the red dot at $r\approx0.67939$. The transitions' representative points are more closer to one another depending on whether the temperature $t$ is closer to $t_g$ or $t_a$.
    The blue point represents the transition  $SBH\longleftrightarrow IBH$, where the (SBH) phase coexist and is transformed to the (IBH) one as $r$ grows and instability ensues. At the black point we assist to the transition $IBH\longleftrightarrow LBH$, here, the two phases coexist and a shift towards the right-hand side operates due to more phase stability. Lastly, the red point is the locus of the transition  $LBH\longleftrightarrow Thermal\;radiation$, where the system is driven to the largely more stable (LBH) phase as $r$ gets smaller. The blue and red transitions acquire the same positive topological charge $\mathcal{Q}_{t}=\mathbf{+1}$, while the third intermediary point exhibits a negative charge of $\mathcal{Q}_{t}=\mathbf{-1}$ which indicates an instability. From an intuitive point of view, it is quite expected that between two positive charges the flow of the vector field $\xi^Q$ will appear as a negative charge, that is, the vectors diverging from the the two positive charges seem to converge on a point midway between them. Thus, it is reasonable to consider the intermediary point as an artifact of the existence of two positive charge next to each other. This unphysical feature can be removed by recalling the Maxwell construction and the equal area law\cite{Barzi:2023zed} which replaces the oscillatory behaviour with a physical plateau connecting the two stable coexistence points.
    
    The green contour permits the computation of the total topological charge of this system of transition points as $\mathcal{Q}_{t}=\mathbf{+1}$. This last remark is rendered trivial if one observes that the flow outside of the green contour is directed outward.
    
    \item Below the temperature $t_g$, ($t<t_{g}$), a single coexistence point (blue dot at $r\approx0.061185$) persists as a stable vortex. The transition $SBH\longleftrightarrow Thermal\;radiation$ occurs with a large shift towards the more stable (SBH) phase as the order parameter $r$ decreases. In this case, the topological charge amounts to $\mathcal{Q}_{t}=\mathbf{+1}$. 
\end{itemize}

At present, we turn our attention to evaluating the topological charge of the equilibrium phases in the subcritical regime. To this purpose, we use the canonical version of the $\eta$-mapping expressed as,

\begin{align}\label{eta_Q}
\begin{cases}
\eta^Q_{r}&=\displaystyle\partial_r\bar{f}_R,\\
\eta^Q_{\theta}&=\displaystyle-\cot\theta \cos\theta.
\end{cases}
\end{align}

\begin{figure}[!ht]
\vspace{0.3cm}
	\centering
	\begin{tabbing}
		\centering
		\hspace{8.3cm}\=\kill
		\includegraphics[scale=.36]{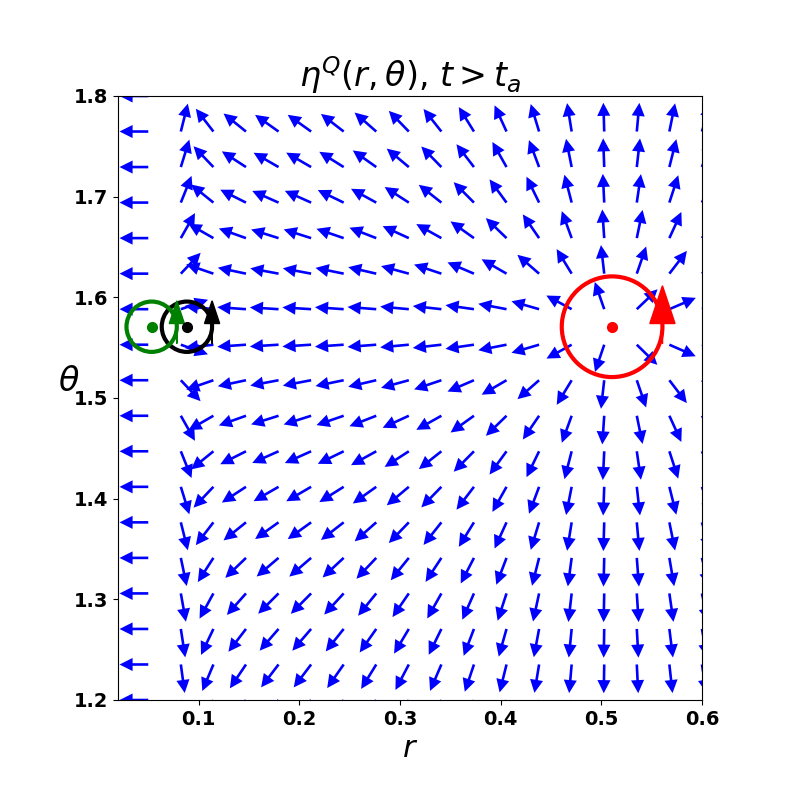}\>
		\includegraphics[scale=.36]{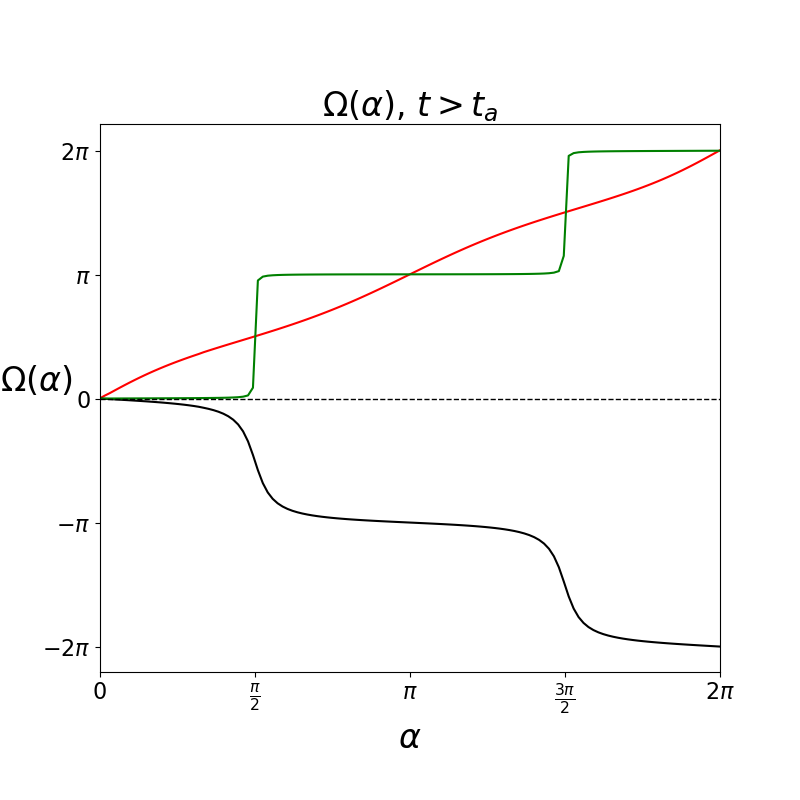}\\
  \includegraphics[scale=.36]{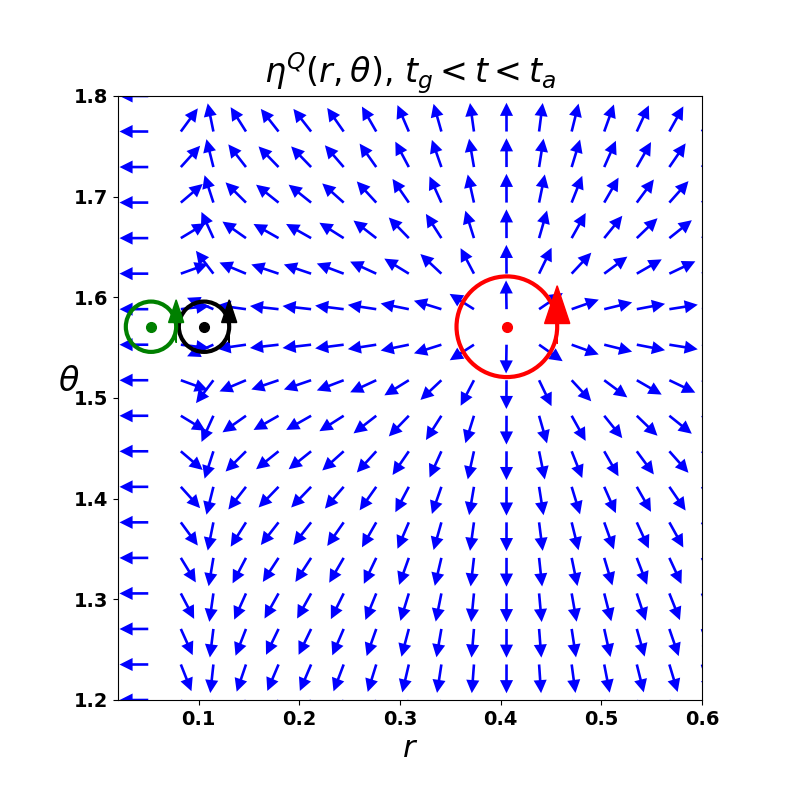}\>
		\includegraphics[scale=.36]{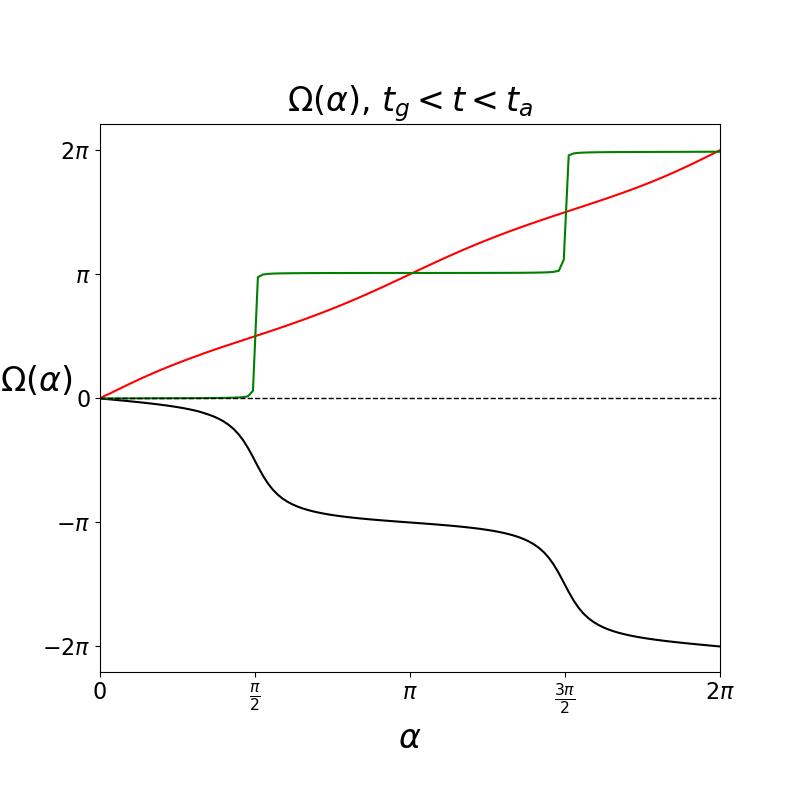}\\
  \includegraphics[scale=.36]{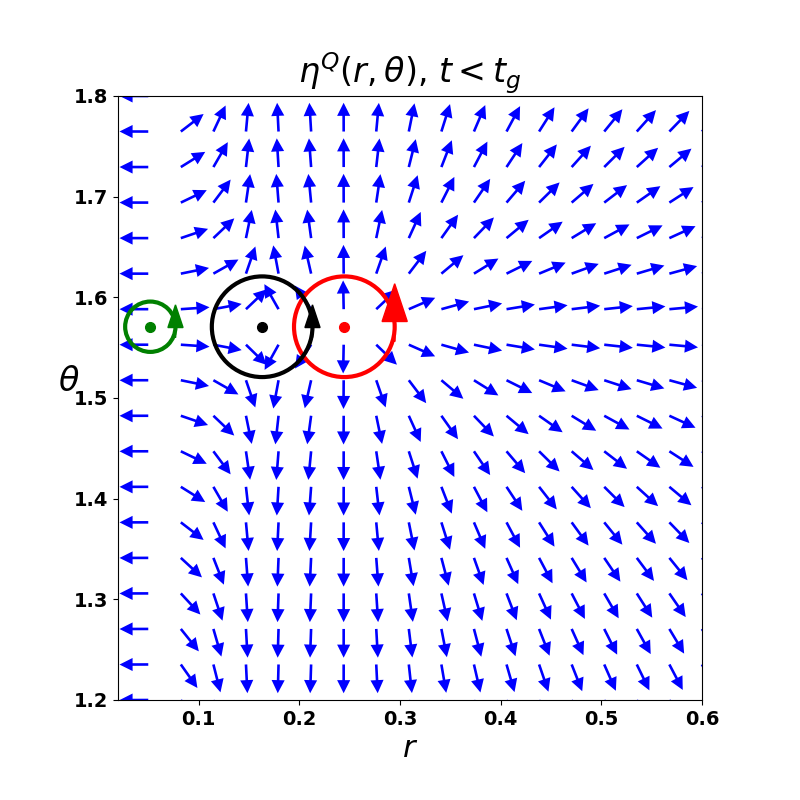}\label{}\>
		\includegraphics[scale=.36]{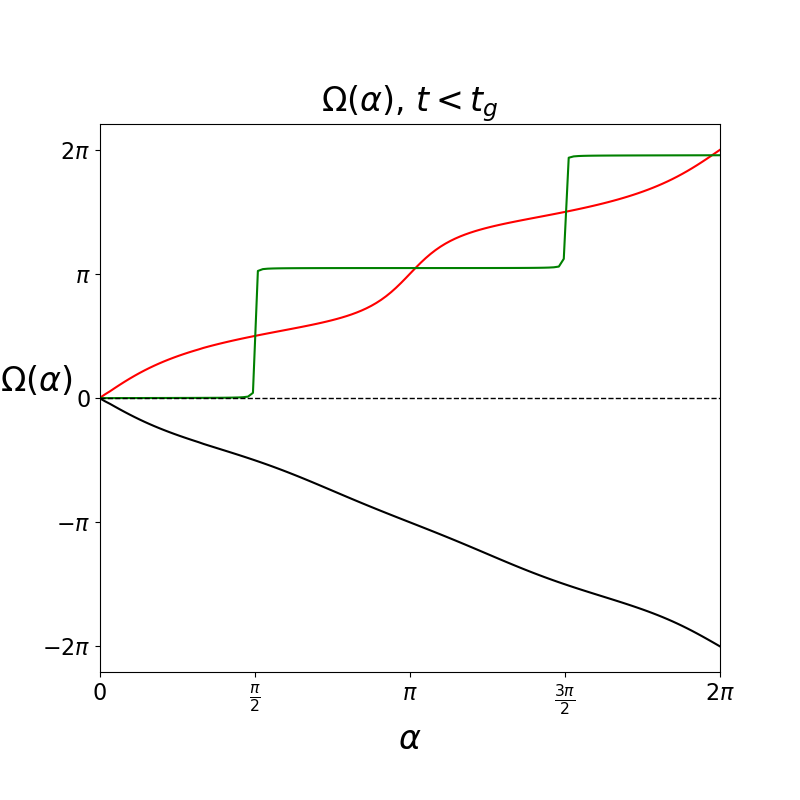}\\
	\end{tabbing}
	\vspace{-1.7cm}
	\caption{ \footnotesize\it Phases points of the charged-flat black hole in the canonical ensemble within Rényi formalism in the subcritical regime for scaled temperatures $t > t_{a}$ (top), $t_{g}<t<t_{a}$ (middle), and $t<t_{g}$ (bottom). {\bf Left:} The normalized flow of $\eta^{Q}$ at $Q<Q_c$. {\bf Righ:} the behavior of
    deflection angle for corresponding black hole phases. In all panels, the electric potential is set to $Q=0.05$.}\label{fig_phase_sub_1}
\end{figure}

In Fig.\ref{fig_phase_sub_1}, we show the (VdW)-like phase systems displayed by the charged-flat black hole in the canonical ensemble below the critical charge, for three different intervals of scaled temperatures. All three systems are composed of the same three phases i.e. (SBH), (IBH) and (LBH) phases. The top panel concerns temperatures above $t_a$, There, the (LBH) phase dominates largely the phase structure with the two other phases closely approaching and almost annihilating one another in favor of the more dominant phase. In other words, (SBH)/(IBH) phases form a tight vortex/anti-vortex pair. From the graphs of deflection angles in the three panels, one readily sees the positive topological charge $\mathbf{\mathcal{Q}_t=+1}$ of the stable (SBH) and (LBH) phases versus the negative value $\mathbf{\mathcal{Q}_t=-1}$ for the unstable (IBH) phase. As the temperature gets below $t_a$ but remains above $t_g$, as seen in the middle panel, the (SBH)/(IBH) vortex pair separates, while the (LBH) vortex approaches the pair, which accentuates the coexistence of the three phases. This tendency continues as the temperature decreases below $t_g$, with now (SBH) being the dominant phase and the (LBH)/(IBH) constituting a more tight nearly annihilating vortex/anti-vortex pair.

\subsection{Topological charge in the critical regime: \texorpdfstring{$Q=Q_c$}.}
\begin{figure}[!ht]
	\centering
	\begin{tabbing}	
		\centering
		\hspace{-1.4cm}
		\includegraphics[scale=.5]{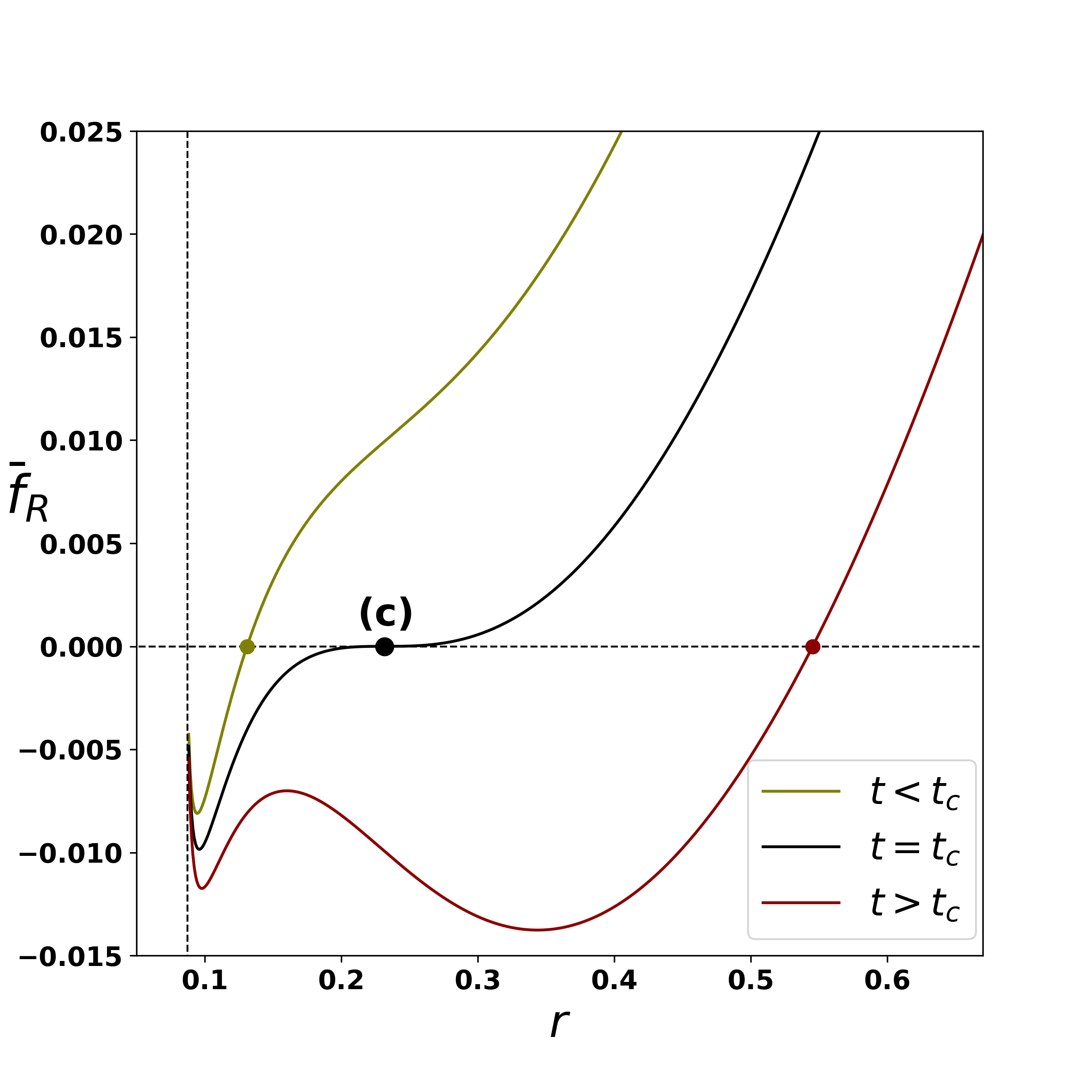}
     \hspace{-0.8cm}
		\includegraphics[scale=.5]{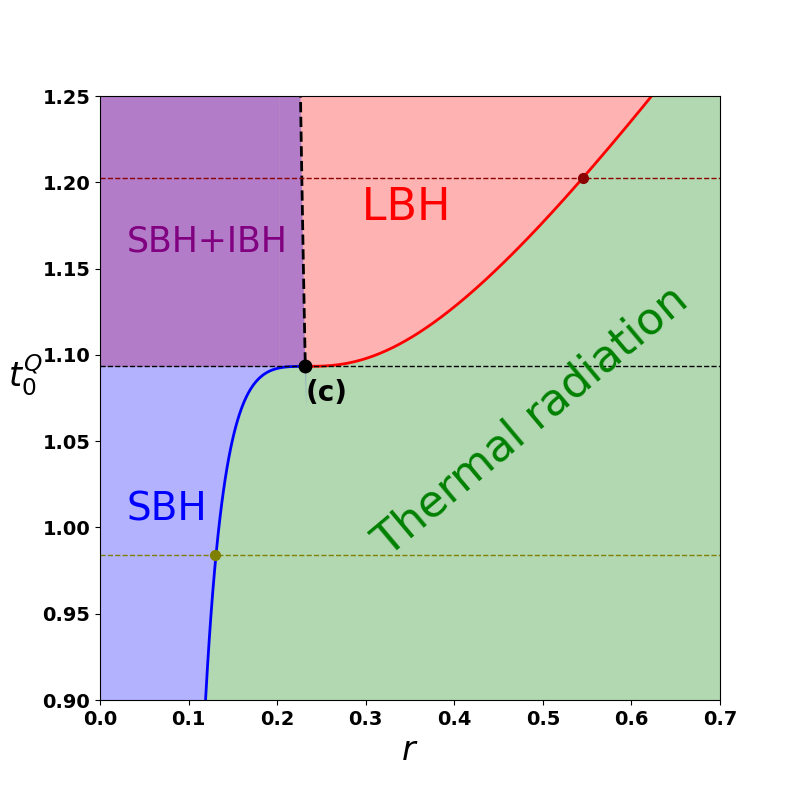}
	\end{tabbing}
	\vspace{-1.2cm}
	\caption{\footnotesize\it {\bf Left:} Behavior of the Bragg-Williams free energy $ \bar{f_R} $, as a function of $r$ at different temperatures $ t $ for the critical regime. {\bf Right:} The coexistence temperature $t_0^Q$ of the small and large black hole phases, as a function of $r$, shows the only possible critical phase transition $SBH\longleftrightarrow LBH$ at the point $(c)$ and the black hole phases for different temperatures chosen in the left panel. Colored
		dots  and horizontal lines correspond to respective colored curves in the left panel.}\label{fig2_Q}
\end{figure}

In Ref.\cite{Promsiri:2020jga}, the authors show that the thermodynamics of charged black holes in asymptotically flat space-time via Rényi statistics unveils a Van-der-Waals phase structure, where a critical behavior appears for specific parameters. Namely, it is worth noting that a second phase transition occurs for the particular value of the electric charge $Q_c$, which is a solution to the equation, $r^{Q_c}_{a}=r^{Q_c}_{g}\approx0.23179$, and the corresponding scaled critical temperature is $t_c\approx1.09334$.
In Fig.\ref{fig2_Q}, we plot on the left the critical Bragg-Williams free energy for different scaled coexistence temperatures above and below $t_c$. 

It shows sets of maxima and minima representing stable and unstable phases. On the right panel, we can see again the coexistence curve and corresponding phase structure. This reveals that two stable phases dominate, namely, the (SBH) phase below the critical point $(c)$, the green olive line on the left panel, where appears a global minimum of $\Bar{f}_R$, and the (LBH) one above $(c)$,  the dark red line on the left panel, where there is also a global minimum. Besides, the intermediate region encountered in the subcritical regime is reduced to the single isolated critical point $(c)$. Thus, in this scheme, the generation and annihilation points coalesce into the critical one prohibiting the appearance of the (IBH) phase at $t=t_c$. On a side note, if we interpret the scaled Rényi temperature as the inverse of the time scale of such phase transitions \cite{Wei2022}, then, $t_c^{-1}=0.91463$  is the time scale of the transition $SBH\longleftrightarrow LBH$ in the critical regime, while in the subcritical regime,  it amounts to $t_a^{-1} + t_g^{-1}=1.57304$. Therefore, the transition $SBH\longleftrightarrow LBH$ is more than one and a half times faster in the critical regime than in the subcritical one, mainly due to the emergence of the (IBH) phase. Additionally, above the critical point a spinodal domain delimited by the dashed black line where a vortex/anti-vortex pair, (SBH)/(IBH), appears due to the spinodal decomposition of the large black hole phase.

\begin{figure}[H]
	\centering
	\begin{tabbing}
		\centering
		\hspace{8.3cm}\=\kill
		\includegraphics[scale=.4]{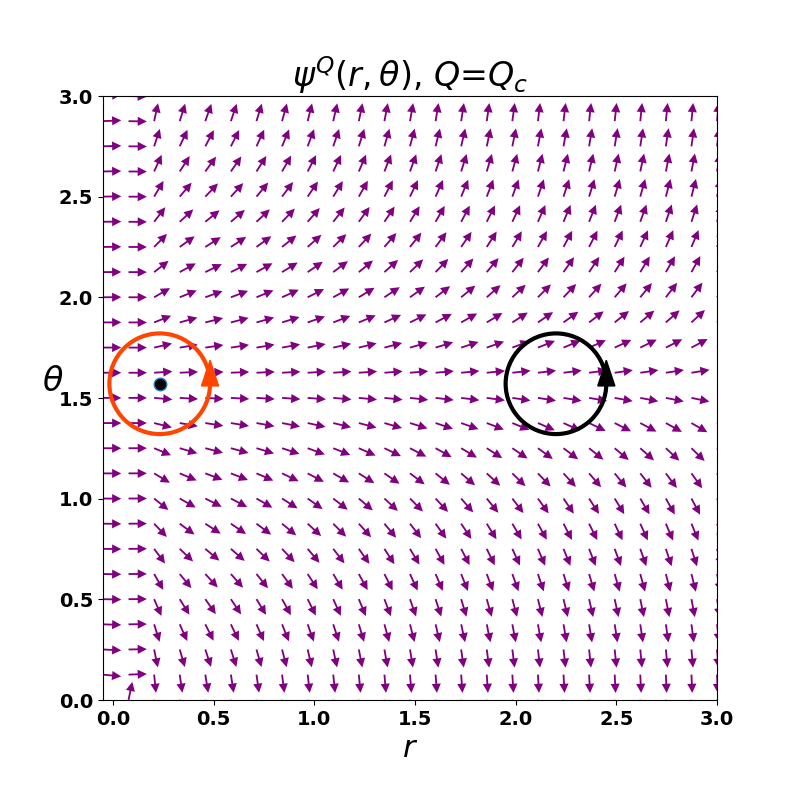}\>
		\includegraphics[scale=.4]{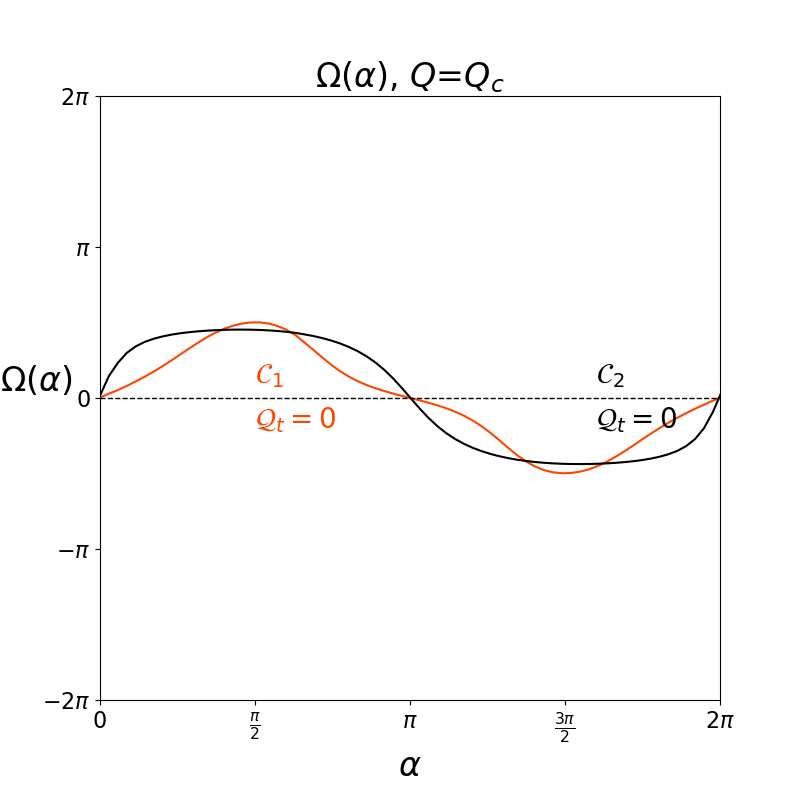}\\
	\end{tabbing}
	\vspace{-1.6cm}
	\caption{\footnotesize Illustration of the topological charge of second order phase transition for a charged-flat black hole in the canonical ensemble within Rényi formalism.}
 \label{fig7}
\end{figure}

Obviously, from Fig.\ref{fig7}, one can evaluate the topological charge of such a critical point, $(c)$, by the $\psi$-mapping to be zero, $\mathcal{Q}_{c}=\mathbf{0}$. As revealed in the subcritical regime, there are two transition points of opposite topological charges, which would explain the null charge of the critical point $(c)$ if we consider the generation and annihilation points, $(g)$ and $(a)$, in the subcritical regime, as a vortex/anti-vortex pair\cite{Ahmed:2022kyv} that annihilates at $Q=Q_c$ or is created as the electric charge decreases below $Q_c$. We notice then that the vortex dynamics of the transition points resembles that of phase points only the first is controlled by the electric charge whereas in the second temperature is the steering parameter.

\begin{figure}[!ht]
	\centering
	\begin{tabbing}
		\centering
		\hspace{8.3cm}\=\kill
		\includegraphics[scale=.36]{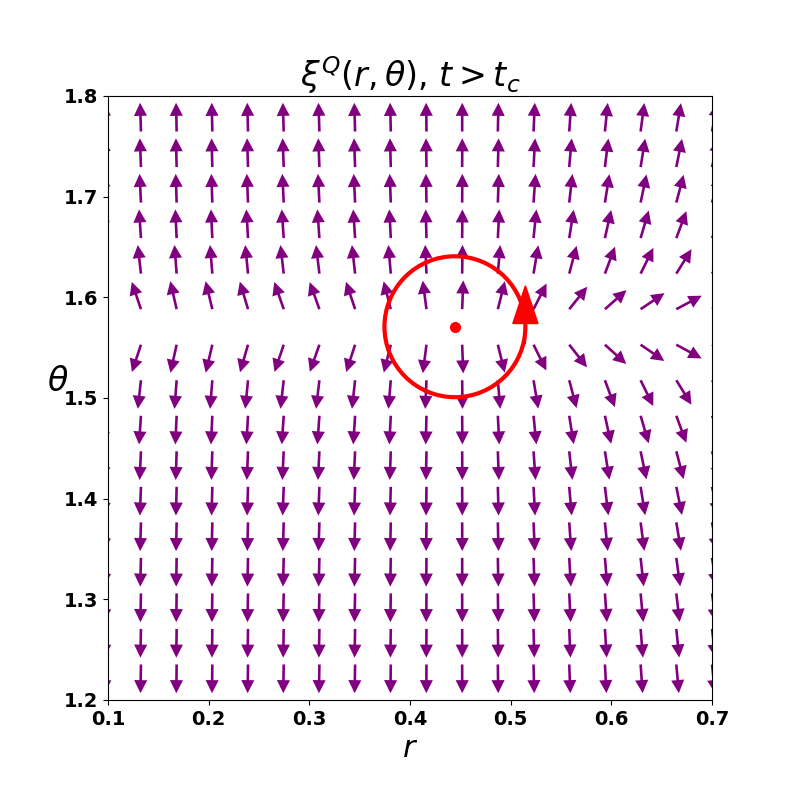}\>
		\includegraphics[scale=.36]{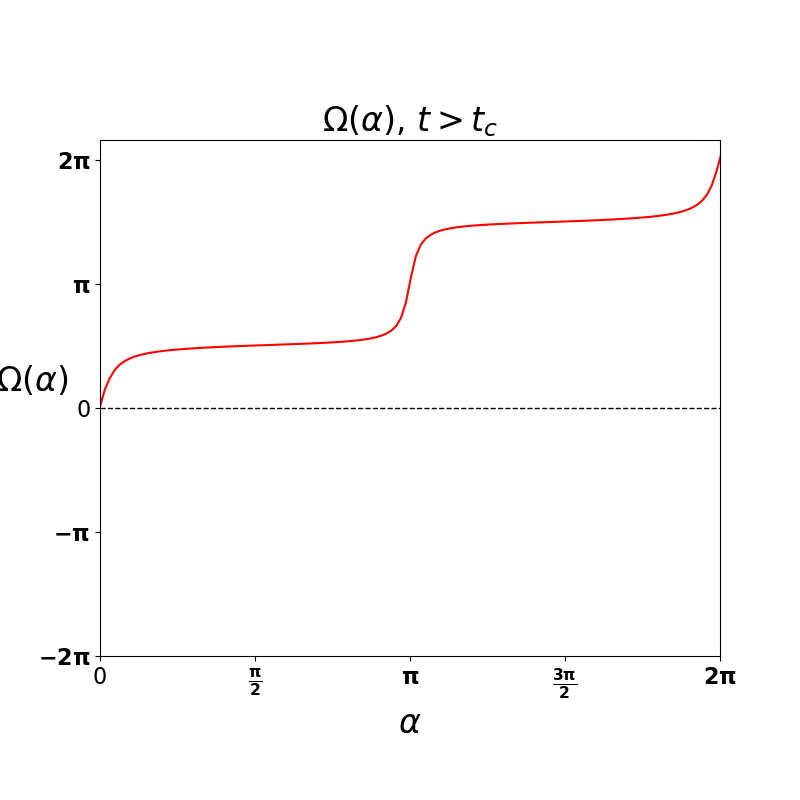}\\
  \includegraphics[scale=.38]{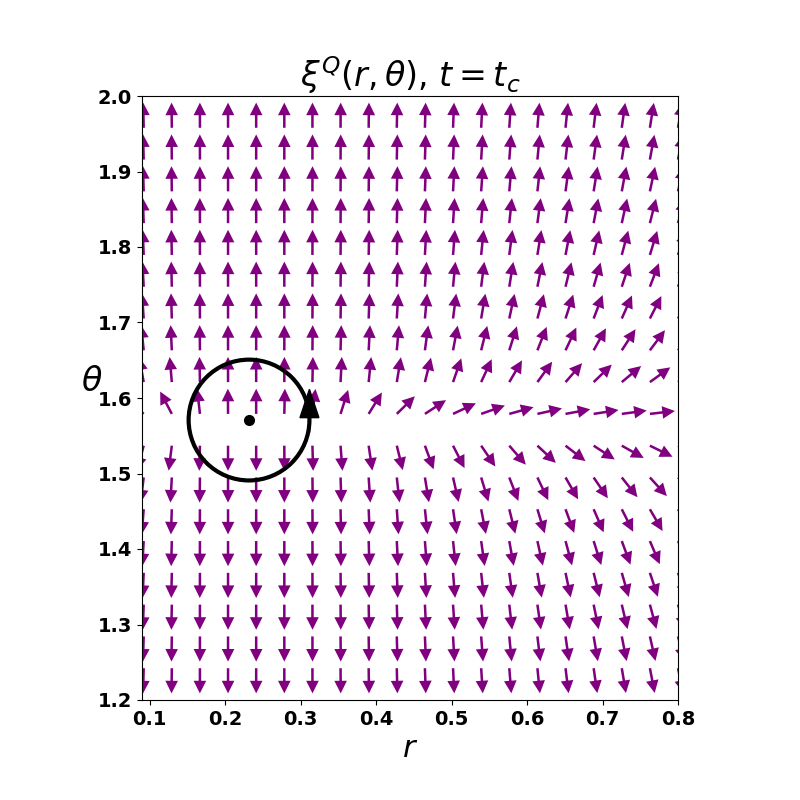}\>
		\includegraphics[scale=.36]{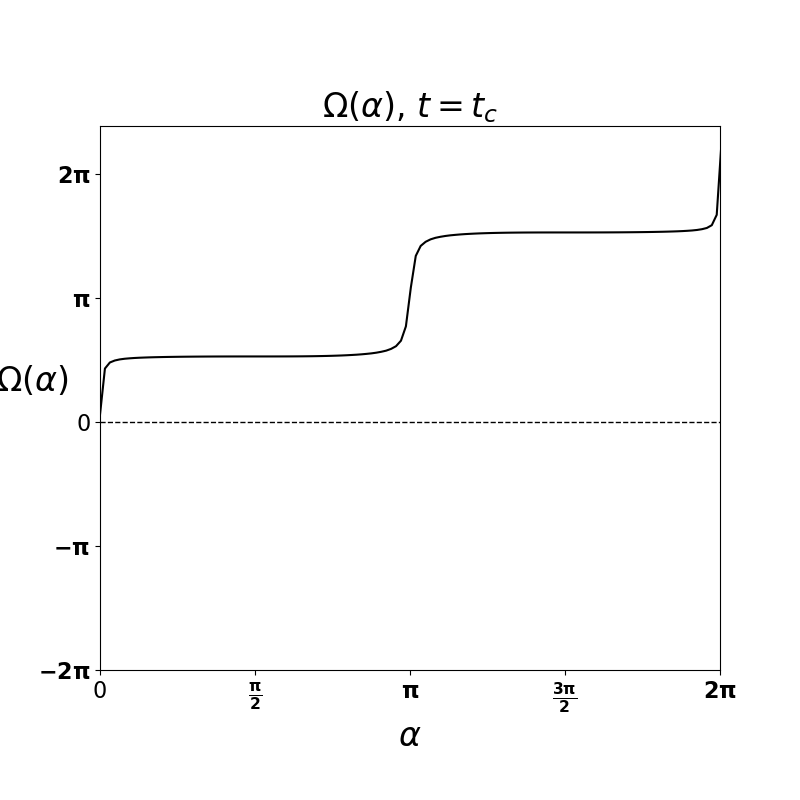}\\
  \includegraphics[scale=.36]{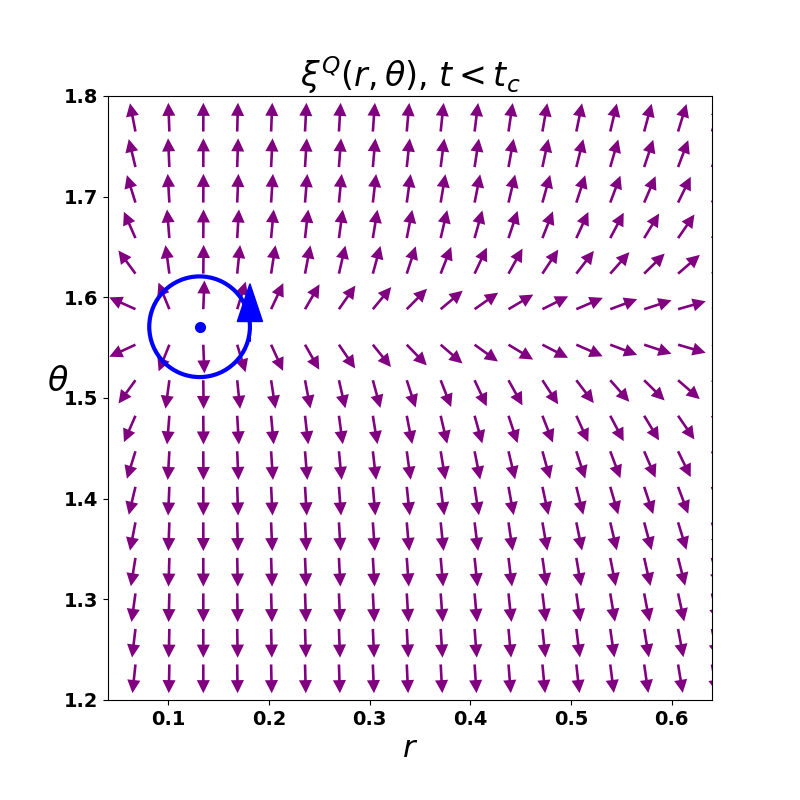}\>
		\includegraphics[scale=.36]{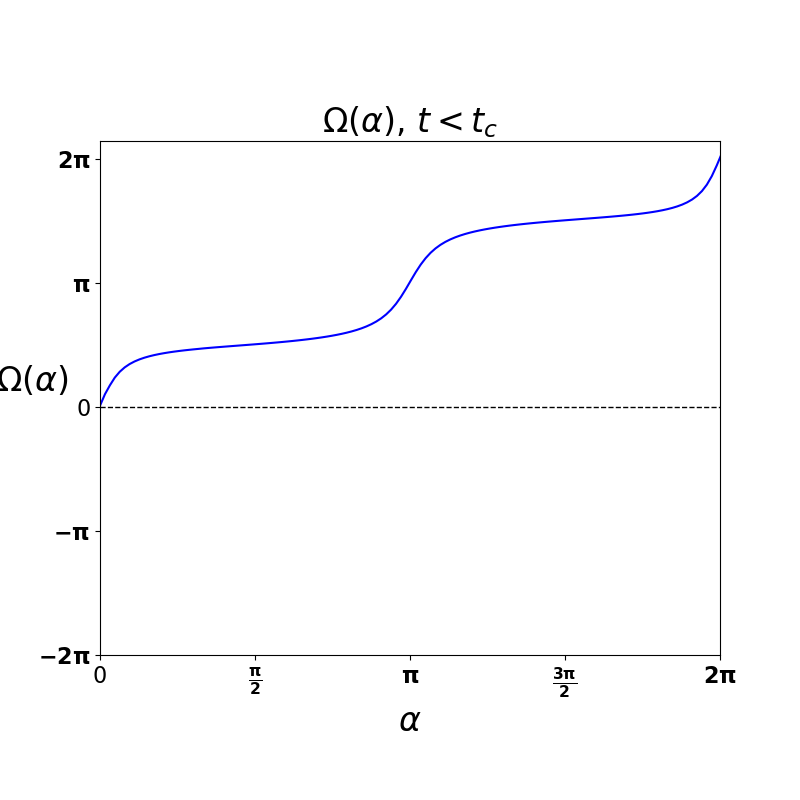}\\
	\end{tabbing}
	\vspace{-1.7cm}
	\caption{\footnotesize\it Phase transitions of the charged-flat black hole in the canonical ensemble within Rényi formalism in the critical regime for scaled temperatures $t > t_{c}$ (top), $t=t_c$ (middle) and $t<t_{c}$ (bottom). {\bf Left:} The normalized flow of $\eta^{Q_c}$. {\bf Right:} the behavior of deflection angle for corresponding black hole phases. In all panels, the electric potential is set to $Q=Q_c$.}\label{fig_trans_cri_Q}
\end{figure}

Exploiting the $\xi$-mapping to gain more insight into the temperature-dependent phase transitions in this regime, we get Fig.\ref{fig_trans_cri_Q}, it says that above and below the critical temperature (red point on the top and blue point on the bottom panels), one coexistence temperature exist and have a positive topological charge
$\mathcal{Q}_{c}=\mathbf{+1}$. The conservation of the topological charge associated with the $\xi$-mapping imposes that the critical point $(c)$ acquires the same positive charge which is indeed confirmed by the direct computation in the middle panel. We have computed two distinct topological charges for the point $(c)$, one with the $\psi$-mapping for being a global transition point, we find a zero charge. The second for is the coexistence point at $t=t_c$ through the $\xi$-mapping, we get $\mathbf{+1}$.

\paragraph{}To probe the topological charge of equilibrium phases for charged-flat black hole and similarly to the subcritical regime, we can define the vector field $\eta^{Q_c}=(\eta^{Q_c}_r,\eta^{Q_c}_\theta)$, employing our Bragg-Williams free energy Eq.(\ref{keyx}) now by taking the electric charge equal to its critical value $Q_c\approx0.087269$,
\begin{equation}
\begin{cases}
\eta^{Q_c}_{r}=\partial_r\bar{f}_R(r,t,Q_c)
\\
\eta^{Q_c}_{\theta}=\displaystyle-\cot\theta \cos\theta
\end{cases}
\end{equation}

\begin{figure}[!ht]
	\centering
	\begin{tabbing}
		\centering
		\hspace{8.3cm}\=\kill
		\includegraphics[scale=.36]{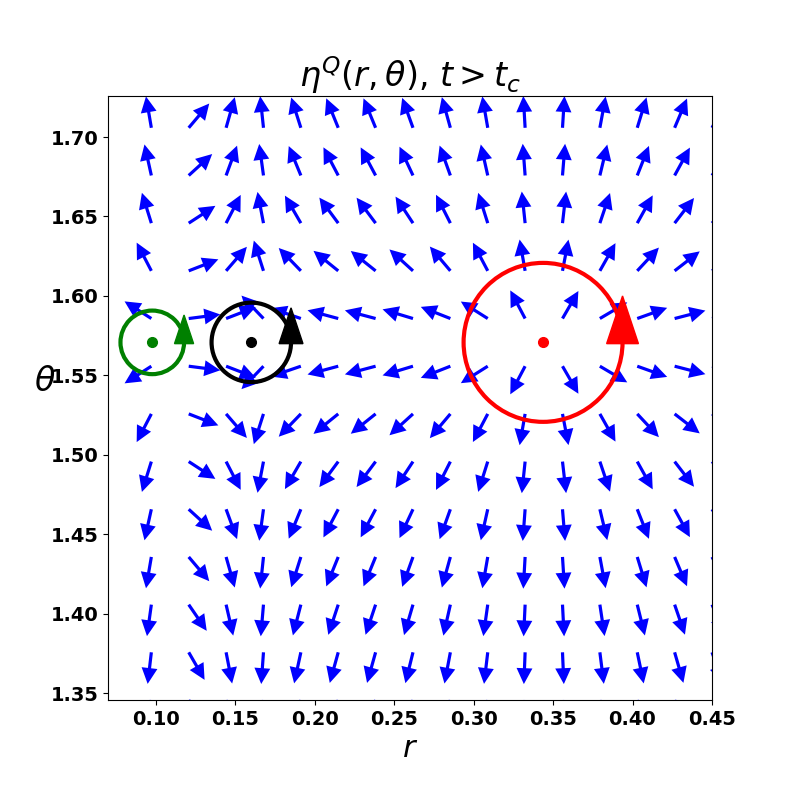}\>
		\includegraphics[scale=.36]{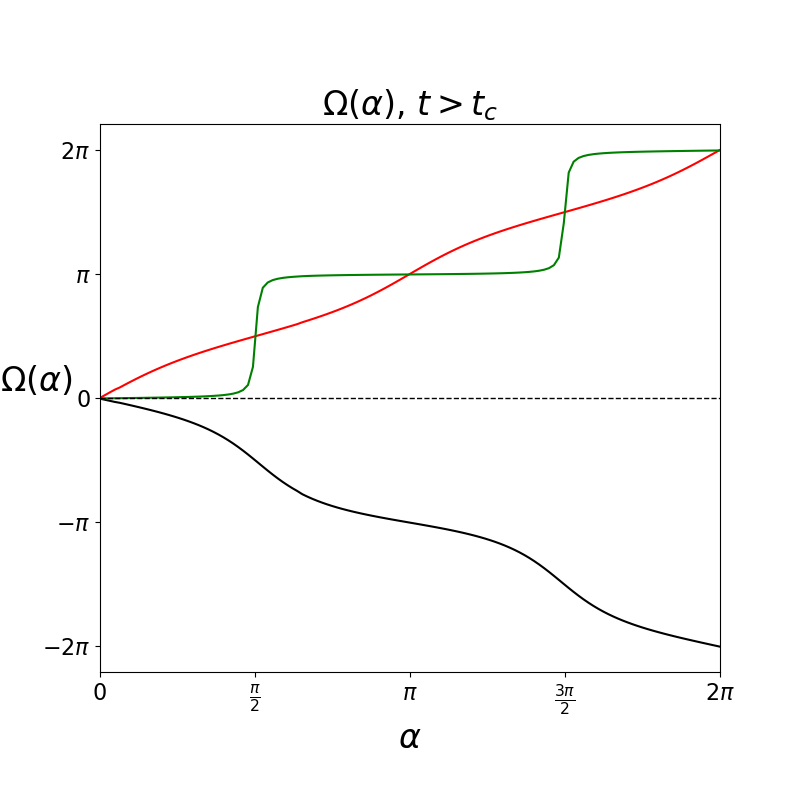}\\
  \includegraphics[scale=.36]{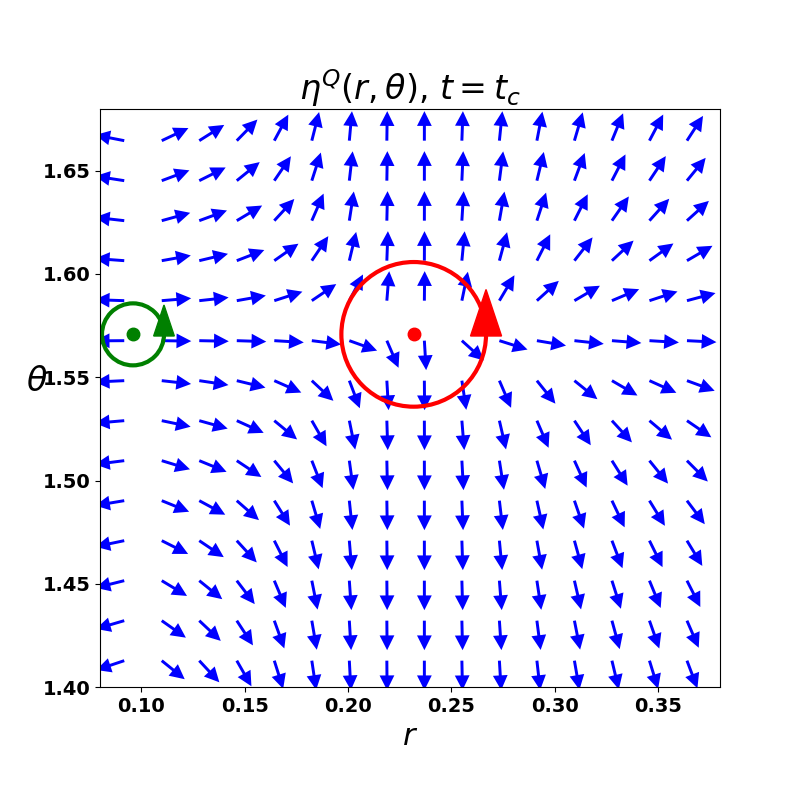}\>
		\includegraphics[scale=.36]{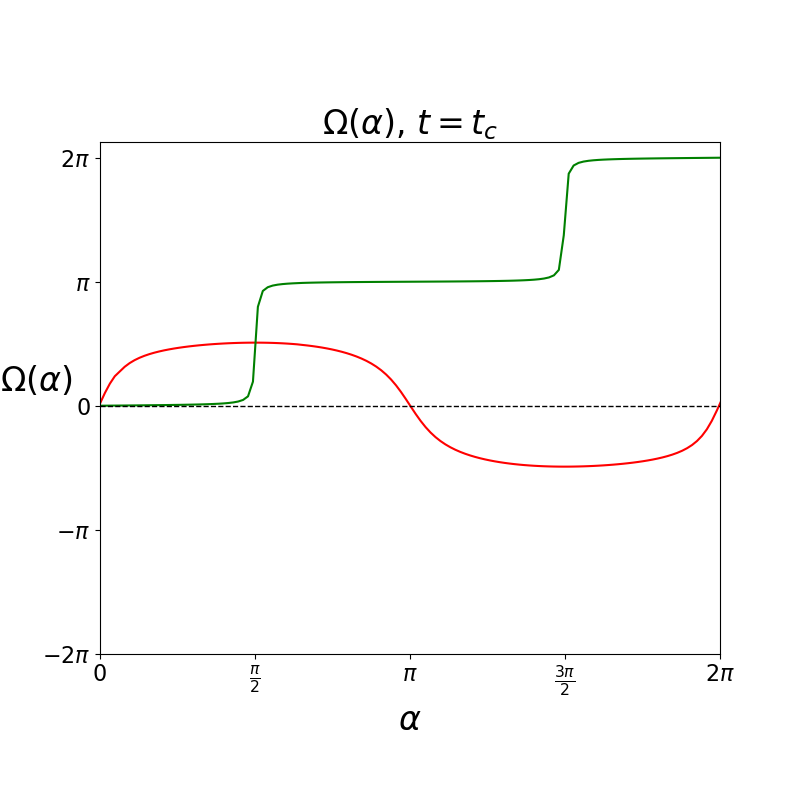}\\
  \includegraphics[scale=.36]{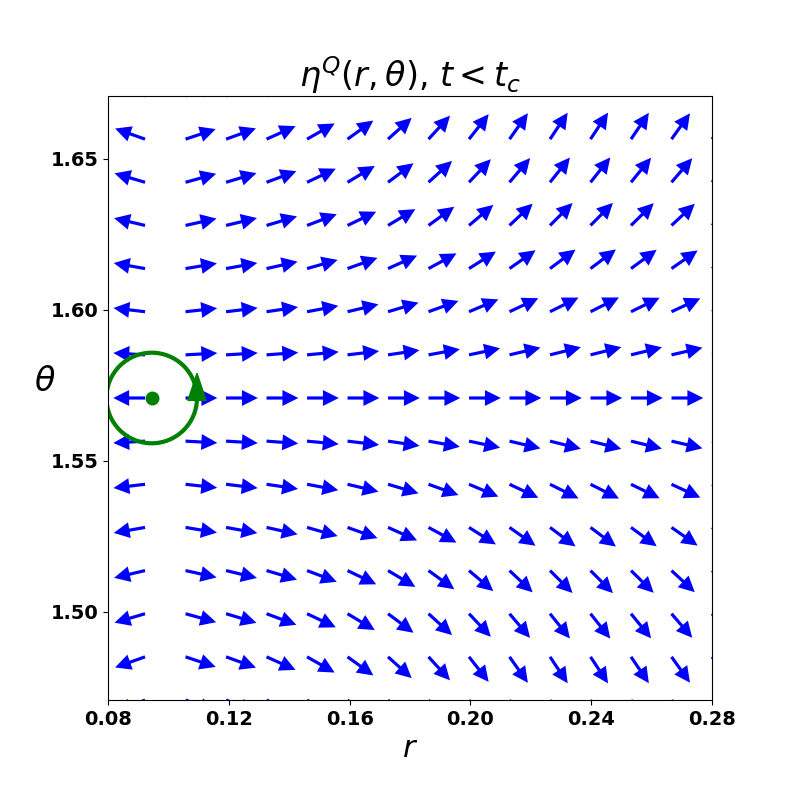}\>
		\includegraphics[scale=.36]{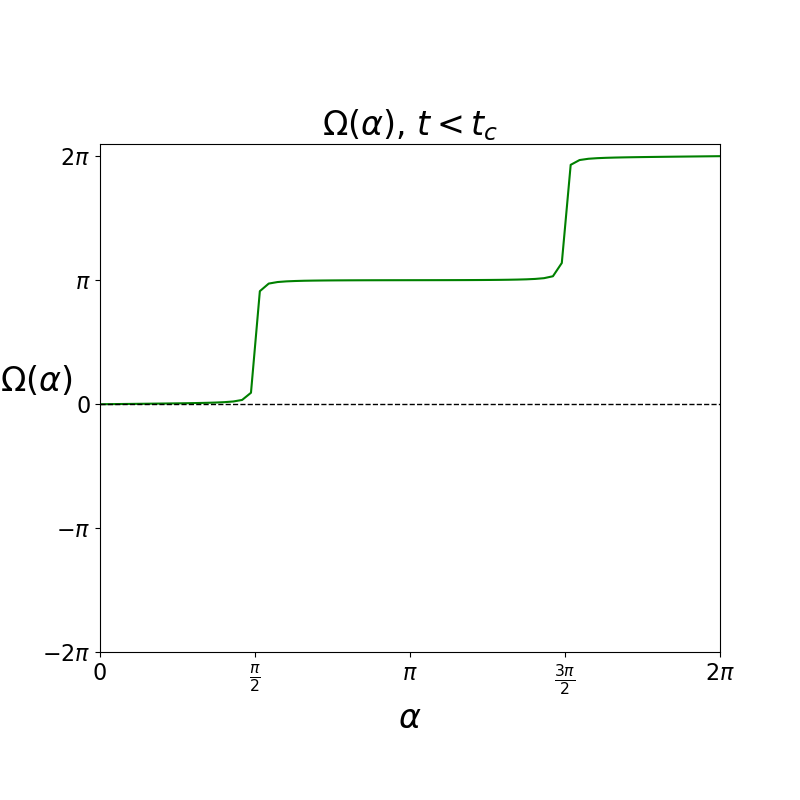}\\
	\end{tabbing}
	\vspace{-1.7cm}
	\caption{\textcolor{blue}{\footnotesize\it Phases of the charged-flat black hole in the canonical ensemble within Rényi formalism in the critical regime for scaled temperatures $t > t_{c}$ (top), $t=t_c$ (middle) and $t<t_{c}$ (bottom). {\bf Left:} The normalized flow of $\eta^{Q_c}$. {\bf Right:} the behavior of deflection angle for corresponding black hole phases. In all panels, the electric potential is set to $Q=Q_c$.}}\label{fig_phases_cri_Q}
\end{figure}

We portray in Fig.\ref{fig_phases_cri_Q} the equilibrium phases of the charged-flat black hole in the critical regime. At the critical charge, three stable phases/vortices can exist depending on the coexistence temperature. Above the scaled critical temperature $t_c=1.09334$, on the top panel, we see three phases, a (LBH) phase (red dot at $r=0.30586$), an (IBH) (black dot at $r=0.17796$) and a (SBH) one (blue dot at $r=0.096842$). All these phases coexist at the same temperature as pointed out earlier by means of the $\xi$-mapping in Fig.\ref{fig_trans_cri_Q} (top panel), without a phase transition. The Topological charges are $\mathbf{+1}$ for the (SBH) and (LBH) phases testifying to their stability, but $\mathbf{-1}$ for the unstable (IBH), the total topological charge of this phase system is  $\mathbf{+1-1+1=+1}$. This indicates a continuing transformation from (SBH) to (LBH) by the intermediary action of the (IBH) phase. At the critical temperature $t=t_c$, two phases coexist, the so-called critical phase $(CR)$ manifests at $r=r_c$ with zero topological charge, and a (SBH) phase with $\mathbf{+1}$. From the necessity of topological charge conservation, the vanishing charge of the critical phase would be problematic if it were the only phase present at $t_c$, however, the appearance of the (SBH) maintains the conservation law, so that the total charge remains equal to $\mathbf{+1}$ as in the top panel. Lastly, below $t_c$, a single (SBH) phase persists (the blue dot at $r=0.094469$) acquiring a positive charge of $\mathbf{+1}$ for its thermodynamic stability. We uncovered three phase systems in the critical regime which transform continually  to one another as the coexistence temperature changes, a step-by-step increase of this temperature reveals the dynamics of this transformation is as follows, increasing the temperature from below $t_c$, shows the emergence of the critical phase, just above $t_c$  we assist at the generation from $(CR)$ of a vortex/anti-vortex pair, (LBH)/(IBH), of opposite charges which get farther from each other as temperature increases. As the (IBH) anti-vortex draws closer to the (SBH) vortex a new pair forms, (SBH)/(IBH).



\subsection{Topological charge in the supercritical regime: \texorpdfstring{$Q>Q_c$}.}

\begin{figure}[!ht]
	\centering
	\begin{tabbing}	
		\centering
		\hspace{8.3cm}\=\kill
		\includegraphics[scale=.4]{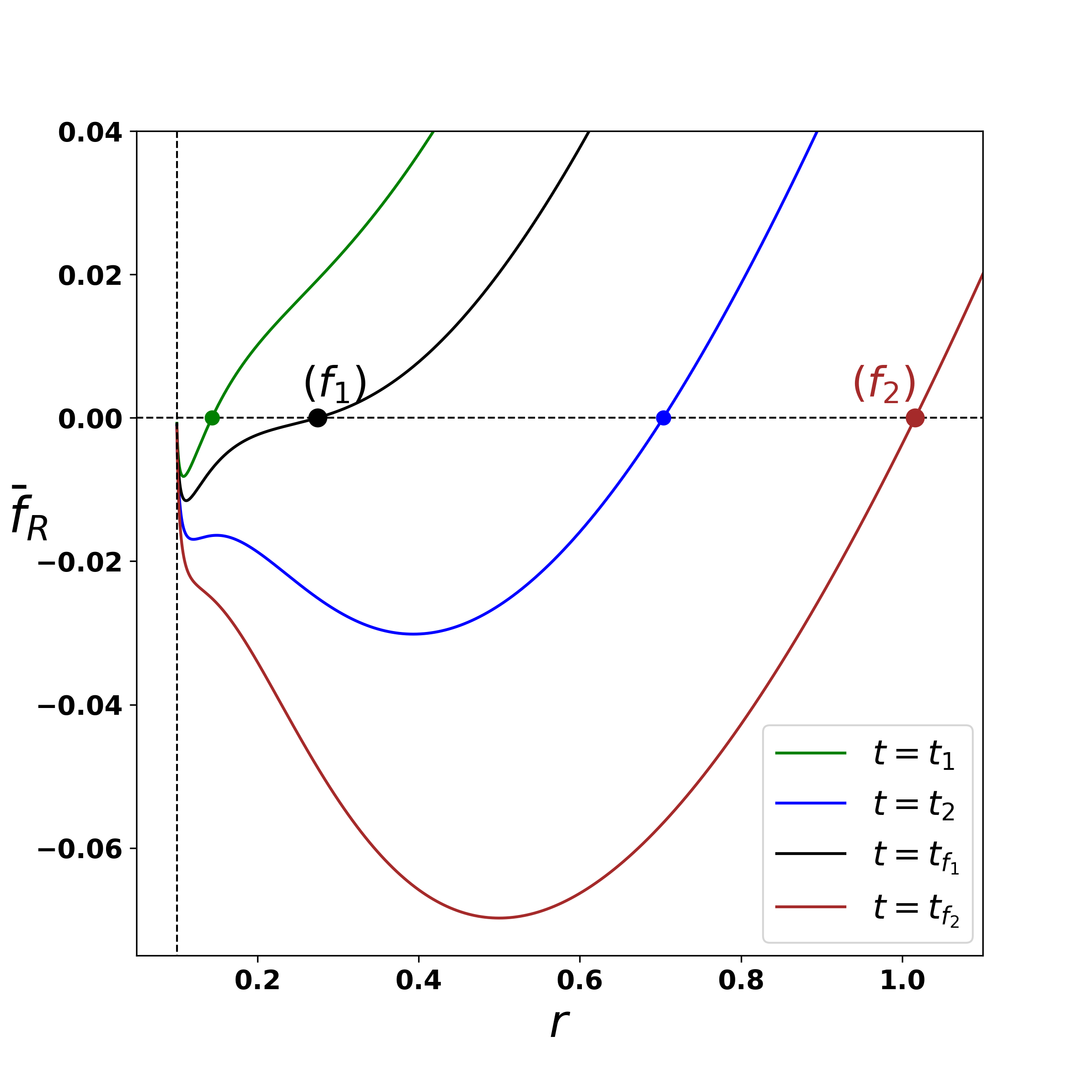}\>
		\includegraphics[scale=.4]{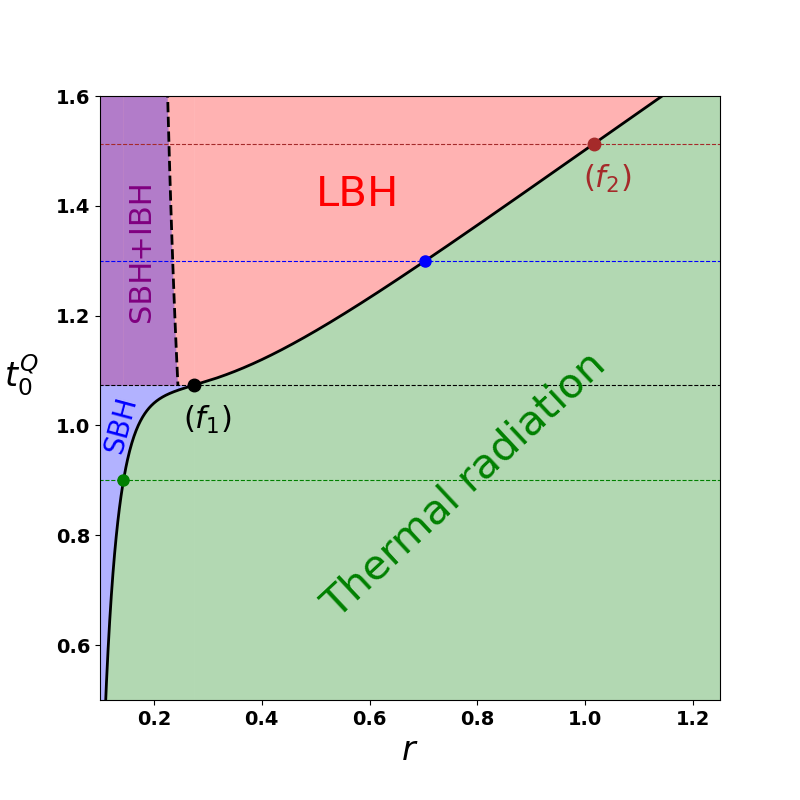}
	\end{tabbing}
	\vspace{-1.2cm}
	\caption{\footnotesize\it {\bf Left:} Behavior of the Bragg-Williams free energy $ \bar{f_R} $, as a function of $r$ at different temperatures $ t $ for the supercritical regime $Q=0.1>Q_c$. {\bf Right:} The coexistence temperature $t_0^Q$ of the supercritical black hole phase, as a function of $r$, shows transition points at temperatures chosen in the left panel. Shown also inflection points $(f_1)$ and $(f_2)$. Colored dots and horizontal lines correspond to respective colored curves in the left panel.}\label{fig4_Q}
 \end{figure}

\paragraph{}Looking at Fig \ref{fig4_Q}, it is obvious that the supercritical regime is characterized by the absence of any global phase transition point of the kind seen in the previous regimes, in particular, there is no generation or annihilation points, indeed, as the temperature increases the black hole goes from a small supercritical black hole state into a large supercritical one. Notable also is the existence of inflection points where, $\frac{\partial^2t_0^Q}{\partial r^2}=0$ and $\frac{\partial t_0^Q}{\partial r}\neq0$, represented by the point $(f_1)$ at $(r=0.27464, \;t=1.07344)$  and $(f_2)$ at $(r=1.01616,\;t=1.51304)$ for $Q=0.1$. Although these are not defects in the $\psi$-mapping used so far to detect global transition features, it is intriguing to note their disappearance as the electric charge is increased above the critical value, which is strangely reminiscent of the behavior of the transition points, $(a)$ and $(g)$, in the critical regime. One may expect that $(f_1)$ and $(f_2)$ are a vortex/anti-vortex pair that is created and annihilated in the flow of some other vector field. The phase structure illustrated on the right panel shows the stability domains of the different black hole states. Small black hole phase dominates for very small order parameter $r$ and temperatures below $t_{f_1}$. At higher temperatures, the (LBH) phase prevails with a growing stability domain as temperature increases. Also appreciable is a small \textcolor{blue}{spinodal} domain indicated by the black dashed line, where the tight pair (SBH)/(IBH) dwells at higher temperatures and small radii.

 \paragraph{}We proceed to investigate the topological structure of the vector fields, $\psi^Q$, and $\eta^Q$, defined through Eq.\eqref{psi_Q} and Eq.\eqref{xi_Q}, respectively, where the electric charge $Q$ is held fixed at a value $Q=0.1>Q_c$. 

\begin{figure}[H]
	\centering
	\begin{tabbing}
		\centering
		\hspace{8.3cm}\=\kill
		\includegraphics[scale=.4]{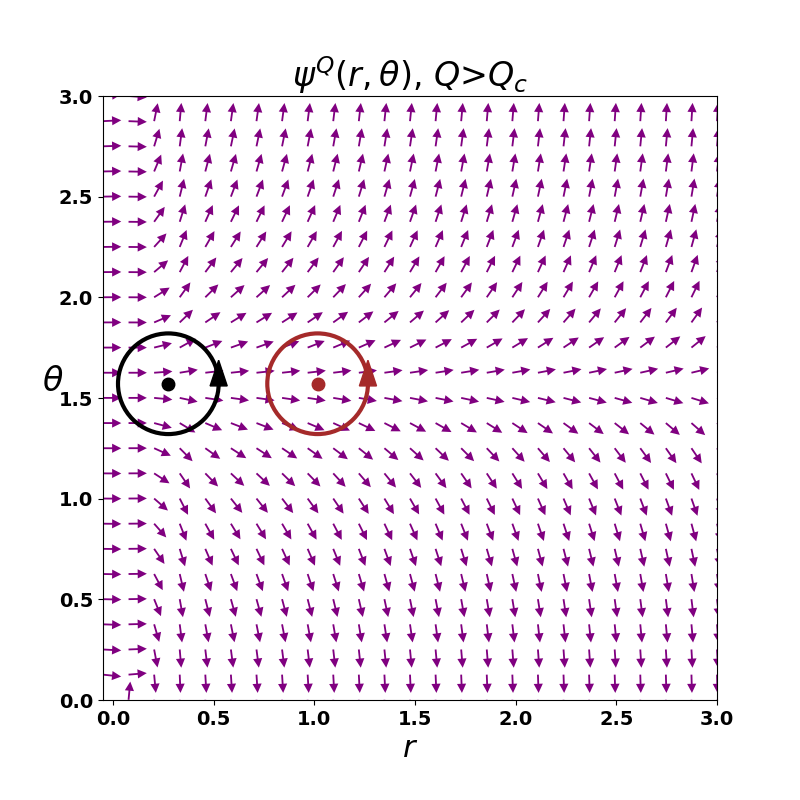}\>
		\includegraphics[scale=.4]{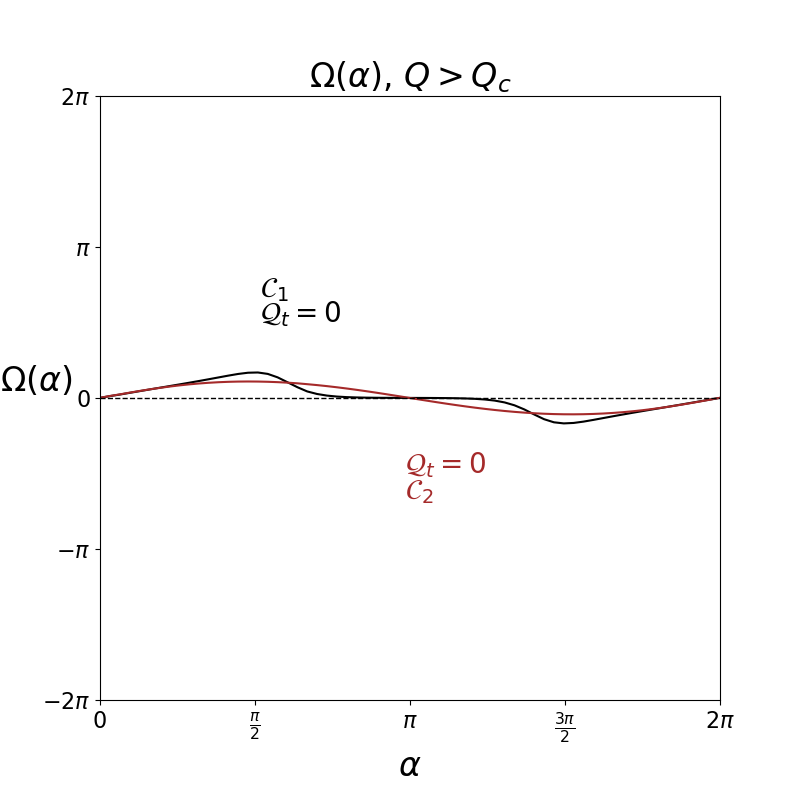}\\
	\end{tabbing}
	\vspace{-1.6cm}
	\caption{\footnotesize\it Topological charge at the inflection point $(f_1)$(black dot)  and $f_2$ (brown dot) in the supercritical regime for a charged-flat black hole in the canonical ensemble within Rényi formalism.}
 \label{fig8}
\end{figure}

\begin{figure}[!ht]
	\centering
	\begin{tabbing}
		\centering
		\hspace{8.3cm}\=\kill
		\includegraphics[scale=.4]{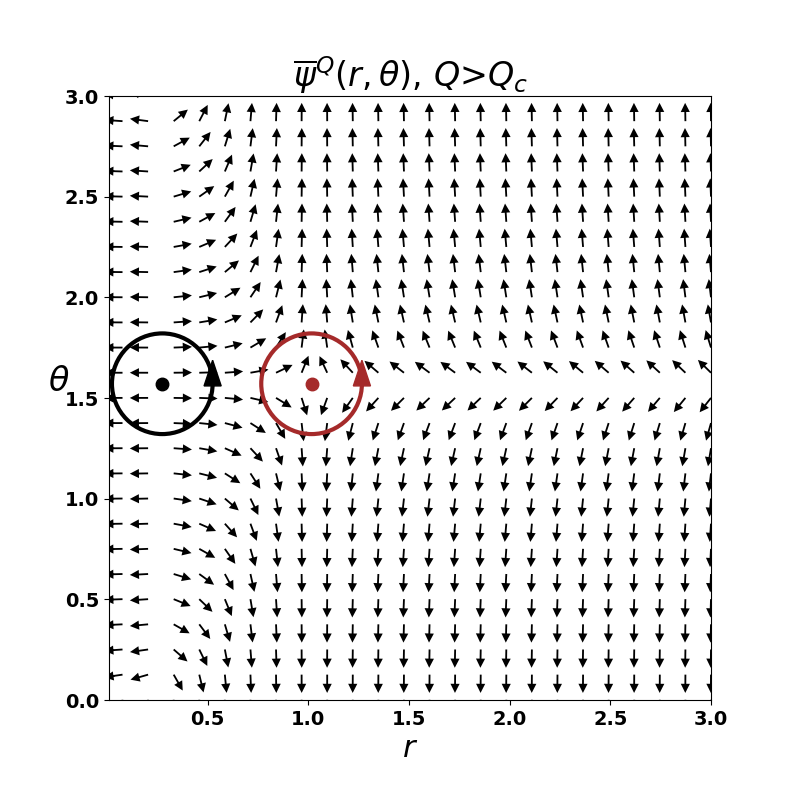}\>
		\includegraphics[scale=.4]{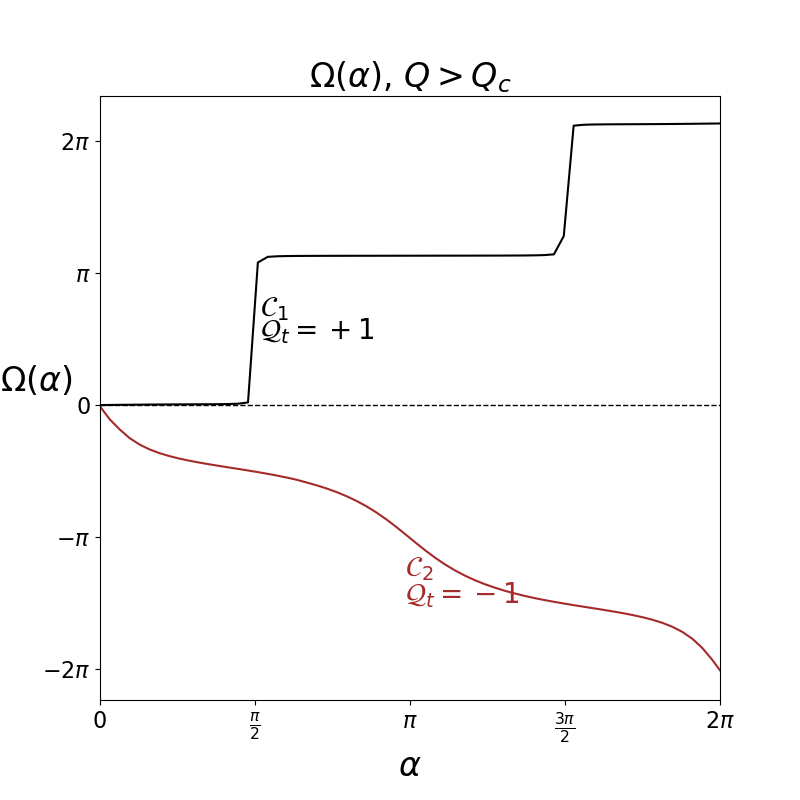}\\
	\end{tabbing}
	\vspace{-1.7cm}
	\caption{\footnotesize\it Topological charge at the inflection points $(f_1)$  and $(f_2$) in the supercritical regime as defects in the flow of the vector field $\overline{\psi}^Q$ for a charged-flat black hole in the canonical ensemble within Rényi formalism.}
 \label{fig9}
\end{figure}

\paragraph{}The computation of the topological charges of the inflection points $(f_1)$ and $(f_2)$, depicted in Fig.\ref{fig8}, shows a vanishing topological charge $\mathcal{Q}_{t}=\mathbf{0}$, similar to the isolated critical point $(c)$. However, the null value of $\mathcal{Q}_{t}$ shows only that these points are not  defects in the flow of $\psi^Q$ as defined in Eq.\eqref{psi_Q}. But they are by definition. i.e. $\small \frac{\partial^2t_0^Q}{\partial r^2}=0$, defects in the flow of the vector field $\overline{\psi}^Q$, defined as the gradient of the scalar field $\frac{1}{\sin(\theta)}\frac{\partial t_0^Q}{\partial r}$, where $t_0^Q$ is the coexistence temperature, with a value of $\mathbf{+1}$ at point $(f_1)$ and $\mathbf{-1}$ at point $(f_2)$, as shown in Fig \ref{fig9}, which confirms our expectation about the dynamics of these inflection points.

\paragraph{}This observation invites the idea of assigning in general, a hierarchy of topological charges to phase transitions and equilibrium phases by considering multiple derivatives of Bragg-Williams free energy $\bar{f}_R$ and the coexistence temperature $t_0$. The aim is to have a more refined classification of the various black hole solutions according to the acquired topological charges. 

\begin{figure}[!ht]
	\centering
	\begin{tabbing}
		\centering
		\hspace{8.3cm}\=\kill
		\includegraphics[scale=.38]{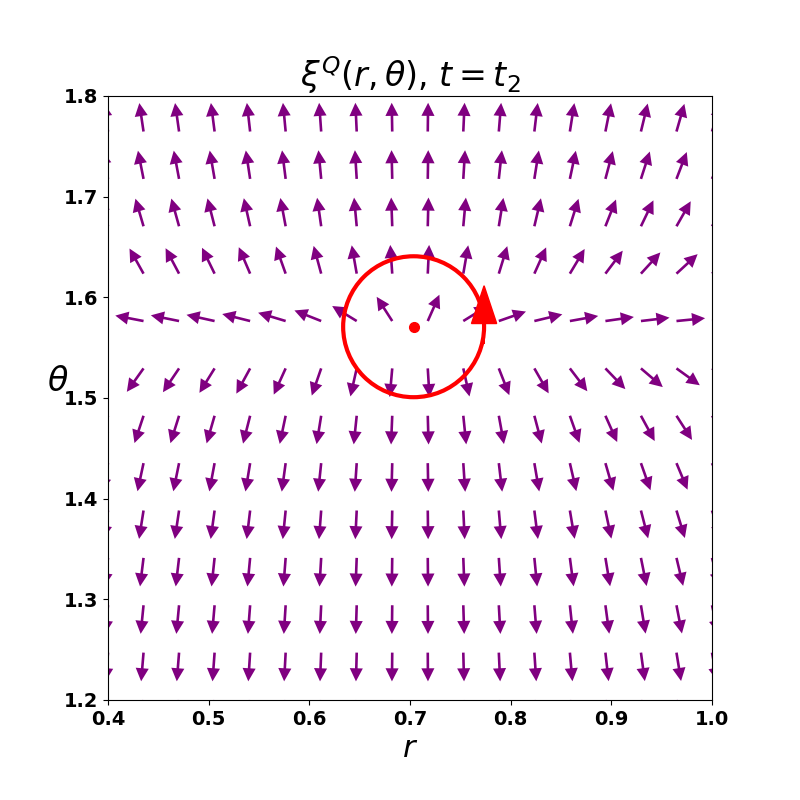}\>
		\includegraphics[scale=.36]{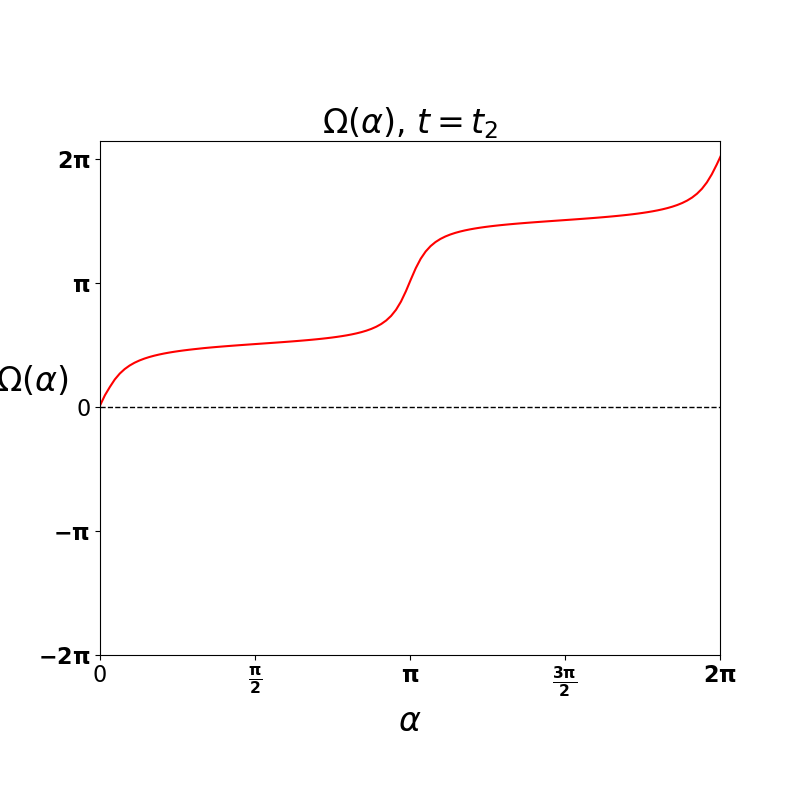}\\
  \includegraphics[scale=.38]{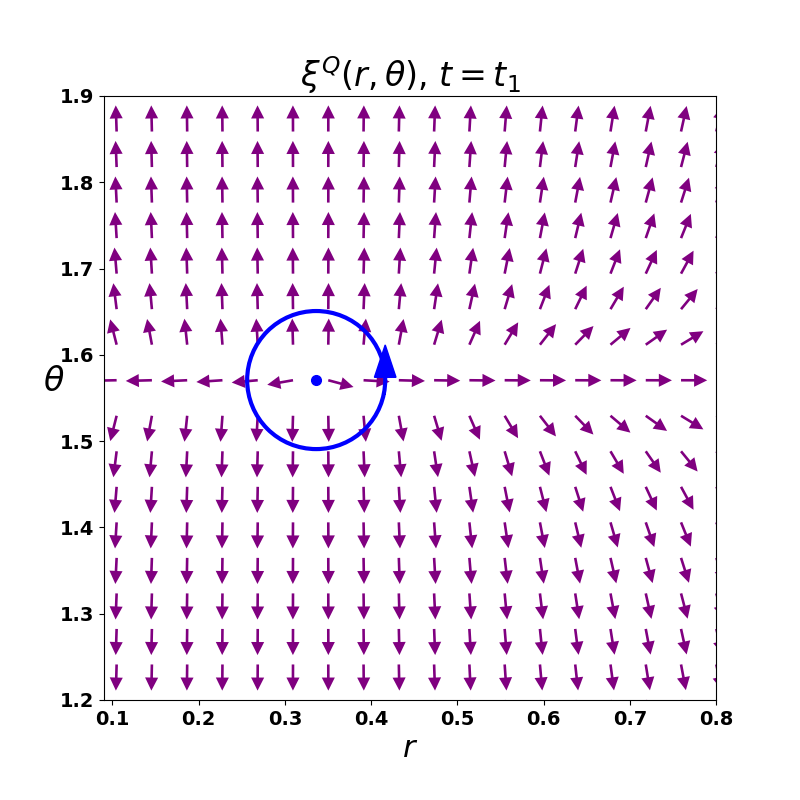}\>
		\includegraphics[scale=.36]{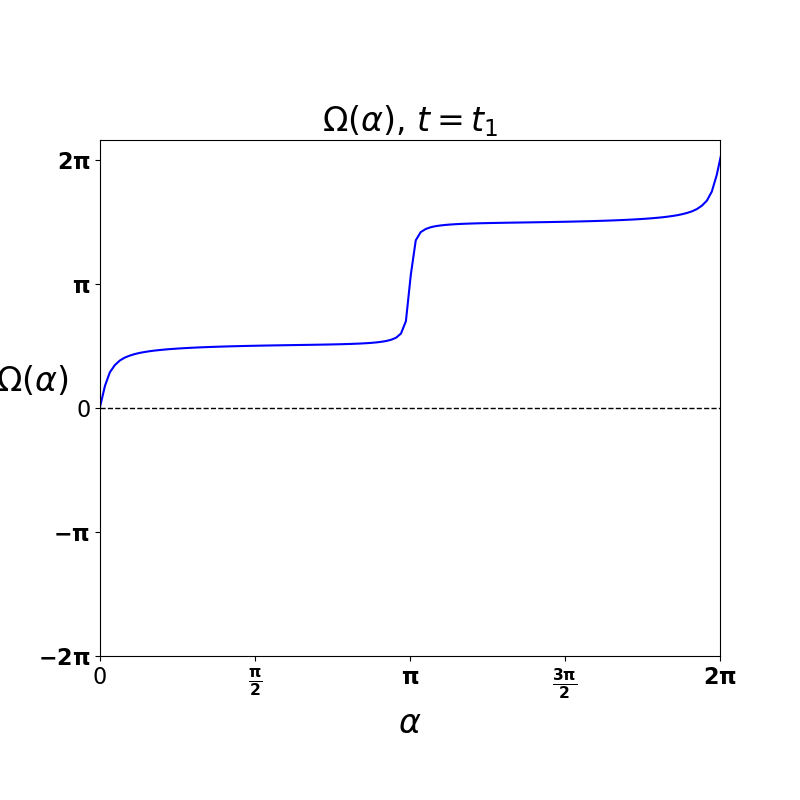}\\
	\end{tabbing}
	\vspace{-1.7cm}
	\caption{\footnotesize\it Phase transitions of the charged-flat black hole in the canonical ensemble within Rényi formalism in the supercritical regime for scaled temperatures $t_{2}=1.3$ (top) and $t_{1}=0.9$ (bottom). {\bf Left:} The normalized flow of $\xi^{Q}$. {\bf Right:} the behavior of deflection angle for corresponding black hole phases. In all panels, the electric potential is set to $Q=0.1$.}\label{fig_xi_sup_Q}
\end{figure}

The topology of the $\xi$-mapping in Fig \ref{fig_xi_sup_Q}, presents two transition points at chosen scaled temperatures above and below the inflection point $(f_1)$. On the top panel, $t_2=1.3$, we find a supercritical transition at $r\approx0.70359$, where the transformation $LBH \longleftrightarrow Thermal\; radiation$ takes place. On the bottom, $t_1=0.9$, a transition appears at $r\approx0.14358$, here, $SBH \longleftrightarrow Thermal\; radiation$ happens. They both get a topological charge equal to $\mathbf{+1}$. By the conservation of the topological charge of the $\xi$-mapping, the inflection points are also expected to have the same charge.
\begin{figure}[!ht]
	\centering
	\begin{tabbing}
		\centering
		\hspace{8.3cm}\=\kill
		\includegraphics[scale=.42]{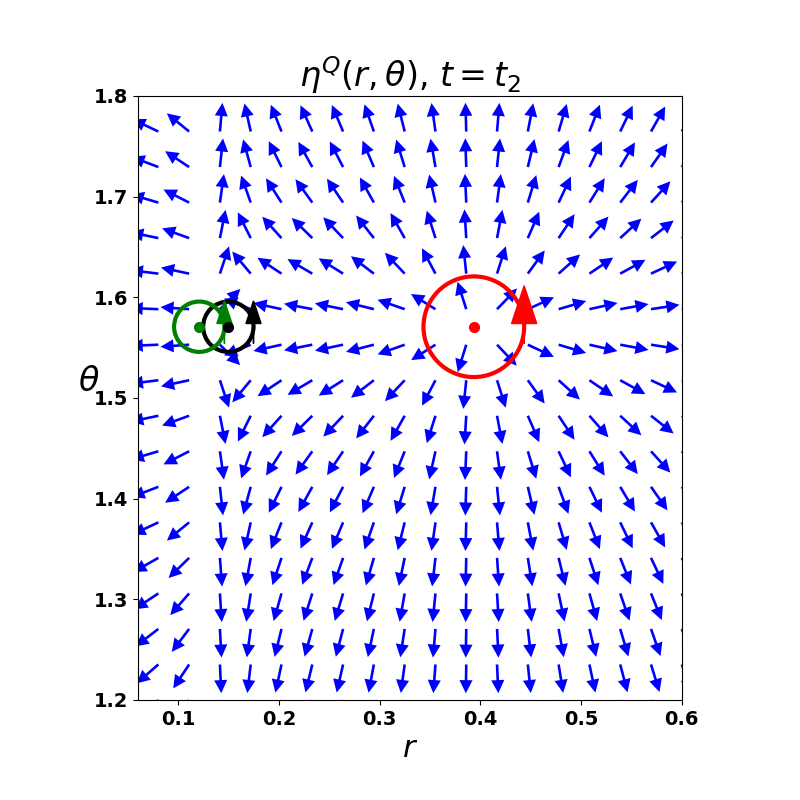}\>
		\includegraphics[scale=.42]{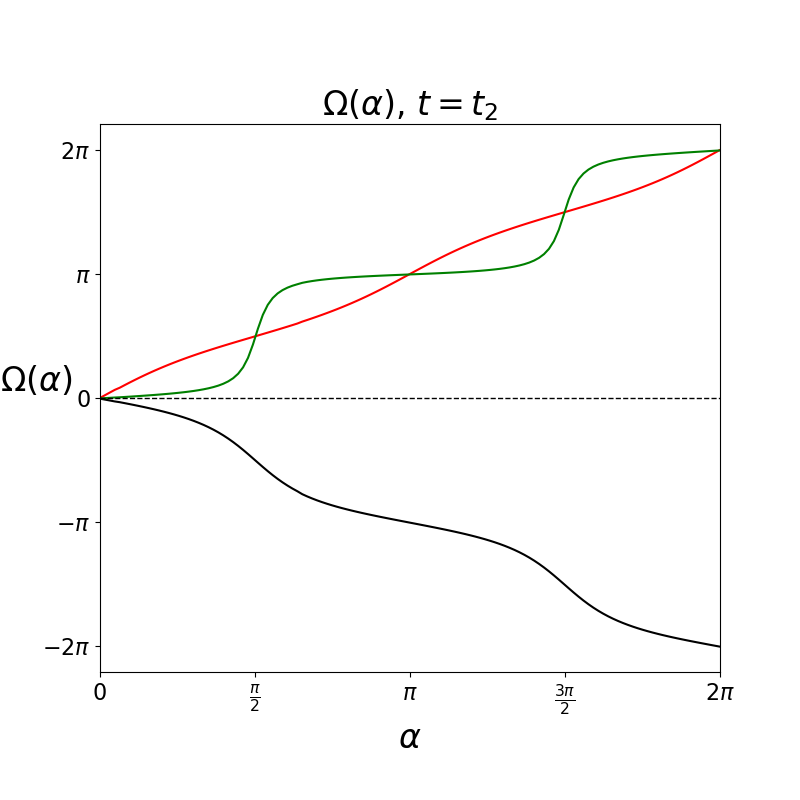}\\
  \includegraphics[scale=.38]{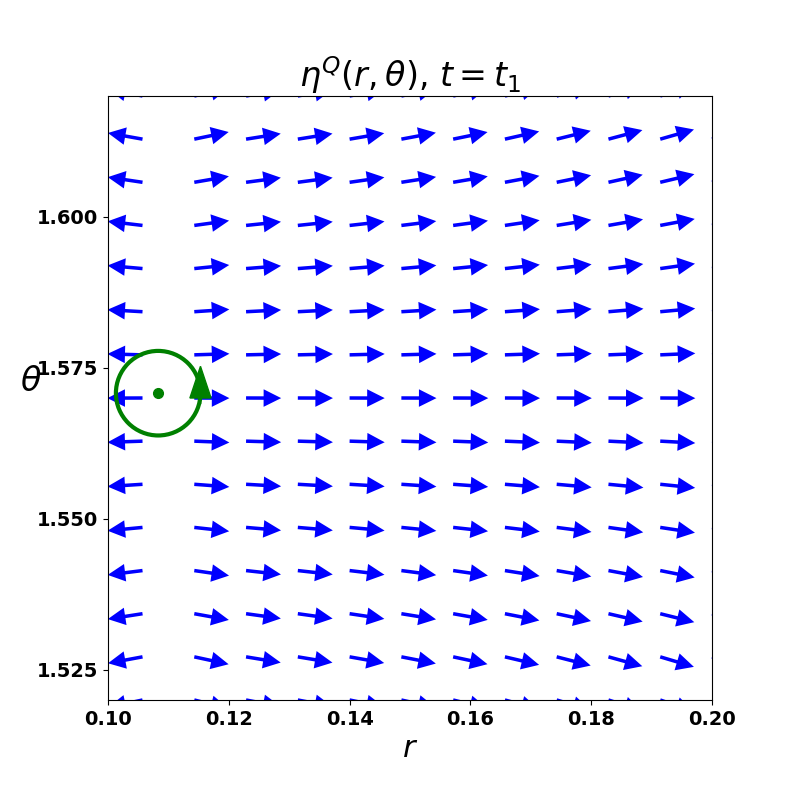}\>
		\includegraphics[scale=.36]{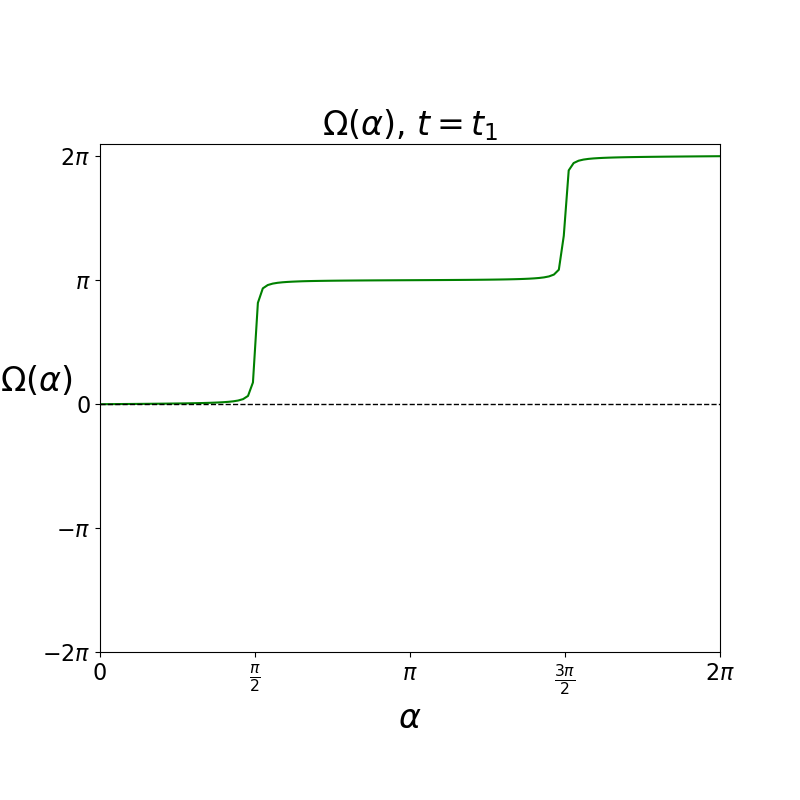}\\
	\end{tabbing}
	\vspace{-1.7cm}
	\caption{\footnotesize\it Phases of the charged-flat black hole in the canonical ensemble within Rényi formalism in the supercritical regime for scaled temperatures $t_{2}=1.3$ (top) and $t_{1}=0.9$ (bottom). {\bf Left:} The normalized flow of $\eta^{Q}$. {\bf Right:} the behavior of deflection angle for corresponding black hole phases. In all panels, the electric potential is set to $Q=0.1$.}\label{fig_phases_sup_Q}
\end{figure}

We finish by inspecting the topological charge of the phase systems above and below the point $(f_1)$. Fig.\ref{fig_phases_sup_Q} presents two phase profiles. At $t=t_2$ (top panel), where the (LBH) phase (red dot at $r=0.39338$) dominates with the existence of a tight (SBH)/(IBH) vortex pair almost annihilating one another (green/black dots at $r=0.12066$ and $r=0.14954 $). The total topological charge is $\mathbf{+1-1+1=+1}$ with (SBH) and (LBH) are the stable phases. The situation changes drastically at $t=t_1$ (bottom panel), only the (SBH) phase exists (the green dot at $r=0.10831$) and acquires a positive topological charge of $\mathbf{+1}$, in a total agreement with the $\eta$-mapping charge conservation requirement.

\clearpage

\section{Conclusion}
In this study, we have built upon the framework introduced in \cite{Yerra:2022coh} and broadened its application to determine the topological charge linked with the Hawking-Page and Van-der-Waals transition points within Rényi statistics. This endeavor is aimed at capturing the nonextensive characteristics of the black hole entropy effect from a topological point of view. To achieve this, we systematically integrate the Rényi statistics formalism into the off-shell Bragg-Williams free energy landscape, allowing us to meticulously examine the phase transition profile of the charged and uncharged black hole. 

We recapitulate in tables  (\ref{table:table_grand_cano}), (\ref{table:table_cano}) and (\ref{table:table_cano_xi}), the classifications introduced by the $\psi$- and $\eta$- and $\xi$-mapping for phase transition points and equilibrium phases in the grand canonical and canonical ensembles.

Of particular interest, our results show that the (VdW) phase transition, exhibited by the charged-flat black hole in the canonical ensemble is of a different topological class than the (HP) one with the conservation of the total topological charge playing a central explanatory role. As is summarized in tables (\ref{table:table_grand_cano}) and (\ref{table:table_cano}), along vertical axes the total topological charge is conserved and amounts to $\mathbf{+1}$ for the (HP) phase transition but vanishes or the (VdW) phase transition. On the contrary, the total topological charge vanishes for equilibrium phase systems associated with (HP) transition but is equal to $\mathbf{+1}$ for the (VdW) phases. This interplay of values may hint into a possible correspondence between the (HP) and (VdW) phase transitions and their associated phase systems, especially knowing that we deal with the same black hole system treated in two statistical ensembles which should not change its underlying physics.

\begin{table}[!ht]
\centering
\resizebox{0.75\textwidth}{!}{%
\begin{tabular}{c|cccc|}
\cline{2-5}
& \multicolumn{4}{c|}{Sch and RN types in grand canonical ensemble}  \\ \cline{2-5} 
& \multicolumn{1}{c|||}{$\psi$-mapping} & \multicolumn{3}{c|}{$\eta$-mapping}\\ 
\hline
\multicolumn{1}{|c|}{$\phi=0$}     & \multicolumn{1}{c|||}{\begin{tabular}[c]{@{}c@{}}$(HP)$\\ $\mathbf{+1}$\end{tabular}} & \multicolumn{1}{c||}{\begin{tabular}[c]{@{}c@{}}BH at $t_{min}$\\  $\mathbf{0}$\end{tabular}} & \multicolumn{1}{c|}{\begin{tabular}[c]{@{}c@{}}\hspace*{0.3cm}SBH\hspace*{0.3cm}\\  $\mathbf{-1}$\end{tabular}} & \begin{tabular}[c]{@{}c@{}}LBH\\ $\mathbf{+1}$\end{tabular}\\ 
\hline
\multicolumn{1}{|c|}{$\phi\neq 0$} & \multicolumn{1}{c|||}{\begin{tabular}[c]{@{}c@{}}$(HP)$\\  $\mathbf{+1}$\end{tabular}} & \multicolumn{1}{c||}{\begin{tabular}[c]{@{}c@{}}BH at $t_{min}$\\  $\mathbf{0}$\end{tabular}}& \multicolumn{1}{c|}{\begin{tabular}[c]{@{}c@{}}SBH\\  $\mathbf{-1}$\end{tabular}} & \begin{tabular}[c]{@{}c@{}}LBH\\  $\mathbf{+1}$\end{tabular} 
\\ \hline
& \multicolumn{1}{c|||}{\begin{tabular}[c]{@{}c@{}}\textcolor{blue}{Total $\mathcal{Q}_{t}$}\\ \textcolor{blue}{\textbf{$\mathbf{+1}$}}\end{tabular}} & \multicolumn{3}{c|}{\begin{tabular}[c]{@{}c@{}}\textcolor{blue}{Total $\mathcal{Q}_{t}$}\\ \textcolor{blue}{\textbf{$\mathbf{0}$}}\end{tabular}}\\
\cline{2-5} 
& \multicolumn{4}{c|}{\begin{tabular}[c]{@{}c@{}}$\mathbf{t\nearrow}$\\$\longrightarrow$\end{tabular}}  \\ 
\cline{2-5}

\end{tabular}%
}
\caption{\footnotesize\it Summary of our results for the $\psi$- and  $\eta$-mapping in the grand canonical ensemble. Total topological charge is indicated in blue.}\label{table:table_grand_cano}
\end{table}

\begin{table}[!ht]
\vspace*{1 cm}
\centering
\resizebox{0.8\textwidth}{!}{%
\begin{tabular}{c|cccccccc|}
\cline{2-9}
& \multicolumn{8}{c|}{\begin{tabular}[c]{@{}c@{}}Reissner-Nordström type in canonical ensemble\end{tabular}}  \\ 
\cline{2-9} 
 & \multicolumn{2}{c|||}{ $\psi$-mapping} & \multicolumn{6}{c|}{ $\eta$-mapping}\\ 
\hline
\multicolumn{1}{|c|}{\begin{tabular}[c]{@{}l@{}}$Q>Q_c$\end{tabular}} & \multicolumn{1}{c|}{\begin{tabular}[c]{@{}c@{}}$(f_1)$\\ \textbf{$\mathbf{0}$}\end{tabular}} & \multicolumn{1}{c|||}{\begin{tabular}[c]{@{}c@{}}$(f_2)$\\ \textbf{$\mathbf{0}$}\end{tabular}} & \multicolumn{3}{c||}{\begin{tabular}[c]{@{}c@{}}SBH\\ \textbf{$\mathbf{+1}$}\end{tabular}} & \multicolumn{1}{c|}{\begin{tabular}[c]{@{}c@{}}SBH\\ \textbf{$\mathbf{+1}$}\end{tabular}}& \multicolumn{1}{c|}{\begin{tabular}[c]{@{}c@{}}IBH\\ \textbf{$\mathbf{-1}$}\end{tabular}}& \multicolumn{1}{c|}{\begin{tabular}[c]{@{}c@{}}LBH\\ \textbf{$\mathbf{+1}$}\end{tabular}}\\ 
\hline
\multicolumn{1}{|c|}{\begin{tabular}[c]{@{}l@{}}$Q=Q_c$\end{tabular}} & \multicolumn{2}{c|||}{\begin{tabular}[c]{@{}c@{}} $\mathbf{(c)}$\\ \textbf{$\mathbf{0}$}\end{tabular}} & \multicolumn{1}{c||}{\begin{tabular}[c]{@{}c@{}} SBH\\ \textbf{$\mathbf{+1}$}\end{tabular}} & \multicolumn{1}{c|}{\begin{tabular}[c]{@{}c@{}}SBH\\ \textbf{$\mathbf{+1}$} \end{tabular}} & \multicolumn{1}{c||}{\begin{tabular}[c]{@{}c@{}}\textbf{CR}\\ \textbf{$\mathbf{0}$} \end{tabular}}& \multicolumn{1}{c|}{\begin{tabular}[c]{@{}c@{}}SBH\\ \textbf{$\mathbf{+1}$} \end{tabular}}& \multicolumn{1}{c|}{\begin{tabular}[c]{@{}c@{}}IBH\\ \textbf{$\mathbf{-1}$} \end{tabular}}& \multicolumn{1}{c|}{\begin{tabular}[c]{@{}c@{}}LBH\\ \textbf{$\mathbf{+1}$} \end{tabular}} \\ 
\hline
\multicolumn{1}{|c|}{\begin{tabular}[c]{@{}l@{}}$Q<Q_c$\end{tabular}} & \multicolumn{1}{c|}{\begin{tabular}[c]{@{}l@{}}$(g)$\\ \textbf{$\mathbf{-1}$}\end{tabular} } & \multicolumn{1}{c|||}{\begin{tabular}[c]{@{}l@{}}$(a)$\\ \textbf{$\mathbf{+1}$}\end{tabular}}& \multicolumn{2}{c|}{\begin{tabular}[c]{@{}l@{}}SBH\\ \textbf{$\mathbf{+1}$}\end{tabular} } & \multicolumn{2}{c|}{\begin{tabular}[c]{@{}l@{}}IBH\\ \textbf{$\mathbf{-1}$}\end{tabular}} & \multicolumn{2}{c|}{\begin{tabular}[c]{@{}l@{}}LBH\\ \textbf{$\mathbf{+1}$}\end{tabular}} \\ 
\hline
& \multicolumn{2}{c|||}{\begin{tabular}[c]{@{}c@{}}\textcolor{blue}{Total $\mathcal{Q}_{t}$}\\ \textcolor{blue}{\textbf{$\mathbf{0}$}}\end{tabular} } & \multicolumn{6}{c|}{\begin{tabular}[c]{@{}c@{}}\textcolor{blue}{Total $\mathcal{Q}_{t}$}\\ \textcolor{blue}{\textbf{$\mathbf{+1}$}}\end{tabular}}\\ 
\cline{2-9} 
& \multicolumn{8}{c|}{\begin{tabular}[c]{@{}c@{}}$\mathbf{t\nearrow}$\\ $\longrightarrow$\end{tabular}} \\
\cline{2-9} 
\end{tabular}
}

\caption{\footnotesize\it Summary of our results for the $\psi$- and  $\eta$-mapping in the canonical ensemble. The total topological charge is indicated in blue.}\label{table:table_cano}

\end{table}
\begin{table}[!ht]
\vspace*{1 cm}\hspace*{-1.5 cm}
\centering
\resizebox{1.1\textwidth}{!}{%
\begin{tabular}{c|ccccc|}
\cline{2-6}
& \multicolumn{5}{c|}{\begin{tabular}[c]{@{}c@{}}Reissner-Nordström type in canonical ensemble\end{tabular}}  \\ 
\cline{2-6} 
 & \multicolumn{5}{c|}{ $\xi$-mapping}\\ 
\hline
\multicolumn{1}{|c|}{\begin{tabular}[c]{@{}l@{}}$Q>Q_c$\end{tabular}}  & \multicolumn{2}{c||}{\begin{tabular}[c]{@{}c@{}}$Thermal\longleftrightarrow SBH$\\ \textbf{$\mathbf{+1}$}\end{tabular}}& \multicolumn{3}{c|}{\begin{tabular}[c]{@{}c@{}}$Thermal\longleftrightarrow LBH$\\ \textbf{$\mathbf{+1}$}\end{tabular}} \\ 
\hline
\multicolumn{1}{|c|}{\begin{tabular}[c]{@{}l@{}}$Q=Q_c$\end{tabular}}& \multicolumn{1}{c||}{\begin{tabular}[c]{@{}c@{}} $Thermal\longleftrightarrow SBH$\\ \textbf{$\mathbf{+1}$}\end{tabular}} & \multicolumn{3}{c||}{\begin{tabular}[c]{@{}c@{}}$Thermal\longleftrightarrow SBH$\\ \textbf{$\mathbf{+1}$} \end{tabular}}& \multicolumn{1}{c|}{\begin{tabular}[c]{@{}c@{}}$Thermal\longleftrightarrow LBH$\\ \textbf{$\quad\mathbf{+1}$} \end{tabular}}  \\ 
\hline
\multicolumn{1}{|c|}{\begin{tabular}[c]{@{}l@{}}$Q<Q_c$\end{tabular}} & \multicolumn{1}{c||}{\begin{tabular}[c]{@{}c@{}}$Thermal\longleftrightarrow SBH$\\ \textbf{$\mathbf{+1}$}\end{tabular} } & \multicolumn{3}{c||}{\begin{tabular}[c]{@{}l@{}}$SBH\longleftrightarrow IBH\longleftrightarrow LBH\longleftrightarrow Thermal$\\ \textbf{$\qquad\;\;\mathbf{+1}\qquad\quad\mathbf{-1}\qquad \quad\mathbf{+1}$}\end{tabular}}  & \multicolumn{1}{c|}{\begin{tabular}[c]{@{}c@{}}$Thermal\longleftrightarrow LBH$\\ \textbf{$\quad\mathbf{+1}$}\end{tabular}} \\ 
\hline
 & \multicolumn{5}{c|}{\begin{tabular}[c]{@{}c@{}}\textcolor{blue}{Total $\mathcal{Q}_{t}$}\\ \textcolor{blue}{\textbf{$\mathbf{+1}$}}\end{tabular}}\\ 
\cline{2-6} 
& \multicolumn{5}{c|}{\begin{tabular}[c]{@{}c@{}}$\mathbf{t\nearrow}$\\ $\longrightarrow$\end{tabular}} \\
\cline{2-6} 
\end{tabular}
}

\caption{\footnotesize\it Summary of our results for the $\xi$-mapping in the canonical ensemble. The total topological charge is indicated in blue.}\label{table:table_cano_xi}

\end{table}

This research study consolidates  also a previous proposal linking the Hawking-Page transition and the possibility of initiating a concurrent topological shift between the black hole and its surrounding space. Moreover, it extends such a proposal to Rényi statistics.
Our outcomes indicate that the values of  $\bm{+1}$/$\bm{-1}$ topological charges are associated with the locally stable/unstable black hole solutions. In addition, the total topological charges are found to be zero for uncharged/charged black holes in the grand canonical ensemble, while in the canonical ensemble a total topological charge of $\mathbf{+1}$ persists, which matches exactly the behavior of uncharged/charged asymptotically AdS black holes and unveils that these two black hole systems and the two ensembles belong to different topological classes.\\

This adds greater support to the proposed idea, highlighting a recognized correlation between the thermodynamic properties of black holes in an asymptotically flat spacetime using Rényi statistics, and the analogous phenomenon occurring in an asymptotically Anti-de-Sitter spacetime through the utilization of Gibbs-Boltzmann statistics in the same statistical ensembles.
\newpage
\appendix
\numberwithin{equation}{section}
\section{Bragg Williams in Rényi formalism in arbitrary $d$-dimensional spacetime}
In this section, we present the expressions for the Bragg-Williams approach in $d$-dimensional spacetime. A straightforward generalization starts from the line element of $d$-dimensional Reissner-Nordstrom  black hole in asymptotically flat spacetime given such as,
\begin{equation}
    ds^2=-f(r)dt^2+\frac{dr^2}{f(r)}+r^2d\omega_{d-2}.
\end{equation}
Where the blackening function stands for, 
\begin{equation}
f(r)=1-\frac{\Omega_d M}{r^{d-3}} + \frac{\Omega_{d}^{2}  \left(d - 2\right)Q^{2} }{8 \left(d - 3\right)r^{2 d-6}},
\end{equation}
where $\Omega_d$ is related to the volume of a $(d-2)$-unit sphere $Vol(S^{d-2})$ by the definition,
\begin{align}\label{omega_d}
\Omega_{d}=&\frac{16\pi}{(d-2)Vol(S^{d-2})}\\
=&\frac{8 \Gamma \left(\frac{d - 1}{2}\right)}{(d - 2) \pi^{\frac{d-3}{2}}},
\end{align}
while $d\omega_{d-2}$ denotes the line element of the unit $(d-2)$-sphere. The mass  $M$ and the electric charge $Q$, depend on the electric potential $\phi$ at the black hole outer horizon of radius $r_h$, as measured by an observer at infinity such as,

\begin{align}
M &=\displaystyle \frac{r_h^{d-3}}{ \Omega_{d}}+ \frac{2  (d - 3)\phi^{2}r_h^{d-3} }{\left(d - 2\right)\Omega_{d}}\label{BH_mass},\\
Q &=\displaystyle \frac{4 \phi r_{h}^{d - 3} \left(d - 3\right)}{\Omega_{d} \left(d - 2\right)}\label{BH_Q}.
\end{align}
In $d$-dimensional spacetime, the Rényi entropy and temperature read,

\begin{equation}
S_R=\displaystyle \frac{\ln{\left(1 + \frac{4 \pi \lambda r_{h}^{d - 2}}{\Omega_{d} \left(d - 2\right)} \right)}}{\lambda}
\end{equation}
\begin{equation}
T_R= \frac{\left(d - 3\right) \left[\Omega_{d} \left(d - 2\right) + 4 \pi \lambda r_{h}^{d - 2}\right] \left[2- d + 2 \phi^{2} \left(d - 3\right)\right]}{4 \pi \Omega_{d} r_{h} \left(d - 2\right)},
\end{equation}

which permits the generalization of the Bragg-Williams free energy to the $d$-dimensional spacetime within Rényi statistics as,

\begin{equation}\label{f_R_phi}
\bar{f_R}(r,t,\phi)=\displaystyle \frac{\lambda r^{d - 3} \left[d-2 - 2 \phi^{2} \left(d - 3\right)\right]- \Omega_{d} t \left(d - 2\right) \ln{\left(1+\frac{ 4 \pi \lambda r^{d - 2}}{\Omega_{d} \left(d - 2\right)} \right)}}{\Omega_{d} \lambda \left(d - 2\right)}
\end{equation}

As in the main text, $r$, $t$ and $\phi$ are free parameters. For $0<\lambda<<1$ and scaling $r\longrightarrow r/\lambda^{\frac{1}{d-2}}$, $t\longrightarrow t\lambda^{\frac{1}{d-2}}$ and $\bar{f_R}\longrightarrow \bar{f_R}/\lambda^{\frac{1}{d-2}}$, we obtain,

\begin{equation}\label{f_R_d}
\bar{f_R}(r,t,\phi)\approx\frac{r^{d-3}\left[8 \pi ^2 t r^{d-1}+(d-2) \Omega_d\left[d-2-4 \pi r t-2 (d-3) \phi^2\right]\right]}{(d-2)^2 \Omega _d^2}
\end{equation}
The Hawking-Page transition point in $d$-dimensional space is determined by satisfying the following conditions, $\bar{f_R}=0$ and $\partial_r\bar{f_R}=0$, or alternatively, by first imposing,
\begin{equation}
\bar{f_R}(r,t_0,\phi)=0 \implies t_0=\frac{\Gamma \left(\frac{d-1}{2}\right) r\left(2 (d-3) \phi ^2-d+2\right)}{\pi ^{\frac{d+1}{2}} r^d-4 \pi  r^2 \Gamma \left(\frac{d-1}{2}\right)},
\end{equation}
then the transition point is given by the minimum of the temperature $t_0$, such that,
\begin{equation}
\frac{\partial t_0}{\partial r}=0 \implies r_{HP}=\displaystyle 2^{\frac{2}{d - 2}} \pi^{\frac{1 - d}{2 \left(d - 2\right)}} \left(\frac{d - 1}{\Gamma\left(\frac{d}{2} - \frac{1}{2}\right)}\right)^{- \frac{1}{d - 2}}
\end{equation}
The generalization of the scalar field $\Psi=\displaystyle\frac{t_0}{sin\theta}$, in the grand canonical ensemble, Eq.\eqref{Psi_t_0_grand_canonical}, to $d$-dimensional spacetime is expressed as,
\begin{equation}
\Psi(r,\theta)=\frac{\displaystyle\Gamma \left(\frac{d-1}{2}\right) \csc (\theta ) r\left[2 (d-3) \phi ^2-d+2\right]}{\pi ^{\frac{d+1}{2}} r^d-4 \pi  r^2 \Gamma \left(\frac{d-1}{2}\right)}
\end{equation}
Similarly, its gradient vector field $\psi$ in the $(r,\theta)$-plan reads,
\begin{equation}\label{topological_field_psi_grand_canonical_d}
\begin{cases}
\psi_r=\partial_r\Psi=\displaystyle\frac{\pi ^{-d-4} \left(2 (d-3) \phi ^2-d+2\right) \Gamma \left(\frac{d-1}{2}\right) \csc (\theta ) \pi ^{\frac{d+3}{2}}\left(4 \pi ^{\frac{d+5}{2}} r^2 \Gamma \left(\frac{d-1}{2}\right)-(d-1) \pi ^{d+2} r^d\right)}{\left(\pi ^{d/2} r^d-4 \sqrt{\pi } r^2 \Gamma \left(\frac{d-1}{2}\right)\right)^2},\\\\
\psi_{\theta}=\partial_{\theta}\Psi=-\displaystyle\frac{\Gamma \left(\frac{d-1}{2}\right) \cos (\theta ) r\left[2 (d-3) \phi ^2-d+2\right]}{\pi ^{\frac{d+1}{2}} r^d-4 \pi  r^2 \Gamma \left(\frac{d-1}{2}\right)\sin^{2}{\left(\theta \right)}}.
\end{cases}
\end{equation}
The generalization of the vector field $\eta$ to arbitrary dimensions follows,
\begin{equation}\label{topological_field_xi_grand_canonical_d}
\begin{cases}
\eta_r=\partial_r\Bar{f}_R\approx\displaystyle \frac{- 4 \pi \Omega_{d} r^{d - 1} t \left(d - 2\right)^{2} + r^{d - 2} \left(d - 3\right) \left(\Omega_{d} \left(d - 2\right) + 4 \pi r^{d - 2}\right) \left(d + 2 \phi^{2}  \left(3 - d\right) - 2\right)}{\Omega_{d} r^{2} \left(d - 2\right) \left(\Omega_{d} \left(d - 2\right) + 4 \pi r^{d - 2}\right)},\\\\
\eta_{\theta}=\displaystyle-\cot\theta \cos\theta.
\end{cases}
\end{equation}%
In the canonical ensemble, we consider the expression of the  extensivity parameter $\lambda$ as a function of the Rényi pressure $p$, using Eq.\eqref{BH_Q}, we give here the generalization of the relation Eq.\eqref{lam_p} in $d$-dimensional spacetime\cite{Barzi:2022ygr} as,
\begin{equation}
    \lambda=\frac{16(d-2)\Omega_{d}\:r_{h}^{d - 2} \:p}{(d-1)(d-3)\left[\displaystyle (d-2)r_h^{2(d-3)}-2(d-3) Q^{2}\right]}.
\end{equation}
\bibliographystyle{unsrt}
\bibliography{RHPtopology.bib} 

\begin{thebibliography}{10}

\bibitem{Yerra:2022coh}
Pavan~Kumar Yerra, Chandrasekhar Bhamidipati, and Sudipta Mukherji.
\newblock {Topology of critical points and Hawking-Page transition}.
\newblock {\em Phys. Rev. D}, 106(6):064059, 2022.

\bibitem{Wei:2022dzw}
Shao-Wen Wei, Yu-Xiao Liu, and Robert~B. Mann.
\newblock {Black Hole Solutions as Topological Thermodynamic Defects}.
\newblock {\em Phys. Rev. Lett.}, 129(19):191101, 2022.

\bibitem{Hawking:1976de}
S.~W. Hawking.
\newblock {Black Holes and Thermodynamics}.
\newblock {\em Phys. Rev. D}, 13:191--197, 1976.

\bibitem{Hawking:1982dh}
S.~W. Hawking and Don~N. Page.
\newblock {Thermodynamics of Black Holes in anti-De Sitter Space}.
\newblock {\em Commun. Math. Phys.}, 87:577, 1983.

\bibitem{Bardeen:1973gs}
James~M. Bardeen, B.~Carter, and S.~W. Hawking.
\newblock {The Four laws of black hole mechanics}.
\newblock {\em Commun. Math. Phys.}, 31:161--170, 1973.

\bibitem{Bekenstein:1973ur}
Jacob~D. Bekenstein.
\newblock {Black holes and entropy}.
\newblock {\em Phys. Rev. D}, 7:2333--2346, 1973.

\bibitem{Hawking:1975vcx}
S.~W. Hawking.
\newblock {Particle Creation by Black Holes}.
\newblock {\em Commun. Math. Phys.}, 43:199--220, 1975.
\newblock [Erratum: Commun.Math.Phys. 46, 206 (1976)].

\bibitem{Bekenstein:1974ax}
Jacob~D. Bekenstein.
\newblock {Generalized second law of thermodynamics in black hole physics}.
\newblock {\em Phys. Rev. D}, 9:3292--3300, 1974.

\bibitem{Srednicki:1993im}
Mark Srednicki.
\newblock {Entropy and area}.
\newblock {\em Phys. Rev. Lett.}, 71:666--669, 1993.

\bibitem{Crossley:2014oea}
Michael Crossley, Ethan Dyer, and Julian Sonner.
\newblock {Super-R\'enyi entropy \& Wilson loops for $ \mathcal{N}=4 $ SYM and
  their gravity duals}.
\newblock {\em JHEP}, 12:001, 2014.

\bibitem{DeHaro:2019gno}
Sebastian De~Haro, Jeroen van Dongen, Manus Visser, and Jeremy Butterfield.
\newblock {Conceptual analysis of black hole entropy in string theory}.
\newblock {\em Stud. Hist. Phil. Sci. B}, 69:82--111, 2020.

\bibitem{Tsallis:2019giw}
Constantino Tsallis.
\newblock {Black Hole Entropy: A Closer Look}.
\newblock {\em Entropy}, 22(1):17, 2019.

\bibitem{Alonso-Serrano:2020hpb}
Ana Alonso-Serrano, Mariusz~P. Dabrowski, and Hussain Gohar.
\newblock {Nonextensive Black Hole Entropy and Quantum Gravity Effects at the
  Last Stages of Evaporation}.
\newblock {\em Phys. Rev. D}, 103(2):026021, 2021.

\bibitem{Headrick:2015gba}
Matthew Headrick, Alexander Maloney, Eric Perlmutter, and Ida~G. Zadeh.
\newblock {R\'enyi entropies, the analytic bootstrap, and 3D quantum gravity at
  higher genus}.
\newblock {\em JHEP}, 07:059, 2015.

\bibitem{Hung:2011nu}
Ling-Yan Hung, Robert~C. Myers, Michael Smolkin, and Alexandre Yale.
\newblock {Holographic Calculations of Renyi Entropy}.
\newblock {\em JHEP}, 12:047, 2011.

\bibitem{Giveon:2015cgs}
Amit Giveon and David Kutasov.
\newblock {Supersymmetric Renyi entropy in CFT$_{2}$ and AdS$_{3}$}.
\newblock {\em JHEP}, 01:042, 2016.

\bibitem{Dong:2016fnf}
Xi~Dong.
\newblock {The Gravity Dual of Renyi Entropy}.
\newblock {\em Nature Commun.}, 7:12472, 2016.

\bibitem{Hirunsirisawat:2022fsb}
Ekapong Hirunsirisawat, Ratchaphat Nakarachinda, and Chatchai Promsiri.
\newblock {Emergent phase, thermodynamic geometry, and criticality of charged
  black holes from R\'enyi statistics}.
\newblock {\em Phys. Rev. D}, 105(12):124049, 2022.

\bibitem{Promsiri:2021hhv}
Chatchai Promsiri, Ekapong Hirunsirisawat, and Watchara Liewrian.
\newblock {Solid-liquid phase transition and heat engine in an asymptotically
  flat Schwarzschild black hole via the R\'enyi extended phase space approach}.
\newblock {\em Phys. Rev. D}, 104(6):064004, 2021.

\bibitem{Nojiri:2021czz}
Shin'ichi Nojiri, Sergei~D. Odintsov, and Valerio Faraoni.
\newblock {Area-law versus R\'enyi and Tsallis black hole entropies}.
\newblock {\em Phys. Rev. D}, 104(8):084030, 2021.

\bibitem{Barzi:2022ygr}
Faical Barzi and Hasan El~Moumni.
\newblock {On R\'enyi universality formula of charged flat black holes from
  Hawking-Page phase transition}.
\newblock {\em Phys. Lett. B}, 833:137378, 2022.

\bibitem{Cimdiker:2022ics}
Ilim \c{C}imdiker, Mariusz~P. Dabrowski, and Hussain Gohar.
\newblock {Equilibrium temperature for black holes with nonextensive entropy}.
\newblock {\em Eur. Phys. J. C}, 83(2):169, 2023.

\bibitem{Barzi:2023mit}
F.~Barzi, H.~El~Moumni, and K.~Masmar.
\newblock {On some phase equilibrium features of charged black holes in flat
  spacetime via R\'enyi statistics}.
\newblock {\em arXiv:2304.04945}, 4 2023.

\bibitem{Cimidiker:2023kle}
Ilim Cimidiker, Mariusz~P. Dabrowski, and Hussain Gohar.
\newblock {Generalized uncertainty principle impact on nonextensive black hole
  thermodynamics}.
\newblock {\em Class. Quant. Grav.}, 40(14):145001, 2023.

\bibitem{Chunaksorn:2022whl}
Phuwadon Chunaksorn, Ekapong Hirunsirisawat, Ratchaphat Nakarachinda,
  Lunchakorn Tannukij, and Pitayuth Wongjun.
\newblock {Thermodynamics of asymptotically de Sitter black hole in dRGT
  massive gravity from R\'enyi entropy}.
\newblock {\em Eur. Phys. J. C}, 82(12):1174, 2022.

\bibitem{Samart:2020klx}
Daris Samart and Phongpichit Channuie.
\newblock {AdS to dS phase transition mediated by thermalon in
  Einstein-Gauss-Bonnet gravity from R\'enyi statistics}.
\newblock {\em Nucl. Phys. B}, 989:116140, 2023.

\bibitem{Gunasekaran:2012dq}
Sharmila Gunasekaran, Robert~B. Mann, and David Kubiznak.
\newblock {Extended phase space thermodynamics for charged and rotating black
  holes and Born-Infeld vacuum polarization}.
\newblock {\em JHEP}, 11:110, 2012.

\bibitem{Kubiznak:2012wp}
David Kubiznak and Robert~B. Mann.
\newblock {P-V criticality of charged AdS black holes}.
\newblock {\em JHEP}, 07:033, 2012.

\bibitem{Belhaj:2015hha}
A.~Belhaj, M.~Chabab, H.~El~Moumni, K.~Masmar, M.~B. Sedra, and A.~Segui.
\newblock {On Heat Properties of AdS Black Holes in Higher Dimensions}.
\newblock {\em JHEP}, 05:149, 2015.

\bibitem{Kubiznak:2016qmn}
David Kubiznak, Robert~B. Mann, and Mae Teo.
\newblock {Black hole chemistry: thermodynamics with Lambda}.
\newblock {\em Class. Quant. Grav.}, 34(6):063001, 2017.

\bibitem{Kubiznak:2014zwa}
David Kubiznak and Robert~B. Mann.
\newblock {Black hole chemistry}.
\newblock {\em Can. J. Phys.}, 93(9):999--1002, 2015.

\bibitem{Wu:2022plw}
Jerry Wu and Robert~B. Mann.
\newblock {Multicritical phase transitions in Lovelock AdS black holes}.
\newblock {\em Phys. Rev. D}, 107(8):084035, 2023.

\bibitem{Altamirano:2013ane}
Natacha Altamirano, David Kubiznak, and Robert~B. Mann.
\newblock {Reentrant phase transitions in rotating anti\textendash{}de Sitter
  black holes}.
\newblock {\em Phys. Rev. D}, 88(10):101502, 2013.

\bibitem{Frassino:2014pha}
Antonia~M. Frassino, David Kubiznak, Robert~B. Mann, and Fil Simovic.
\newblock {Multiple Reentrant Phase Transitions and Triple Points in Lovelock
  Thermodynamics}.
\newblock {\em JHEP}, 09:080, 2014.

\bibitem{Bravetti:2012hd}
Alessandro Bravetti, Davood Momeni, Ratbay Myrzakulov, and Hernando Quevedo.
\newblock {Geometrothermodynamics of higher dimensional black holes}.
\newblock {\em Gen. Rel. Grav.}, 45:1603--1617, 2013.

\bibitem{Chabab:2019mlu}
M.~Chabab, H.~El~Moumni, S.~Iraoui, and K.~Masmar.
\newblock {Phase transitions and geothermodynamics of black holes in dRGT
  massive gravity}.
\newblock {\em Eur. Phys. J. C}, 79(4):342, 2019.

\bibitem{Bhattacharya:2017hfj}
Krishnakanta Bhattacharya and Bibhas~Ranjan Majhi.
\newblock {Thermogeometric description of the van der Waals like phase
  transition in AdS black holes}.
\newblock {\em Phys. Rev. D}, 95(10):104024, 2017.

\bibitem{Liu:2014gvf}
Yunqi Liu, De-Cheng Zou, and Bin Wang.
\newblock {Signature of the Van der Waals like small-large charged AdS black
  hole phase transition in quasinormal modes}.
\newblock {\em JHEP}, 09:179, 2014.

\bibitem{Chabab:2016cem}
M.~Chabab, H.~El~Moumni, S.~Iraoui, and K.~Masmar.
\newblock {Behavior of quasinormal modes and high dimension RN\textendash{}AdS
  black hole phase transition}.
\newblock {\em Eur. Phys. J. C}, 76(12):676, 2016.

\bibitem{Simovic:2019zgb}
Fil Simovic and Robert~B. Mann.
\newblock {Critical Phenomena of Born-Infeld-de Sitter Black Holes in
  Cavities}.
\newblock {\em JHEP}, 05:136, 2019.

\bibitem{Chabab:2020xwr}
M.~Chabab, H.~El~Moumni, and J.~Khalloufi.
\newblock {On Einstein-non linear-Maxwell-Yukawa de-Sitter black hole
  thermodynamics}.
\newblock {\em Nucl. Phys. B}, 963:115305, 2021.

\bibitem{Ali:2020bgc}
Md~Sabir Ali, Sushant~G. Ghosh, and Sunil~D. Maharaj.
\newblock {Effective thermodynamics and critical phenomena of rotating
  regular-de Sitter black holes}.
\newblock {\em Class. Quant. Grav.}, 37(18):185003, 2020.

\bibitem{Li:2020khm}
Ran Li and Jin Wang.
\newblock {Thermodynamics and kinetics of Hawking-Page phase transition}.
\newblock {\em Phys. Rev. D}, 102(2):024085, 2020.

\bibitem{Ali:2023wkq}
Md~Sabir Ali, Hasan El~Moumni, Jamal Khalloufi, and Karima Masmar.
\newblock {Born-Infeld-AdS black hole phase structure: Landau theory and free
  energy landscape approaches}.
\newblock {\em 2303.11711}, 3 2023.

\bibitem{Li:2020nsy}
Ran Li, Kun Zhang, and Jin Wang.
\newblock {Thermal dynamic phase transition of Reissner-Nordstr\"om Anti-de
  Sitter black holes on free energy landscape}.
\newblock {\em JHEP}, 10:090, 2020.

\bibitem{Nguyen:2015wfa}
Phuc~H. Nguyen.
\newblock {An equal area law for holographic entanglement entropy of the AdS-RN
  black hole}.
\newblock {\em JHEP}, 12:139, 2015.

\bibitem{ElMoumni:2018fml}
H.~El~Moumni.
\newblock {Revisiting the phase transition of
  AdS-Maxwell\textendash{}power-Yang\textendash{}Mills black holes via AdS/CFT
  tools}.
\newblock {\em Phys. Lett. B}, 776:124--132, 2018.

\bibitem{Li:2018aax}
Shou-Long Li and Hao Wei.
\newblock {Holographic Entanglement Entropy and Van der Waals transitions in
  Einstein-Maxwell-Dilaton theory}.
\newblock {\em Phys. Rev. D}, 99(6):064002, 2019.

\bibitem{Duan:1984ws}
Yishi Duan.
\newblock {THE STRUCTURE OF THE TOPOLOGICAL CURRENT}.
\newblock {\em SLAC-PUB-3301}, 3 1984.

\bibitem{Duan:1998kw}
Yi-Shi Duan, Sheng Li, and Guo-Hong Yang.
\newblock {The bifurcation theory of the Gauss-Bonnet-Chern topological current
  and Morse function}.
\newblock {\em Nucl. Phys. B}, 514:705--720, 1998.

\bibitem{Fu:2000pb}
Li-Bin Fu, Yi-Shi Duan, and Hong Zhang.
\newblock {Evolution of the Chern-Simons vortices}.
\newblock {\em Phys. Rev. D}, 61:045004, 2000.

\bibitem{Guo:2020qwk}
Minyong Guo and Sijie Gao.
\newblock {Universal Properties of Light Rings for Stationary Axisymmetric
  Spacetimes}.
\newblock {\em Phys. Rev. D}, 103(10):104031, 2021.

\bibitem{Bargueno:2022vkf}
Pedro Bargue\~no.
\newblock {Light rings in static and extremal black holes}.
\newblock {\em Phys. Rev. D}, 107(10):104029, 2023.

\bibitem{Cunha:2017qtt}
Pedro V.~P. Cunha, Emanuele Berti, and Carlos A.~R. Herdeiro.
\newblock {Light-Ring Stability for Ultracompact Objects}.
\newblock {\em Phys. Rev. Lett.}, 119(25):251102, 2017.

\bibitem{Cunha:2020azh}
Pedro V.~P. Cunha and Carlos A.~R. Herdeiro.
\newblock {Stationary black holes and light rings}.
\newblock {\em Phys. Rev. Lett.}, 124(18):181101, 2020.

\bibitem{Fan:2022bsq}
Zhong-Ying Fan.
\newblock {Topological interpretation for phase transitions of black holes}.
\newblock {\em Phys. Rev. D}, 107(4):044026, 2023.

\bibitem{Fang:2022rsb}
Chaoxi Fang, Jie Jiang, and Ming Zhang.
\newblock {Revisiting thermodynamic topologies of black holes}.
\newblock {\em JHEP}, 01:102, 2023.

\bibitem{Wei:2021vdx}
Shao-Wen Wei and Yu-Xiao Liu.
\newblock {Topology of black hole thermodynamics}.
\newblock {\em Phys. Rev. D}, 105(10):104003, 2022.

\bibitem{Yerra:2022alz}
Pavan~Kumar Yerra and Chandrasekhar Bhamidipati.
\newblock {Topology of black hole thermodynamics in Gauss-Bonnet gravity}.
\newblock {\em Phys. Rev. D}, 105(10):104053, 2022.

\bibitem{Ahmed:2022kyv}
Moaathe~Belhaj Ahmed, David Kubiznak, and Robert~B. Mann.
\newblock {Vortex-antivortex pair creation in black hole thermodynamics}.
\newblock {\em Phys. Rev. D}, 107(4):046013, 2023.

\bibitem{Ali:2023zww}
Md~Sabir Ali, Hasan El~Moumni, Jamal Khalloufi, and Karima Masmar.
\newblock {Topology of Born-Infeld-AdS Black Hole Phase Transition}.
\newblock {\em arXiv:2306.11212}, 6 2023.

\bibitem{Sadeghi:2023aii}
Jafar Sadeghi, Saeed Noori~Gashti, Mohammad~Reza Alipour, and Mohammad Ali~S.
  Afshar.
\newblock {Bardeen black hole thermodynamics from topological perspective}.
\newblock {\em Annals Phys.}, 455:169391, 2023.

\bibitem{Gogoi:2023qku}
Naba~Jyoti Gogoi and Prabwal Phukon.
\newblock {Topology of thermodynamics in R-charged black holes}.
\newblock {\em Phys. Rev. D}, 107(10):106009, 2023.

\bibitem{Zhang:2023uay}
Ming Zhang and Jie Jiang.
\newblock {Bulk-boundary thermodynamic equivalence: a topology viewpoint}.
\newblock {\em JHEP}, 06:115, 2023.

\bibitem{Wu:2023sue}
Di~Wu and Shuang-Qing Wu.
\newblock {Topological classes of thermodynamics of rotating AdS black holes}.
\newblock {\em Phys. Rev. D}, 107(8):084002, 2023.

\bibitem{Wu:2022whe}
Di~Wu.
\newblock {Topological classes of rotating black holes}.
\newblock {\em Phys. Rev. D}, 107(2):024024, 2023.

\bibitem{Bai:2022klw}
Ning-Chen Bai, Lei Li, and Jun Tao.
\newblock {Topology of black hole thermodynamics in Lovelock gravity}.
\newblock {\em Phys. Rev. D}, 107(6):064015, 2023.

\bibitem{Liu:2022aqt}
Conghua Liu and Jin Wang.
\newblock {Topological natures of the Gauss-Bonnet black hole in AdS space}.
\newblock {\em Phys. Rev. D}, 107(6):064023, 2023.

\bibitem{Chatzifotis2023}
Nikos Chatzifotis, Panagiotis Dorlis, Nick~E. Mavromatos, and Eleftherios
  Papantonopoulos.
\newblock Thermal stability of hairy black holes.
\newblock {\em Physical Review D}, 107:084053, 4 2023.

\bibitem{Maldacena:1997re}
Juan~Martin Maldacena.
\newblock {The Large N limit of superconformal field theories and
  supergravity}.
\newblock {\em Adv. Theor. Math. Phys.}, 2:231--252, 1998.

\bibitem{Witten:1998qj}
Edward Witten.
\newblock {Anti-de Sitter space and holography}.
\newblock {\em Adv. Theor. Math. Phys.}, 2:253--291, 1998.

\bibitem{Witten:1998zw}
Edward Witten.
\newblock {Anti-de Sitter space, thermal phase transition, and confinement in
  gauge theories}.
\newblock {\em Adv. Theor. Math. Phys.}, 2:505--532, 1998.

\bibitem{Cai:2007wz}
Rong-Gen Cai, Sang~Pyo Kim, and Bin Wang.
\newblock {Ricci flat black holes and Hawking-Page phase transition in
  Gauss-Bonnet gravity and dilaton gravity}.
\newblock {\em Phys. Rev. D}, 76:024011, 2007.

\bibitem{Gursoy:2010jh}
Umut Gursoy.
\newblock {Continuous Hawking-Page transitions in Einstein-scalar gravity}.
\newblock {\em JHEP}, 01:086, 2011.

\bibitem{Zhang:2015wna}
Shao-Jun Zhang.
\newblock {Hawking\textendash{}Page phase transition in new massive gravity}.
\newblock {\em Phys. Lett. B}, 747:158--163, 2015.

\bibitem{Adams:2014vza}
Allan Adams, Daniel~A. Roberts, and Omid Saremi.
\newblock {Hawking-Page transition in holographic massive gravity}.
\newblock {\em Phys. Rev. D}, 91(4):046003, 2015.

\bibitem{Banados:2016hze}
M\'aximo Ba\~nados, Gustavo D\"uring, Alberto Faraggi, and Ignacio Reyes.
\newblock {Phases of higher spin black holes: Hawking-Page, transitions between
  black holes and a critical point}.
\newblock {\em Phys. Rev. D}, 96(4):046017, 2017.

\bibitem{Su:2019gby}
Bing-Yu Su, Yuan-Yuan Wang, and Nan Li.
\newblock {The Hawking\textendash{}Page phase transitions in the extended phase
  space in the Gauss\textendash{}Bonnet gravity}.
\newblock {\em Eur. Phys. J. C}, 80(4):305, 2020.

\bibitem{Promsiri:2020jga}
Chatchai Promsiri, Ekapong Hirunsirisawat, and Watchara Liewrian.
\newblock {Thermodynamics and Van der Waals phase transition of charged black
  holes in flat spacetime via R\'enyi statistics}.
\newblock {\em Phys. Rev. D}, 102(6):064014, 2020.

\bibitem{dilaton}
Hasan El~Moumni, Karima Masmar, and Safaa Mazzou.
\newblock Critical phenomena of charged dilatonic black holes through r{\'e}nyi
  statistics approach.
\newblock {\em International Journal of Modern Physics D}, 31(5):22500407, 3
  2022.

\bibitem{Majhi:2016txt}
Bibhas~Ranjan Majhi and Saurav Samanta.
\newblock {P-V criticality of AdS black holes in a general framework}.
\newblock {\em Phys. Lett. B}, 773:203--207, 2017.

\bibitem{Li:2014ixn}
Gu-Qiang Li.
\newblock {Effects of dark energy on P\textendash{}V criticality of charged AdS
  black holes}.
\newblock {\em Phys. Lett. B}, 735:256--260, 2014.

\bibitem{DEMAMI2023116316}
F.~Demami, H.~{El Moumni}, K.~Masmar, and S.~Mazzou.
\newblock Thermodynamics and phase transition structure of charged black holes
  in $f(r)$ gravity background from r{\'e}nyi statistics perception.
\newblock {\em Nuclear Physics B}, 994:116316, 2023.

\bibitem{PhysRevE.83.061147}
T.~S. Bir\'o and P.~V\'an.
\newblock Zeroth law compatibility of nonadditive thermodynamics.
\newblock {\em Phys. Rev. E}, 83:061147, Jun 2011.

\bibitem{Wei2022}
Shao~Wen Wei, Yu~Xiao Liu, and Robert~B. Mann.
\newblock {Black Hole Solutions as Topological Thermodynamic Defects}.
\newblock {\em Physical Review Letters}, 129(19):1--7, 2022.

\bibitem{Barzi:2023zed}
F~Barzi Id, H~El, Moumni Id, and K~{Masmar Id}.
\newblock {On some phase equilibrium features of charged black holes in flat
  spacetime via R{\'{e}}nyi statistics}.
\newblock {\em arXiv:2304.04945v1}, 2023.

\end{thebibliography}
\end{document}